\documentclass[traditabstract, twocolumn]{aa} 

\usepackage{graphicx}
\usepackage{natbib}
\usepackage{morefloats}
\usepackage{txfonts}
\usepackage{color}

\usepackage{hyperref}

\newcommand{\ecs}{erg cm$^{-3}$ s$^{-1}$}
\newcommand{\inten}{erg cm$^{-2}$ s$^{-1}$}

\newcommand{\cmt}{cm$^{-3}$}
\newcommand{\cms}{cm$^{-2}$}
\newcommand{\msun}{M$_\odot$}
\newcommand{\modotyr}{M$_\odot$~yr$^{-1}$}
\newcommand{\gm}{$\Gamma_{\rm mech}$}
\newcommand{\gsu}{$\Gamma_{\rm surf}$}
\newcommand{\go}{$G_0$}
\newcommand{\zsun}{$Z_\odot$}
\newcommand{\mum}{$\mu$m}
\newcommand{\xcp}{$x_{{\rm C}^+}$}
\newcommand{\xc}{$x_{{\rm C}}$}
\newcommand{\xo}{$x_{{\rm O}}$}
\newcommand{\xco}{$x_{\rm CO}$}
\newcommand{\xhcop}{$x_{\rm HCO^+}$}
\newcommand{\thco}{$^{13}$CO}
\newcommand{\NH}{N_{\rm H}}

\usepackage{epsfig}

\begin{document}


   \title{Diagnostics of the molecular component of PDRs with mechanical heating II: line intensities and ratios }

   \subtitle{}

   \author{M. V. Kazandjian\inst{1}, R. Meijerink\inst{2,1}, I. Pelupessy\inst{1}, F. P. Israel\inst{1}, M. Spaans\inst{2}}

   \institute{Sterrewacht Leiden, PO Box 9513, 2300 RA Leiden, The Netherlands\\
     \email{mher@strw.leidenuniv.nl}
       \and 
       Kapteyn Astronomical Institute, PO Box 800, 9700 AV Groningen, The Netherlands
     }

   \date{Received Xxxxxxxxx xx, xxxx; accepted Xxxxx xx, xxxx}

   \abstract{ CO observations in active galactic nuclei and
     star-bursts reveal high kinetic temperatures.  Those environments
     are thought to be very turbulent due to
     dynamic phenomena such as outflows and high supernova rates.\\
     We investigate the effect of mechanical heating on atomic
     fine-structure and molecular lines, and their ratios.  We try to
     use those ratios as a diagnostic to constrain the amount of
     mechanical heating in an object and also study its significance
     on
     estimating the H$_2$ mass.\\
     Equilibrium photo-dissociation models (PDRs hereafter) were used
     to compute the thermal and chemical balance for the clouds.  The
     equilibria were solved for numerically using the optimized
     version of the Leiden PDR-XDR code.  Large velocity gradient
     calculations
     were done as post-processing on the output of the PDR models using RADEX.\\
     High-$J$ CO line ratios are very sensitive to mechanical heating
     (\gm~hereafter). Emission becomes at least one order of magnitude
     brighter in clouds with $n \sim 10^5$~\cmt~and a star formation
     rate of 1~\modotyr~(corresponding to \gm~$ = 2 \times
     10^{-19}$~\ecs).  Emission of low-$J$ CO lines is not as
     sensitive to \gm, but they do become brighter in response to \gm.
     Generally, for all of the lines we considered, \gm~increases
     excitation temperatures and decreases the optical depth at the
     line centre.  Hence line ratios are also affected, strongly in
     some cases.  Ratios involving HCN are a good diagnostic for \gm,
     where the HCN(1-0)/CO(1-0) increases from 0.06 to 0.25 and the
     HCN(1-0)/HCO$^+$(1-0) increase from 0.15 to 0.5 for amounts of
     {\gm~equivelent to 5\% of the surface heating rate}.  Both ratios
     increase to more than 1 for higher \gm, as opposed to
     being much less than unity in pure PDRs.\\
     The first major conclusion is that low-$J$ to high-$J$ intensity
     ratios will yield a good estimate of the mechanical heating rate
     (as opposed to only low-$J$ ratios). The second one is that the
     mechanical heating rate should be taken into account when
     determing $A_V$ or equivalently $N_{\rm H}$, and consequently the
     cloud mass. Ignoring \gm~will also lead to large errors in
     density and radiation field estimates.

}
\authorrunning{Kazandjian {\it et. al}}
\titlerunning{Line Diagnostics of PDRs with Mechanical Heating}
\maketitle
\section{Introduction}
The study of molecular gas in external galaxies dates back to the
mid-seventies, with the detection of ground-state emission from CO
(the most abundant {molecules} after hydrogen) in {a small number of 
bright nearby galaxies.  At present observations CO and many other 
molecules exist for a very large number of galaxies, near and far. It 
is important to be able to interpret the emission in the various
lines from those galaxies, since that gives us insight
in the physics dominating the interstellar medium in the star forming 
regions of these extra-galactic sources.}

For decades, line observations had to be done from the ground in a frequency
range limited by atmospheric opacity, so that for most molecular
species only the {low} transitions were accessible.  {Level transitions at 
higher rest frame frequencies, were only possible for 
distant galaxies for which the high-frequency lines were
red-shifted into atmospheric windows accessible from the ground.
In the past few years the {\it Herschel} Space 
Observatory \citep{Pilbratt2010} operating outside the Earth's 
atmosphere has provided direct observations of spectral lines at 
frequencies hitherto impossible or hard to access.

By way of example we mention the determination of extensive $^{12}$CO rotational 
transitions ladder in galaxies such as M82 
\citep{loenen10, panuzzo2010, kamentzky2012} and Mrk231 \citep{vanderwerf10, gozalez12}.  
{\it Herschel} ran out of coolant in April 2013, but at about the same
time, the Atacama Large Millimeter Array (ALMA) became
operational. With ALMA, a large fraction of the important
submillimeter spectrum is still accessible, at vastly superior
resolution and sensitivity, allowing detection and measurement of
diagnostic molecular line transitions largely out of reach until then.}

{Conducting} detailed studies of the physical properties of the molecular gas {of 
close-by star-forming galaxies} involves a challenging inversion problem, where 
resultant line intensities are used to constrain gas densities, molecular content, kinetic 
temperatures and the nature and strengths of the radiation field exciting the gas.
{In order to solve} this problem, it is necessary to get a clearer understanding of
the underlying physics and phenomena characterizing specific regions
such as galaxy centres, including our own.

A good starting point to analyze molecular gas emission is the
application of the so-called large-velocity-gradient (LVG) models {
\citep{sobolev1960}.  This assumes an escape probability formalism for photons in different geometries which 
simplifies solving for the radiative transfer significantly.  
LVG models have been widely used by the ISM community with some other basic assumptions to 
estimate the molecular density of the gas, species abundances and the kinetic 
temperature \citep[][among others]{henkel83, Jansen1994, Hogerheijde2000, radex, despotic}.
These models provide only an insight for the physical and chemical properties of the clouds;
the actual nature of the source of energy cannot be determined using those LVG models, see for 
example \cite{Israel2009a,Israel2009b}.  The next level of sophistication over 
LVG modelling involves the application of photon-dominated region (PDR) models \citep{cloudy, Hollenbach1999, 
petit06, rolling07, bisbas12}.  These models self-consistently solve for the thermal and chemical 
structure of clouds irradiated by UV photons.

In PDRs, energy sources other than UV photons could dominate the thermal and chemical balance. In the vicinity of 
an active galactic nucleus (AGN), PDR models can be augmented by models for X-ray dominated region
\citep[XDRs][]{Maloney1996,bradford03,meijerink2005-1,Papadopoulos2011, bayet11-1, meijerink11}.}

In both these models, the underlying assumption is that the thermal balance is
dominated by radiation.  The physical situation in galaxy centers, star-bursts and 
dense cores \citep{pineda10} is, however, more complicated.  There are other processes, such as 
mechanical feedback that may also excite the gas mechanically \citep{loenen2008, Ossenkopf02-a, 
Ossenkopf02-b, mvk12}.

{Although these models are much more sophisticated than LVG models, a simplified comparison of many 
PDR codes\footnote{In the comparison benchmark the chemistry involved 4 elements (H, He, O and C) and 30 species.  For more details on the benchmarks and the codes used see \texttt{http://goo.gl/7Hf6mD}}
\citep{rolling07} has shown that they shed a statistical view on the underlying
processes. This is particularly true in the transition zone from atomic 
to molecular gas, where an order of magnitude difference between the various quantities 
compared in the models is not uncommon. Such discrepancies are mainly due to the uncertainties
 in the chemical reaction constants, which in turn influence the reaction rates, abundances and 
thermal balance \citep{vasyunin04}.  In addition to those uncertainties observations of extra-galactic 
sources have spatial resolution limitations. For example the resolution of {\it Herschel} for the 
nucleus of NGC 253 is on the order of 1 kpc.  The surface area covered by such a beam size contains 
a large number of clouds.  In modelling the nucleus of such galaxies one might need to consider 
two or more PDRs simultaneously.
Although considering more than one PDR component improves the fits significantly, the increased
number of free parameters usually has a negative impact on to the statistical significance 
of those fits.  This is particularly valid whenever the number of lines being fitted is low.

Here we follow the modelling of paper I \citep{mvk12} where we studied the effect 
of mechanical heating (\gm) by considering its impact on the thermal and 
chemical structure (abundances, column densities and column density ratios of species) of
PDRs.  Hence our basic modelling premise will be the same in this paper.  Namely, 
an 1D semi-infinite plane-parallel geometry is adopted. It is assumed that the slab
is illuminated with a FUV source from one side.  Another major assumption is that the 
clouds are in an equilibrium state.  Since equilibrium is assumed, we consider a 
simplified recipe in accounting for mechanical feedback. For simplicity the contribution 
of mechanical heating to the total heating budget, is added in an ad-hoc fashion
 uniformly throughout the cloud.

Our approximation of the effect of mechanical heating by a single homogeneous heating term is a
simplification.  In practice, the mechanical energy which could be liberated by supernova events 
or gas outflows, is deposited locally in shock fronts.  In these fronts, which are
the interaction surfaces between high speed flows and the ambient medium, the energy is not 
necessarily distributed uniformly throughout the cloud volume.  On the other hand, this energy 
will eventually cascade to smaller spatial scales and thermalize en route to equilibrium.  The 
efficiency of the ``thermalization'' is conservatively taken to be 10\%.  Consequently the 
approximation we adopt may be less applicable to systems where the dynamical time-scales are 
comparable to the thermal and chemical time-scales; this occurs for example in clouds in the inner 
kpc of galaxy centres.  Our choice for the ranges in mechanical heating explored is based primarily on 
estimates by \cite{loenen2008}. They found that fits for the line ratios of the first rotational 
transition ($J = 1-0$) of the molecules HCN, HNC and HCO$^+$ are greatly enhanced by using PDR models 
which included ``additional'' heating. They attributed this extra heating to dissipated turbulence 
and provided an estimates for it.
The major conclusion of Paper I was that even small amounts of mechanical heating, as low
as 1\% of the surface UV heating, has significant effects on the molecular abundances
and column densities.  Those effects are mainly manifested as enhanced CO abundances which
could increase by up to a factor of two. Although this might not seem a significant effect, the column densities 
of the high density tracers such as HCN and HNC increase (or decrease) by an order of magnitude
depending on the amount of \gm.}

The aim of this paper is to understand both the ground-state and the
more highly excited states of molecular gas in galaxy centers and to
determine whether turbulence or shocks can make a major contribution
to the molecular emission. {In other words, we thus extend the work
  done in Paper I, which focused on the chemical abundances and column
  densities only, by studying the signature of mechanical feedback on
  selected atomic and molecular emission lines.}

The models presented in this paper also apply 
to other regions where the gas is, e.g., (1) heated by young stellar objects 
(YSO's), (2) stirred up turbulently by the fast motions of stars, or (3) violently 
heated by supernovae.  {Since we assume equilibrium, applying our models to those 
regions is of course an approximation}. In all cases,
non-negligible amounts of mechanical energy may be {eventually} injected into the
ISM \citep{sb99}. Part of this energy is then converted to {mechanical} heating,
particularly important in so-called star-burst galaxies.

Since the amount of turbulent energy absorbed by the ISM is a priori
unknown, we explore a wide range of possibilities of turbulent heating
{contributions} to PDRs. In our approach, the additional heating
self-consistently modifies the emission.  {In this paper we
provide two new estimates of mechanical heating rates and re-enforce
our assumptions of Paper I (see the methods section below).}

Although we introduce an extra free parameter (the amount of absorbed 
turbulent energy), the basic molecular {abundances and} gas parameters are 
self-consistently determined by the equilibria that we solve for in the PDR models.  In the 
following we explore, {using those 1D equilibrium PDR models}, the effect of mechanical 
heating (\gm) on atomic and molecular line intensities for a range of : densities, 
FUV flux (\go), metallicities and column densities. {In doing so we} aim to find good 
diagnostics for \gm~{and to check for the usefulness of 
such PDR models with an additional ``ad-hoc'' heating term in constraining mechanical heating.}

\section{Methods \label{sec:methods}}

{A PDR is primarily characterized by its gas number density, the FUV
  flux of the environment and its spatial extent (usually measured in
  $A_V$ or alternatively the column density of H).} The conversion of
$A_V$ to $N_{\rm H}$ is given by $N_{\rm H} = 1.87\cdot 10^{21} A_V
(Z_\odot/Z)$~cm$^{-2}$. In PDRs, the main heating sources are
typically the {FUV} photons irradiating the cloud surface.  {In
  addition to FUV heating}, cosmic-rays penetrating the molecular
interior of clouds {also contribute to heating the gas}.  To this, we
now apply increased amounts of (mechanical) heating which might be due
to absorbed turbulent energy.  We {discuss in
  section-\ref{subsec:mech-heating} the details of the inclusion} of
mechanical heating into the PDR models.

\begin{figure}[!tbh]
  \centering
  \includegraphics[scale=1.0]{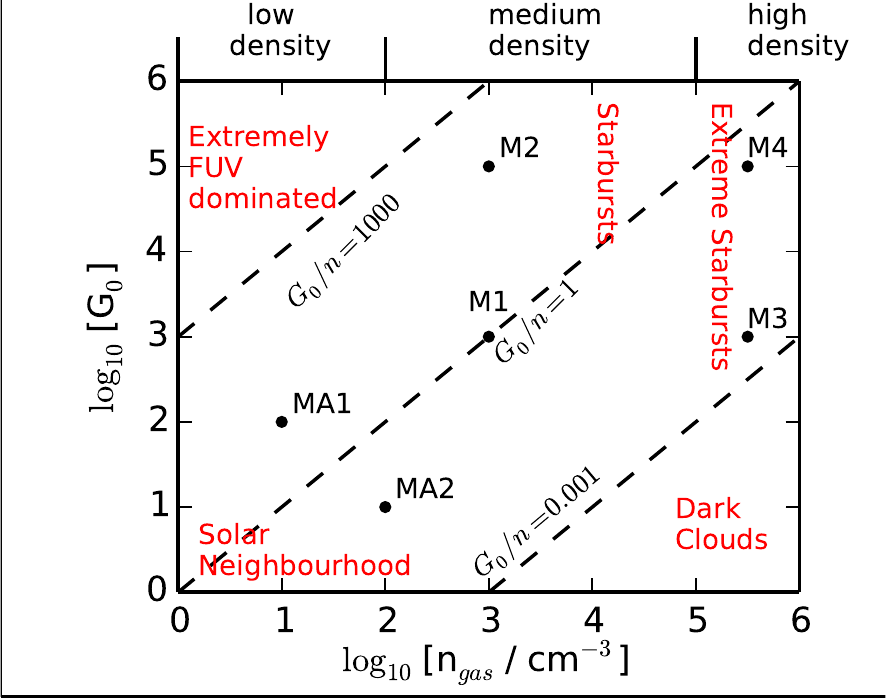}
  \caption{{Diagram indicating different regiemes in the $(n,G_0)$ parameter space. The 
  labeled points correspond to our reference models used throughout the paper (see Table-\ref{tbl:refModels})}.
  \label{fig:n-go-diagram}}
\end{figure}

{In modelling a PDR, we used an optimized version of the PDR code described  
by \cite{meijerink2005-1,meijerink2007-1}.  For a detailed description of the code used, we refer the
reader to the methods section of Paper I and to \cite{meijerink2005-1}.  The 
ISM is modelled as a homogeneous cloud of uniform gas density illuminated by a UV source from 
the side. For simplicity, the cloud is assumed to be an equilibrium plane-parallel 
semi-infinite slab.  The thermal state and the chemical abundances of all the species  
within the cloud are solved for self-consistently.  For more 
details see \ref{subsec:rad-tran}. We explore a parameter space similar to that in Paper I 
where $ 1 < n < 10^6$~\cmt~ and $ 1 < G_0 < 10^6$.  In Fig-\ref{fig:n-go-diagram} we show a schematic
diagram (a template grid) for the ($n$, \go) parameter space where we highlight some situations
in which the ISM could be.  We devide the grid into three grades in density: low ($ 0 < n < 10^2$~\cmt), 
medium ($ 10^2 < n < 10^5$~\cmt) ahd high ($ 10^5 < n < 10^6$~\cmt).  In addition to those two fundamental
parameters specifically for a PDR, we study the response of a cloud's emission on increasing amounts of \gm~for 
different depths $A_V$.  We specifically consider emission in 
atomic fine-structure lines of [OI], [CII], and [CI] in addition to the various molecular line
transitions of CO, \thco, HCN, HNC, HCO$^+$, CN, and CS.

A full range of possible ``extra'' heating rates is explored. From pure PDRs} where heating is 
dominated by the FUV source, to regions where the heating budget is dominated 
by turbulence.  This allows us not just to constrain the effect of 
turbulent heating (as we will demonstrate throughout the paper), but 
also to improve on estimates of molecular cloud column densities in 
cases where turbulent heating contributions are ignored.

A difference with respect to the approach in Paper I lies in the choice of the 
\gm~{parametrization}.  Based on the conclusions of Paper I, we decided that
in a scheme for probing the effects of \gm~on a grid of PDR models, it
is more convenient to include it in the heating budget as a per
unit mass {term, rather than a} per unit volume {one} (see next section). 

In this paper, we consider higher H column densities than included in
Paper I, expanding coverage from columns corresponding to $A_V =
10$~mag to columns corresponding to $A_V = 30~{\rm mag}$.  The main
constraint on this depth is the limit in the interpolation tables used
in the PDR code for computing the self-shielding of CO.  In general,
the properties deep inside the molecular cloud ($A_V > 10$ mag) are constant.  This 
fact can be exploited to approximate the cloud properties at even higher $A_V$ values.

We note that all figures shown in main text of this paper correspond 
to PDRs of solar metallicity.  We have, in fact, also considered other
metallicities, including those as low as $Z = 0.1 Z_\odot$ which
characterize the most metal-poor dwarf galaxies as well as $Z = 2 Z_\odot$
{typical to galaxy centers}.  At any fixed $A_V$,
the corresponding H column density ($N_{\rm H} = N({\rm H}) + 2
N({\rm H}_2)$) is taken to depend inversely on the cloud metallicity
in a linear {fashion}.  We illustrate this as follows. PDRs with the lowest
metallicity and highest $A_V = 30~{\rm mag}$ considered will have
$\NH = 5.6\times 10^{23}$~\cms~compared to $\NH = 9.4\times
10^{21}$~\cms~for clouds with a $Z = 2 Z_\odot$~and an $A_V = 10~{\rm
  mag}$.  Figures corresponding to non-solar metallicity conditions 
can be found in the Appendix.

\subsection{Mechanical heating} \label{subsec:mech-heating}

A major conclusion of Paper I was that mechanical heating must not be
neglected in calculating heating-cooling balances. Addition of a
modest amount of mechanical heating to the cloud volume, corresponding
to no more than a small fraction of the UV surface heating, already
suffices to alter the chemistry of the PDR significantly.

The {PDR model grids in Paper I were parametrized} by $n$ (first
axis - horizontal axis), \go~ (second axis - vertical axis) and \gm~(per 
unit volume, third axis). The parameter space was sampled by picking 
equidistant points in log space for each axis. The dis-advantage of such 
a representation is that for all the models each grid (as a function 
of $n$ and \go) has the same amount of \gm~independent of $n$.  For 
example, a cloud with $n = 1$~\cmt~would have the same amount of \gm~added 
as one with $n = 10^6$~ \cmt. What might be a huge amount for the former cloud would be
negligible for the latter.  It is thus preferable to parametrize \gm~adaptively 
for each density level. In the following, the third axis
is replaced by the new {parametrization} of \gm.  This new
{parametrization is defined using the symbol $\alpha$ where} :

\begin{equation}
  \alpha = \frac{\Gamma_{\rm mech}}{\Gamma_{\rm surf}} \sim \frac{\Gamma_{\rm mech}}{\Gamma_{\rm photo}}
  \label{relation}
\end{equation}

\noindent where $\Gamma_{\rm surf}$~ is the total heating rate at the surface of
the PDR (at $A_V = 0$~ mag), and $\Gamma_{\rm photo}$ is the
photo-electric heating rate. The expression relating it to $n$ and \go~ is
$\Gamma_{\rm photo} = \epsilon_0 G_0 n$, where $\epsilon$ is the heating 
efficiency. The treatment of 
mechanical heating in terms of the total surface heating is to a high
degree of accuracy equivalent to a {parametrization} as a function of
$\Gamma_{\rm photo}$. $\Gamma_{\rm photo}$ is linearly proportional to $n$. For a certain 
section in $\alpha$ (in the third axis) each model thus has a \gm~proportional 
to its density.  We note that this new {parametrization} is also equivalent to
saying that \gm~is added per unit mass, since multiplying $n$ by $m_H$ (the mass of a 
hydrogen atom) corresponds to a mass density.

\begin{figure}[!tbh]
  \begin{minipage}[b]{1.2\linewidth} 
    \includegraphics[scale=0.6]{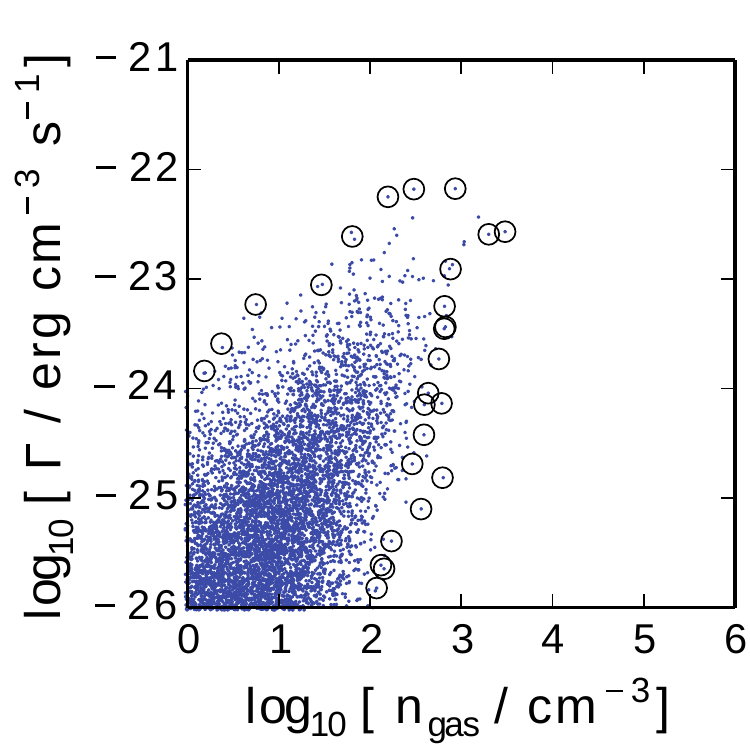}
    \includegraphics[scale=0.6]{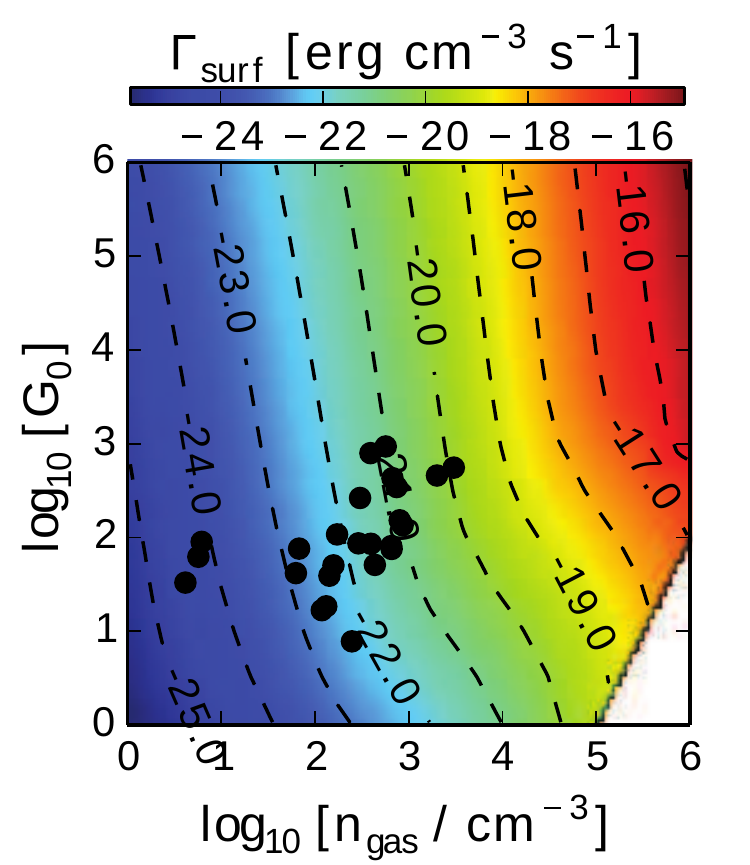}
  \end{minipage}
  \caption{ (Left) Mechanical heating rates applied to the SPH simulation of a dwarf 
    galaxy {\citep{intiPhdT, inti2009-1}}. The maximum heating rate is $\sim 10^{-22}$~ \ecs. (Right) 
    Surface heating (\gsu) for zero mechanical heating ($\alpha = 0$) for 
    Z~ $=$~ \zsun. The heating rates range from $\sim 10^{-22}$~ to 
    $\sim 10^{-20}$~ \ecs~ at $n \sim 10^3$~ \cmt. An SPH particle with 
    $n \sim 10^3$~ \cmt and a maximum \gm~(in the simulation) would have an 
    $\alpha =$~ 1 and 0.1 if its \go~ is 1 and $10^3$~ respectively. The black points
    in the right panel correspond to the the boundary points in $n$ and \gm~
    distribution (plotted as open circles in the left panel) .\label{fig:surfaceHeating}} 
\end{figure}

To make this clear, we will now consider an example.  The case $\alpha
= 0$~corresponds to a situation where no mechanical heating is
present in the PDR (what we call this a reference or pure PDR model).  {On the 
other hand, the} case $\alpha = 1$ {represents a model where 
the} amount of mechanical heating added to the reference model, is
equivalent to the heating at its surface.  We {now} point the
reader to the right panel of Figure-\ref{fig:surfaceHeating}, which is
a grid of reference models (i.e a section in $\alpha = 0$).  For example, the
surface heating for a model with $n = 10^4$~ \cmt~ and \go~ $= 10^3$ is
$\sim 10^{-19}$~ \ecs.  In a grid with $\alpha = 0.01$, this model would
have an amount \gm$= 0.01 \times 10^{-19} = 10^{-21}$~ \ecs~ added to its 
heating budget.  More such examples are summarized in Table-\ref{tbl:gmechValues}.

\begin{table}[h]  
\centering
\begin{tabular}{c c c c c}
  \hline
     $n$      & $G_0$ &             \gm           & $\alpha$ & SFR density \\
  (cm$^{-3}$) &       & (erg~ cm$^{-3}$~ s$^{-1}$) &          &(M$_\odot$~yr$^{-1}$ kpc$^{-3}$)\\
  \hline
  \hline
  10$^2$      & 1.0    & $3.5 \times 10^{-27}$     & $1 \times 10^{-4}$   & $< 0.01$\\
  10$^4$      & 10$^3$ & $2.1 \times 10^{-22}$     & $2 \times 10^{-3}$   & $\sim 2$\\
  10$^4$      & 10$^4$ & $2.1 \times 10^{-21}$     & $4 \times 10^{-3}$   & $\sim 20$\\
  10$^{4.5}$  & 10$^5$ & $2.1 \times 10^{-20}$     & $1 \times 10^{-2}$   & $\sim 200$\\
  10$^6$      & 10$^5$ & $3.4 \times 10^{-17}$     &   0.3               & $> 1000$\\
  \hline
\end{tabular}
\caption{Mechanical heating rates for different densities and FUV 
  luminosities for models whose metallicity is Z~$=$~\zsun. These models
  span the full density and UV ranges we have explored.  Almost the full
  range of mechanical heating rates are also covered, from quiescent 
  discs (first row) to highly turbulent star-bursts (last row).
  \label{tbl:gmechValues}}
\end{table}

A simple recipe (also adopted in Paper I) for estimates of \gm~in a
star-burst is presented in {\cite{loenen2008}}.  One of the main
assumptions concerns the fraction of the energy of a super-nova {(SN hereafter)} event
transferred into heating the ISM ($\eta_{\rm trans}$), which was
assumed to be 10\%.  However, $\eta_{\rm trans}$ is not well known in
general. The amount of turbulent energy transferred into heating the
ISM, (\gm), is also related to $\eta_{\rm trans}$.  For this reason we
consider values of $\alpha$ ranging for 0 to 1. This range would cover most 
heating rates that would be a result from the full range of $\eta_{\rm trans}$.  
As an example, $\alpha$ could be related to the local SFR in a
star-burst. In general, a higher SFR, would result in a high SN rate,
thus a larger $\alpha$ as shown in Table-\ref{tbl:gmechValues}.

{Moreover, we used the mechanical luminosity curves in \cite{sb99} as
  an independent method of estimating the amount of \gm~that can be
  disposed into the ISM.  In \cite{sb99} stellar population synthesis
  models are used to predict spectrophotometric properties of active
  star-forming regions. Figures-111 to 114 in that paper, provide
  predictions for the mechanical luminosity over a time-span of 1 Gyr,
  both for an instantaneous star-burst
  ($L_{\rm mech}=10^{40}-10^{41}$~erg/s) and a continuous one
  ($L_{\rm mech}=10^{42}$~erg/s). The mass of the stars formed during
  these two scenarios is $10^6$~\msun.  If we assume that this occurs
  in a box whose size is 100 pc (the same spatial scale used in
  \cite{loenen2008}), we obtain a \gm$\sim 10^{-20}$~\ecs~and
  \gm~$\sim 10^{-22}$ -- $10^{-21}$\ecs~for the continuous and the
  instantaneous star-burst scenarios respectively. In computing those
  estimates, we assumed that the mechanical luminosity is fully
  absorbed by the ISM since they occur over a time-span of at least 50
  Myr, which is much longer than the chemical time-scale.

Our final attempt to estimate \gm~relied on extracting the mechanical feedback from
softened particle hydrodynamics simulations of dwarf galaxies \citep{inti2009-1}. There
the mechanical heating is computed self-consistently in the SPH code. In the left panel
of Figure-\ref{fig:surfaceHeating}, we show the mechanical heating rate (per unit volume) 
as a function of the gas density of the SPH particles. It is quite comforting to note that,
although the previous two methods were quite simplistic, the estimated mechanical
heating rates are close to those obtained by the SPH simulation for the same density
range.}

\subsection{Radiative Transfer} \label{subsec:rad-tran}

Two methods were used {in} computing the radiative transfer of the
atomic and molecular lines. Those of the atomic fine-structure
(hereafter FS) lines were computed self-consistently within the PDR
code (see \cite{meijerink2005-1} for the details).  For these lines
the temperature gradient within the slab has been taken into
account.

In computing the emission of the molecular species the statistical
equilibrium non-LTE radiative transfer code RADEX \citep{radex} was used. 
{In order to account for clouds with different
depths, the various emission were computed for different $A_V$. This is
equivalent to saying that we considered a semi-finite slab illuminated
by a UV source from one side (the primary UV source).  In using this approximation
the UV flux from a source which could be on the other side of the truncated slab 
model, is ignored.  Accounting for that second source would be important in the 
case when $A_V$ is low ($< 1$ mag) and when the lines are optically thin.  Although the
the lowest $A_V$ we considered was 5 mag, the worst case scenario even for high 
$A_v$ is when the secondary UV source (from the other side) has the same strength
 as the primary.  In this case, ignoring the secondary UV source would reduce the emission 
by at least half the amount.  For our purposes, since we would be looking 
at line ratios and at clouds with $A_V > 5$~mag, we expect this approximation to be 
satisfactory for our purposes.  Hence, for simplicity we assume that the 
semi-infinite slab is illuminated by a UV source from one side only.  There are PDR 
codes which allow for a secondary UV source, for example \cite{petit06}. An even 
more advanced PDR code which allows for an arbitrary 3D geometry has been developed by \cite{bisbas12}.}

The large velocity gradient (LVG) approximation upon which RADEX is based
does not take into account a PDR temperature gradient.  For the
molecular species, this is not crucial, for two reasons: firstly,
molecules are only abundant in regions beyond $A_V \gtrsim 5$~ mag. In
those regions the molecular abundances are at least two orders of
magnitude higher than those close to the surface of the PDR ($A_V = 0$~
mag).  Secondly, deep into the cloud where $A_V \gtrsim 5$~ mag
temperatures are practically constant, very different from the steep
temperature gradients close to the PDR surface.

Although the temperature is almost constant in the molecular zone, we
tried to account for the contribution of the temperature gradient of
the atomic and radical zones by using a weighted average of the
quantities required to compute the emission.  The gas kinetic
temperature and the densities of the colliding species (mainly H$_2$)
were weighted against the density of the species of interest.  The
weighted temperatures are $\sim$~3\% higher than the saturated kinetic
temperatures in the molecular zone.  This is due to higher kinetic temperatures 
at the surface compared to those in the molecular zone. In contrast the weighted 
density of the colliding species was about 5\% lower than those in the
molecular zone. This reflects the fact that H$_2$ has a much lower abundance 
{near the surface of the cloud} compared to that at the {molecular zone}.  These two 
counteracting effects, higher temperature
and lower abundances result in a 1\% increase in the emission compared
to the case where no weighting was done.

The densities of the colliders are also weighted with respect to the
abundances of species whose {emission} are being computed.  The
colliders considered were H$_2$, H$^+$, He and e$^-$.  Finally, a
background radiation field corresponding to that of the current day 
CMB is used by {RADEX.  
One of the assumptions we adopted, which cloud effect the outcome of the 
line ratios of \thco~to CO, is the elemental abundance of C and $^{13}$C.  Since we 
opted to keep the number of parameter low for this exploratory investigation, we chose 
C / $^{13}$C$~= 40$ for all our models, even for models with higher and lower metallicities. 
The actual ratio could be as high as 90 and as low as 20 in our galaxy \citep{savage02}.  Since 
we picked a lower value, we expect that variations over the possible range 
would only decrease the emission of \thco; hence leading for line ratios involving \thco/CO to decrease.}

One of the main advantages offered by the use of RADEX (one-zone
approximation) over a calculation based on the temperature gradient is
the much greater speed of computation.  For each model, the LVG
approximation requires only one computation of the population
densities.  In contrast, it would have been necessary to compute the
population densities for each slab of the discretized PDR when taking 
the temperature gradient into account.

{It still remains to decide what micro-turbulence line width to use.  Since most 
molecular line emission emanates from the shielded 
region where $A_V > 5$ mag and where the gas is relatively cold, the line widths
there are expected to be low.  Again, here we resort to the simulations 
of \cite{inti2009-1}, where for such clouds, the velocity dispersion is on the 
order of few km/s.  Hence, for simplicity we used a micro-turbulence 
line-width of 1 km/s for all our models. We considered other line widths as well but the effect 
on the line ratios was negligible.  We note that this line width should not be 
confused with the measured (observed) line-width.  The measured line-width would 
be due to the contribution of multiple clouds along the line of sight.  For extra-galactic 
sources, this line-width would be much larger than the micro-turbulence 
line-width.}

As a check on the validity of using LVG models, we compared the
emission grids computed with the LVG approximation to those computed
by \cite{meijerink2007-1}.  Although the grids in \cite{meijerink2007-1} have
been computed taking into account the temperature gradient in the PDR
, they agree quite well with the ones we computed here with
{RADEX.  As an independent check, the grids without \gm~agreed quite
well with the ones computed by \cite{kaufmann99}.}

\section{Results}

Atomic fine-structure and molecular emission lines are studied in
order to see how {mechanical feedback could} affect their emission {and 
their various line ratios}.  Here we present results
from the reference models summarized in Table-\ref{tbl:refModels}, and
highlight some emission grids.  The {grids for 
different column densities} are presented in the Appendix. 
{We start by presenting
results of the atomic lines, followed by those for molecular species.  For each of the species 
considered we highlight the main chemical and radiative factors which effect their emission.  Once 
we have characterized the effect of mechanical feedback on the emission of the reference models and the model
grids, we proceed by highlighting trends in the line ratios which could be used as diagnostics for 
mechanical feedback in the ISM.}

\begin{table}[h]  
  \centering
  \begin{tabular}{c c c}
    \hline
    Model Name & $\log n$ (cm$^{-3}) $ & $\log G_0$\\
    \hline
    \hline
    MA1 & 1.0 & 2.0 \\
    MA2 & 2.0 & 1.0 \\
    \hline
    M1  & 3.0 & 3.0 \\
    M2  & 3.0 & 5.0 \\
    M3  & 5.5 & 3.0 \\
    M4  & 5.5 & 5.0 \\
    \hline
  \end{tabular}
  \caption{This is a list of the reference models for typical PDRs used in this paper.
    The models in the top part (MA1 and MA2) are particularly useful for studying the 
    line emission of the atomic species (the FS lines).  Since molecular line emission 
    is too low to be relevant in these two low-density models, higher density models are 
    also considered (M1 to M4). All these models are identical to the reference models 
    from Paper I and \cite{meijerink2005-1}.
    M1 and M2 correspond to low-density clouds in star-bursting galaxy centres, whereas 
    M3 and M4 correspond to much denser clouds where excitation of high density gas tracers 
    such as HCN are possible.  {These models are also shown in our template grid in Fig-\ref{fig:n-go-diagram}.}
    \label{tbl:refModels}}
\end{table}

\subsection{Atomic species - intensities}

\begin{figure}[!tbh]
  \centering
  \includegraphics[scale=0.45]{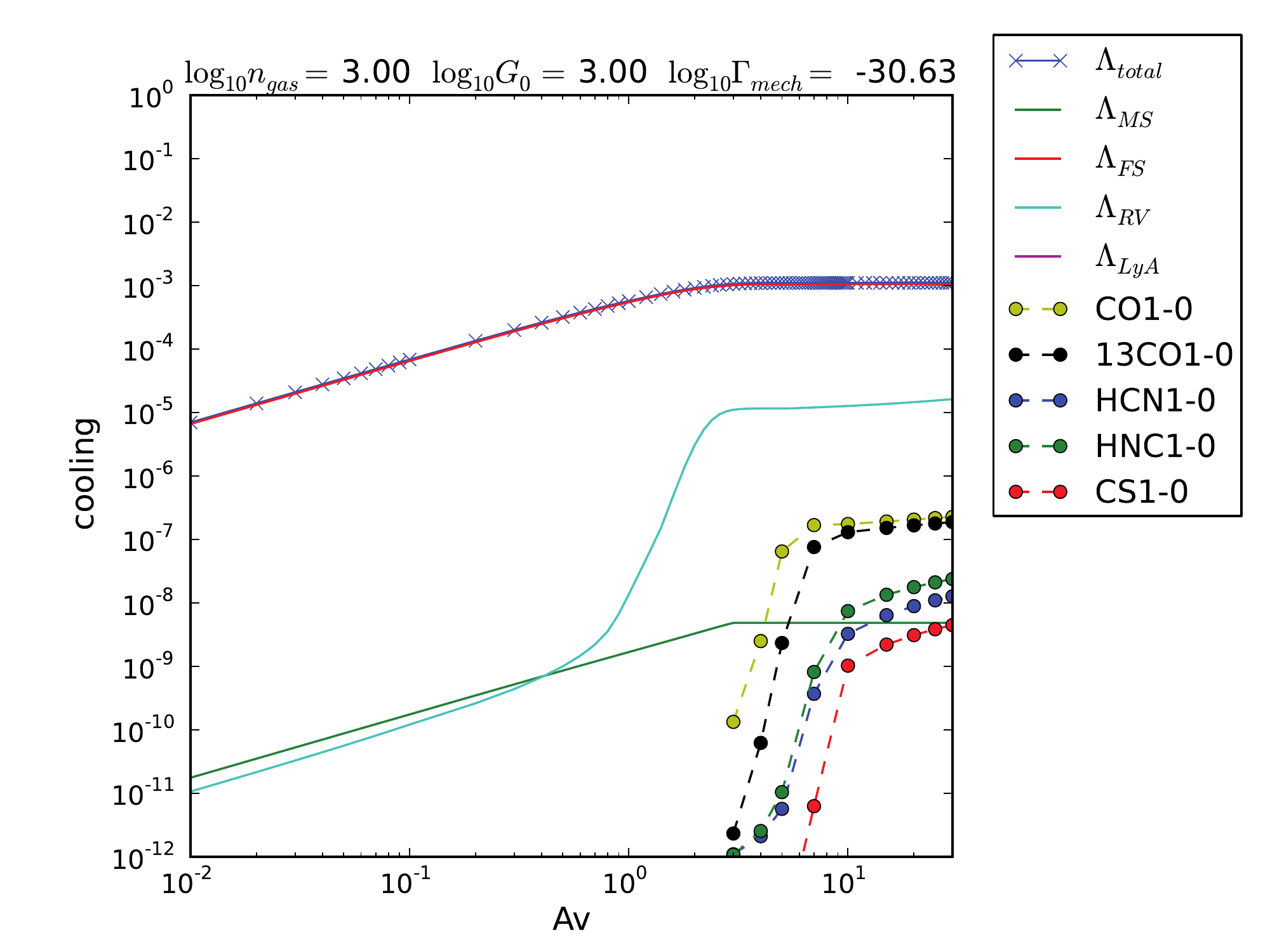}
  \caption{Integrated cooling rates (in \inten) $\int_0^{A_v} \Lambda(A'_v) dA'_v$
    as a function of A$_V$ for the various cooling mechanisms included
    in the PDR code.  The curves for the total cooling (blue with crosses)
    and the FS cooling curve overlap. The remaining solid curves (MS
    for meta-stable line cooling, RV for ro-vibrational cooling of H$_2$,
    and LyA for the Lyman-alpha cooling) are shown for completeness.
    The dotted curves (with filled circles) are the first rotational
    transitions for the molecular species computed using LVG models.\label{fig:typicalCooling} }
\end{figure}

FS lines such as [OI] 63~\mum~ and [CII] 158~\mum~ are the dominant
coolants in PDRs.  High temperatures $(> 100$K) are necessary to have
{a bright emission} of those lines.
Thus most of the emission originates
from the {surface layer of the PDR where $\NH < 10^{21}$~\cms}.  The temperatures 
drop with increasing $\NH$ as such
column densities are optically thick for FUV photons
\citep{hollenbach2009}. 
In Figure-\ref{fig:typicalCooling} we show the integrated cooling rate
as a function of depth ($A_V$) for M1.  From a qualitative point of
view, the cooling budgets for the remaining models are similar.  The
purpose of this figure is to illustrate the steep dependence of the FS
cooling as a function of $A_V$.  We also see here that the cooling due
to the lowest transition of the molecular species considered adds up
to less than $\sim$1\% of the total cooling.  For simplicity, we do
not include this small contribution in the thermal balance,
albeit it is not quite negligible.

\begin{table}[h]  
\centering
\begin{tabular}{c c c c c}
  \hline
  line & $\nu$ (GHz)   & $\lambda$ (\mum) & $n_{cr}$ (\cmt) & $E / k_b$ (K)\\
  \hline
  \hline
  Atomic \\
  \hline
  $[{\rm OI}]$  & 4759 & 63  & $5.0 \times 10^5$ (H) & 228 \\
  $[{\rm CI}]$  & 809  & 369 & $3.0 \times 10^2$ (L) & 63  \\
  $[{\rm CI}]$  & 492  & 609 & $2.0 \times 10^3$ (M) & 24  \\
  $[{\rm CII}]$ & 1897 & 158 & $2.8 \times 10^3$ (M) & 92  \\
  \hline
  \hline
  Molecular \\
  \hline
  CO(1-0)       &  115  &  2602  &  $2.2 \times 10^3$ (M) & 5.53\tablefootmark{a} \\
  CO(2-1)       &  230  &  1301  &  $1.2 \times 10^4$ (M) & 16.6 \\
  CO(3-2)       &  346  &  867.6 &  $3.8 \times 10^4$ (M) & 33.2 \\
  CO(4-3)       &  461  &  650.7 &  $9.1 \times 10^4$ (M) & 55.3 \\
  CO(6-5)       &  691  &  433.9 &  $3.0 \times 10^5$ (H) & 116  \\
  CO(7-6)       &  806  &  371.9 &  $4.6 \times 10^5$ (H) & 155  \\
  CO(10-9)      &  1151 &  260.4 &  $1.1 \times 10^6$ (H) & 304  \\
  CO(16-15)     &  1841 &  162.9 &  $3.5 \times 10^6$ (H) & 751  \\
  \hline
  $^{13}$CO(1-0) &  110  &  2722  &  $1.9 \times 10^3$ (M) & 5.29\tablefootmark{a} \\
  $^{13}$CO(2-1) &  220  &  1361  &  $1.0 \times 10^4$ (M) & 15.9 \\
  $^{13}$CO(3-2) &  330  &  907.5 &  $3.3 \times 10^4$ (M) & 31.7 \\
  $^{13}$CO(6-5) &  661  &  453.8 &  $2.6 \times 10^5$ (H) & 111  \\
  \hline
  HCN(1-0)      &  88.6 &  3384  &  $2.0 \times 10^6$ (H) & 4.25\tablefootmark{b}  \\
  HCN(4-3)      &  354  &  846.3 &  $2.1 \times 10^8$ (H) & 42.5 \\
  \hline
  HNC(1-0)      &  90.7 &  3309  &  $3.6 \times 10^5$ (H) & 4.35\tablefootmark{b} \\
  HNC(4-3)      &  362  &  827.6 &  $2.1 \times 10^7$ (H) & 43.5 \\
  \hline
  HCO$^+$(1-0)  &  89.2 &  3363  &  $2.1 \times 10^5$ (H) & 4.28\tablefootmark{c} \\
  HCO$^+$(4-3)  &  356  &  841.0 &  $9.3 \times 10^6$ (H) & 42.8 \\
  \hline
  CS(1-0)       &  45.0 &  6123  &  $5.4 \times 10^4$ (M) & 2.35\tablefootmark{d} \\
  CS(4-3)       &  196  &  1531  &  $3.0 \times 10^6$ (H) & 23.5 \\
  \hline
  CN(1$_{1/2}$-0$_{1/2}$) &  113   &  2651 &  $1.8 \times 10^6$ (H) & 5.43\tablefootmark{e}\\
  CN(2$_{3/2}$-1$_{3/2}$) &  226   &  1325 &  $4.7 \times 10^6$ (H) & 16.3   \\
  \hline
\end{tabular}
\tablefoot{\\
  \tablefoottext{a}{From \cite{yang2010}}\\
  \tablefoottext{b}{From \cite{Dumouchel2010, Green1974}}\\
  \tablefoottext{c}{From \cite{Botschwina93}}\\
  \tablefoottext{d}{From \cite{Lique2006}}\\
  \tablefoottext{e}{From \cite{Klisch1995, lique2010}}\\
}
\caption{Critical densities and the transition (excitation) energies for
  the atomic FS lines and the molecular lines.  The critical densities
  for the molecular lines displayed were computed for a kinetic
  temperature of 50K; whereas the one for the FS lines were 
  computed at 500K.  The values of $n_{cr}$ are shown to give the 
  reader an idea of the regime for the densities where the 
  species would be at LTE or not. The symbols ``L'' (low), ``M'' (medium), ``H'' (high) indicate
  the range in density where the $n_{cr}$ of the corresponding transition
  lies.\label{tbl:atomSpecsProp}}.
\end{table}

\noindent Although most of the FS emission emanates from the PDR's
surface section, the molecular zone also contributes.  This 
{contribution of the molecular zone} to the FS cooling is important when considering line
ratios. It depends on the location of the C$^+$/C/CO transition zone.
A fast transition (as a function of $A_V$) from C$^+$ to C,
would result in a lower abundance of C$^+$ (thus lower emission).
Usually, the $A_V$ where the C$^+$/C/CO transition zone occurs is
closer to the surface for PDRs with high densities than clouds with
low densities.  This phenomenon is described in more detail in
{\cite{mvk12}}.  For example, in low-density PDRs the abundances of
C$^+$ (\xcp) decreases slightly from $\sim 10^{-4}$ (at the surface) to
$\sim 10^{-5}$ in the molecular zone.  However, a much greater
decrease in \xcp~is observed in the molecular zone of {medium and} high-density
PDRs; there \xcp~decreases to $\sim 10^{-10}$.  {Consequently}, we expect to
find higher contributions to C$^+$ cooling from the molecular zones in
low- density clouds than in high-density clouds.

A similar analysis applies to the other FS lines of C and O.  For
example {in clouds which are highly dominated by UV radiation
(see Figure-\ref{fig:n-go-diagram})} atomic abundances remain
relatively high in the molecular zone.  At the surface $10^{-6}<$
\xcp $< 10^{-4}$ and \xc~$\sim 10^{-4}$ due to the high flux of FUV
photons which can penetrate deep into the clouds. Hence recombination 
is less effective in locking the atoms into molecules.  At densities
$n>10^3$\cmt, the FUV photons are blocked by higher columns of CO
resulting in \xcp $< 10^{-8}$ and \xc $< 10^{-6}$.

{The intensity of the emission depends mainly on the chemical
  abundances of the species in question (discussed in the previous two
  paragraphs) and strongly on the kinetic temperature of the gas. Here
  we} investigate the effect of \gm~on {the emission, from the atomic
  and molecular zones, by looking at its effect on the kinetic
  temperature of the gas.}  In general {we expect the emission} to be
enhanced as \gm~is introduced into a PDR.  {At the surface, mechanical
  feedback increases the temperatures at most by} a factor of three,
even for the highest \gm~considered.  {As an example, in} the low
\go$/n$ ($\sim 0.01-10$) models such as MA1 and MA2, the surface
temperature increases from 110 K to 300 K without mechanical heating
($\alpha=0$, see Eq. \ref{relation} and Fig. \ref{fig:surfaceHeating})
and from 2200 K to 4500 K for maximum \gm ($\alpha=1.0$),
respectively. {Note that the surface of low density PDRs is a lot more
  sensitive to \gm~compared to the high density ones.  Moreover,} the
temperature increase in the molecular zone is much greater.  For
instance {in the low density PDRs, we} see a boost from $\sim 10$K to
100K in MA2 {and MA1 whenever} \gm~$=0.1$\gsu.

{We now} turn our attention to the contribution of the molecular zone to the
integrated luminosity of the FS lines.  In pure PDRs ($\alpha = 0$) {which 
are in the} \go/$n \gg 1$ regime, at least 2\% of the integrated
luminosity comes from depths with A$_V > 5$~ mag.
Since \gm~increases temperatures in the molecular zone, we expect the
FS line emission to increase as well (this is explained in more detail
in section-\ref{subsub:molspec}).  For instance, when $\alpha = 0.25$~
{the contribution of the molecular zone increases to} $\sim 10$\% for 
clouds with A$_V=10$~ mag.  {For clouds with higher A$_V$ ($\sim$~30 mag) this 
contribution increases further up to 50\%.}

\begin{figure}[!tbh]
  \begin{minipage}[b]{1.0\linewidth} 
    \center
  \includegraphics[scale=1.0]{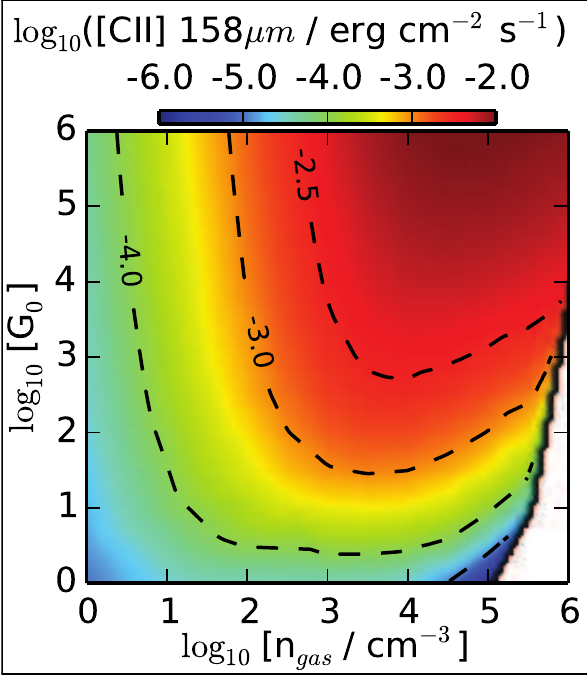}
  \includegraphics[scale=1.0]{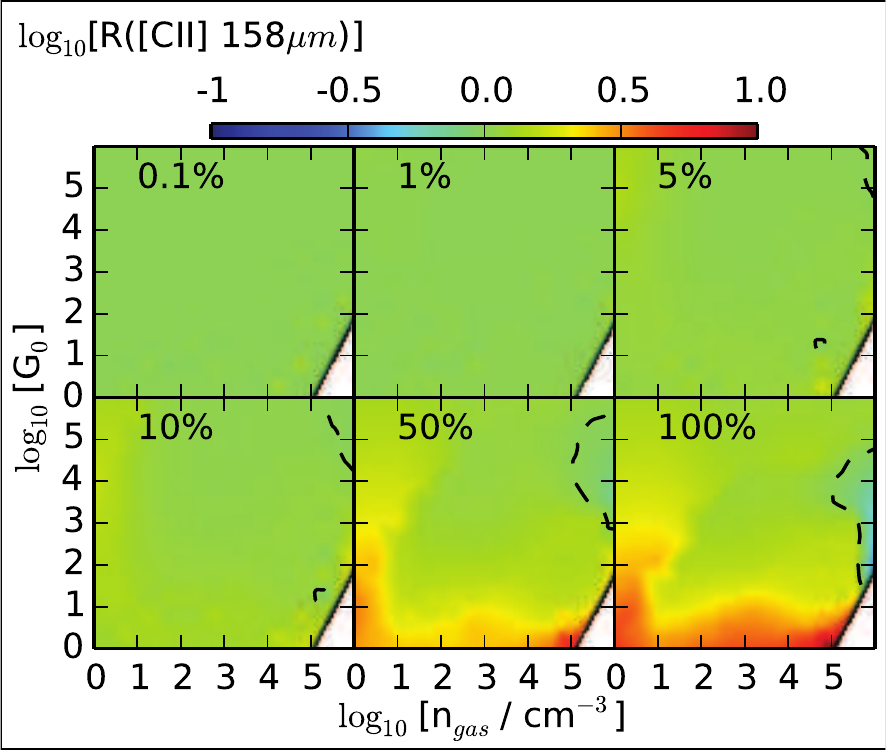}
  \end{minipage}
  \caption{[{\bf top}] The reference grid for the [CII] 158~\mum~ line
    {emission} for PDRs without mechanical heating. [{\bf bottom}]
    [CII] 158 \mum~ line {emission} corresponding to different value of
    $\alpha$ {labelled} at the top of each panel. Each model in the grid
    has an additional amount of heating introduced to its energy
    budget.  The added \gm~is in terms of a percentage of the surface
    heating (as explained in the methods section).  Each grid shows
    the percentage change (increase or decrease) in emission relative
    to that in the reference grid in the {\bf top} panel.  For
    instance, when \gm~$=$~ 0.5\gsu~ ($\alpha = 0.5$) (panel with the
    50\% label), the emission in MA1 is enhanced by a factor of
    $\sim 3$.  A decrease in emission is observed only when
    \gm~$>$~ 0.5\gsu~ to the right of the contour line at M4.  We
    {define the relative change as 
    $R = I(\alpha)/I(\alpha=0)$, where $I(\alpha$) is the emission
    intensity of the line at a specific value of $\alpha$. 
    Here and in all such subsequent plots the dashed 
    contour line traces $R = 1$.  On this line the emission with 
    and without extra heating are the same. In other words, models on 
    this line experience no change in the emission because of \gm. 
    In general 'redder' regions correspond to enhanced emission, whereas 
    'bluer' regions indicated regions where emission are suppressed.}
\label{fig:C+-1-0-grids} }
\end{figure}

In the top panel of Figure-\ref{fig:C+-1-0-grids}, we show the
emission grid of the [CII] 158~\mum~ FS line.  This grid agrees very
well with that shown in Figure-3 in \cite{kaufmann99}.  In the lower
panel, we {illustrate the relative changes in the emission of that same
grid as a function of $\alpha$. We} see that the [CII] 158~\mum~ FS line 
emission depends weakly on \gm~whenever it is below 0.5~\gsu~ ($\alpha < 0.5$).
{Whenever $\alpha \ge 0.5$ low density PDRs such as} MA1 and MA2 show
an increases in the emission by up to a factor of two.  The mid- and
high-density PDRs {whose} \go$~ >10^2$ are hardly effected.  The emission
of [OI] 63~\mum, [CI] 369~\mum~, and [CI] 609~\mum~
exhibit a stronger dependence on \gm.  {These} grids are shown in
Figure-\ref{fig:fs-lines-app} in Appendix-A.

As a basis for discussion, we first highlight some of the main
features of the FS emission grids in the absence of any \gm~
contribution. (1) Emission intensities increase as \go~ increases (see
the top panels of Figure-\ref{fig:C+-1-0-grids} and
Figure-\ref{fig:fs-lines-app}).  This is caused by the ability of the
FUV photons to penetrate deeper into the cloud at fixed $n$ but higher
\go, thus leading to a thicker atomic region. (2) Emission peaks near
the critical densities {of the lines} ($n_{cr}$ hereafter). {Those $n_{cr}$ are around
$10^3$~\cmt~with the exception of the critical density of the [OI] line which 
lies in the high density region. Both excitation} energy $E_{ul}/k_B$ and $n_{cr}$ {for
all the lines are listed} in Table-\ref{tbl:atomSpecsProp}.  (3) The emission intensities
range from 10$^{-6}$~\inten~ to 10$^{-2}$~\inten~, spanning four orders of
magnitude.  This is significantly brighter than the molecular emission 
(see section-\ref{subsub:molspec}), which peaks at $10^{-5}$~\inten.  The range
in the intensities of the atomic fine-structure lines is narrower than 
that of the molecular lines. (4) When \gm~is introduced {the }
emission in enhanced for $n < n_{cr}$; the opposite is observed for 
$n > n_{cr}$.  This is particularly valid for the neutral {and atomic species
(see the bottom panels of Figure-\ref{fig:C+-1-0-grids} and Figure-\ref{fig:fs-lines-app}).}

{Here we focus on the last point mentioned in the previous paragraph. Particularly we 
try to determine the reason causing the different behaviour of the emission for 
$n < n_{cr}$ vs $n > n_{cr}$}.  The boost in the emission could be due to an 
increase in temperature, or in abundance, or both.  By {analysing} the chemistry 
we see that the dominant reactions remain the same (up to {$\alpha \sim 0.1$) for $n < n_{cr}$.  
Moreover there are no fundamental changes in the abundances for $n < n_{\rm cr}$.  This ties
the emission boost to the increasing amounts of \gm~that raise the gas 
temperature, particularly in the molecular region.}

Let us now consider the part of the parameter space in the grid where
$n > n_{\rm cr}$. The emission tends to decrease whenever
$\alpha > 0.1$.  The only exception to this occurs in the emission grid 
for the [OI] 63~\mum~ line (see rightmost panel in the bottom row of
Figure-\ref{fig:fs-lines-app}).  {At} such densities, O
maintains a much higher abundance than C$^+$ in the molecular zone.
{Hence the} strongest decrease is seen for the two neutral carbon lines [CI]
369~\mum~ and [CI] 609~\mum~ (see left and middle panels of
Figures-\ref{fig:fs-lines-app}) {whereas the opposite is observed for [OI]} 63~\mum~.  
We will discuss each of those below.

The high density region of the [CII] 158~\mum~ grid reveals a decrease
in the emission only in extreme cases ($\alpha > 0.5$).  This is seen
most clearly to the right of the {dashed} line in the panels for $\alpha
=$ 0.1, 0.5 and 1 in Figure-\ref{fig:C+-1-0-grids}).  This is simply
because C$^+$ becomes un-abundant at {high densities} ($n \gtrsim 10^5$~\cmt).

The emission of [OI] 63~\mum~ is more interesting but less trivial to explain,
since it is the result of an interplay between the cooling due to the
FS lines and the additional \gm.  We have already mentioned that [OI]
63~\mum~ is the dominant cooling line in all the models.  As \gm~is
added more cooling is required to maintain a thermal balance.  This
occurs by increasing the gas temperature throughout the PDR which in
turn boosts the total cooling rate by increasing the [OI] 63~\mum~
emission. At the densities of interest here ($n>10^4$\cmt) and with
$\alpha > 0.5$, \xo~ decreases from 10$^{-4}$ to 10$^{-8}$ in the
molecular zone due to the dominant reaction of O with OH.
Notwithstanding the steep decrease in \xo, the [OI] 63~\mum~ emission
increases because of the much higher temperatures.

{Finally, the reduction} in the [CI] emission (at high densities) when
\gm~is introduced, is due to \xc~{decreasing} by an order of magnitude
{in the atomic region of the PDR}; dropping from $\sim 10^{-4}$ to $\sim 10^{-5}$ where most 
of the emission comes from.  We discuss this for {an extreme PDR with
conditions typical for an extreme star-burst} ($n = 10^6$~ \cmt~ and \go~ $=10^6$). {There} the 
emission decrease is greatest.  This decrease is {due to the reduced production rate (by one order 
of magnitude) of} C through the neutral-neutral reaction of H with CH.  Moreover, higher abundances 
of H$_2$ in the atomic region enhance the radiative association reaction of 
C with H$_2$, leading to a faster destruction of C.  {These} two processes results in a reduction
of \xc~throughout the PDR {leading to lower} emission.

The different {dependence} on \gm~{throughout the grids of the lines}
encourage us to look in detail at {the} various combinations of line ratios
in an attempt to identify effective diagnostics for \gm.  We first
summaries the most important model emission features that play a role
in determining the ratios:
\begin{itemize}
\item For pure PDRs (without \gm) we can safely say that the majority
  of the FS {emission} are from the surface of the cloud and up to $A_V
  = 10$~mag.  Particularly, [OI] 63~\mum~ and [CII] 158~\mum~ saturate
  at $A_V \sim 5$~mag.
\item When \gm~is introduced, the molecular zone contributes
  increasingly to the integrated emission of the lines, especially
  those of [CI].  When $n < n_{cr}$ more than half of the emission in
  these lines is from the molecular zone.  Hence, line ratios might
  depend on the column densities of the PDRs considered.
\item {As} \gm~{increases, emission} of the FS {lines for clouds with $n < n_{\rm cr}$ increases;
  whereas it} tends to decrease above that density (except for [OI]).
\end{itemize}

\subsection{Atomic species - line Ratios}
\begin{figure}[!tbh]
  \centering
  \includegraphics[scale=0.55]{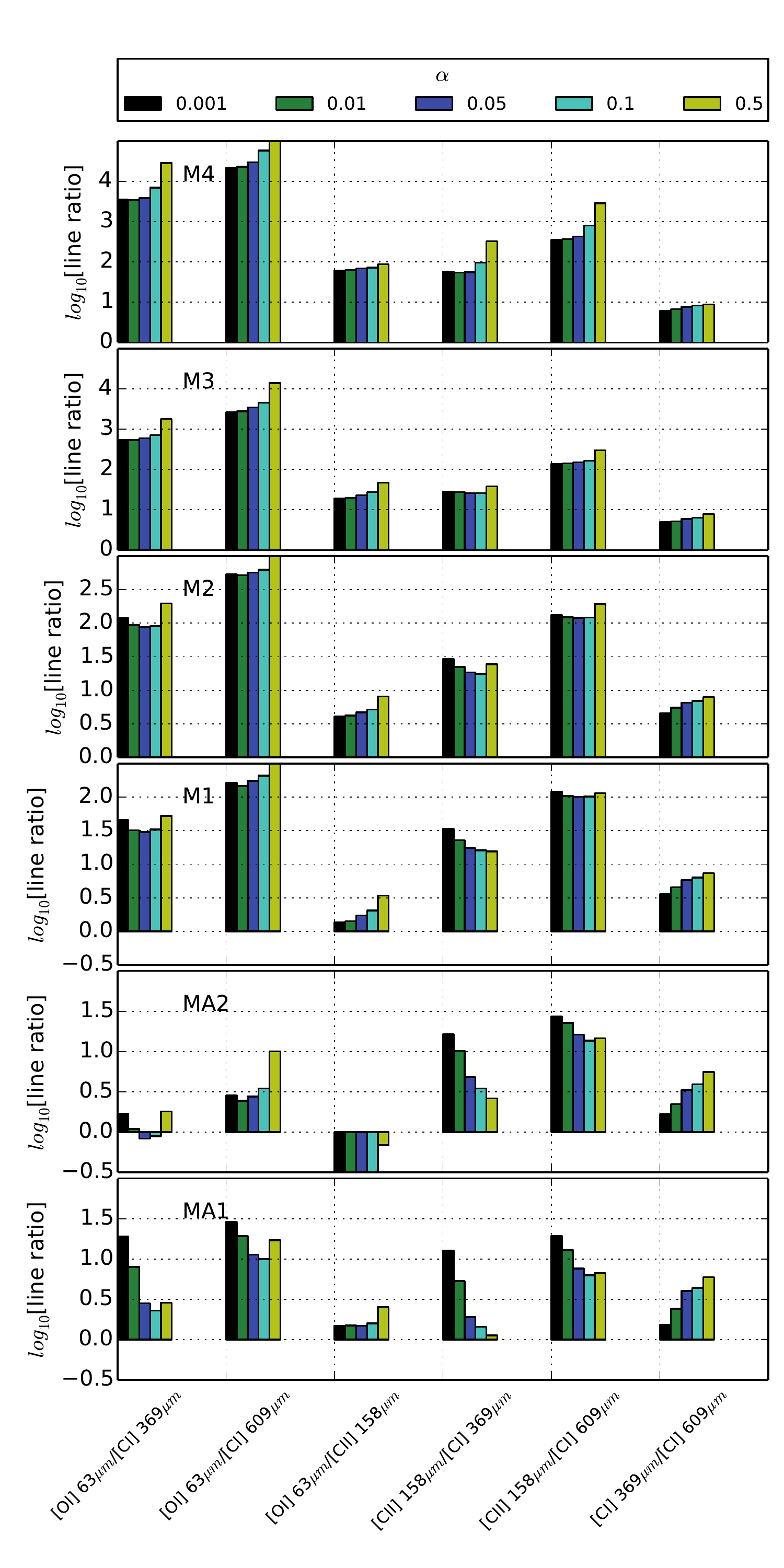}
  \caption{Fine-structure line ratios for different values of \gm~($Z=$ \zsun) for 
    the reference models.
    \label{fig:bar-plot-lineRatios-atomic-z-1.0}
  }
\end{figure}

{So far we have touched upon the intensities of the emission
of FS lines.  What actually matters is the relative enhancement of one emission 
line compared to another (i.e the line ratio); this is a commonly used technique
since understanding line ratios sheds insight on the underlying excitation 
mechanism(s) of such emission.}
In Figure-\ref{fig:bar-plot-lineRatios-atomic-z-1.0} we show (for the
reference models) all the possible combinations of line-ratios of 
the FS {lines} that we considered.  The analogous figures for metallicities 
either higher or lower than solar are presented in the Appendix 
(see Figure-\ref{fig:bar-plot-lineRatios-atomic-z-other}).  We note that the
line ratios apply to fixed cloud sizes ($A_V = 10$~mag).

The very low-density models (MA1 and MA2) show a distinctive response
to \gm~in comparison to the {rest of the models}.  Only in these two
models, some of the line ratios (expressed on log-scales) {change sign} i.e. 
the ratio {changes} from being below unity to above unity or {vice-versa} due to 
one line becoming brighter than the other.

For instance, in MA1,
the ionized to neutral atomic carbon line ratio [CII] 158\mum/[CI]
369\mum~shows a very nice dependence on \gm.  It decreases from $\sim
15$ to unity as \gm~increases. A similar behaviour is observed for 
MA2, but the ratio saturates to $\sim 3$ at high $\alpha$.
Another interesting line ratio is that of neutral oxygen to neutral
carbon ([OI] 63\mum/[CI] 369\mum), which exhibits a very strong
response to \gm.  The ratio decreases as $\alpha$ increases.  The
reason for this is the fact that the [CI] 369\mum~ line is enhanced
(and becomes stronger than the [OI] 63\mum~ line) due to the low
energy of the [CI] 369\mum~ transition (24 K) compared to the 228 K of
the [OI] 63\mum~ line.  Hence, the [OI] 63\mum~ emission remains
restricted to the surface, whereas the total emission of [CI] 369\mum~
gets a significant contribution from the deeper molecular interior.
This line ratio decreases by {approximately a} factor of ten for MA1 (from $\sim 20$ to $\sim 3$)
when \gm~is as low as 5\% of the surface heating.  A less distinctive behaviour is observed
in MA2, where the line ratio decreases from $\sim 2$ to $\sim 1$ and
increases again to above 2 for $\alpha = 0.5$.
In MA1, the ratio of the principal cooling lines, neutral oxygen to
singly ionized carbon ([OI] 63\mum/[CII] 158\mum), has a weak
dependence on \gm.  It increases only for extreme mechanical heating
rates corresponding to $\alpha = 0.5$.  More interestingly, in MA2 the
ratio has a stronger dependence of \gm~approaching unity (from $\sim 0.3$) as $\alpha$ increases. This
is explained by the fact that as $\alpha$ increases, temperatures
increase as well. Since \xo~ is about 100 times higher in the
molecular zone than \xc, the [OI] 63\mum~ emission increases.  This
occurs despite the fact that the transition energy for the [CII]
158~\mum~ line is less than half that of the [OI] 63~\mum~ line.  A
similar behaviour is observed in M1, but there the ratios increase from
1 to $\sim 3$.

The neutral carbon-carbon line ratio [CI] 369\mum/[CI] 609\mum~ is
particularly interesting since it involves FS lines of the same atomic
species; {hence the line ratio} depends purely on the radiative properties 
of the lines and neither on the chemistry nor the column density.  In MA1, we
see a steady increase in the ratio from 1 to $\sim 5$ for $\alpha =
0.05$.  This {can be easily explained. As} \gm~causes temperatures to rise,
the upper levels become more populated, so that the third level
involving the [CI] 369~\mum~ line becomes brighter than [CI] 609\mum.
{For} a detailed discussion {on level populations} see section-\ref{subsub:molspec}.

The C[II] 158\mum/[CI] 609\mum~and [OI] 63\mum/[CI] 609\mum~ratios
do not exhibit any very interesting {dependence} on \gm. They are 
shown just for completeness.

The high-density models M1 to M4 exhibit a slow monotonous (either increasing or
decreasing) {dependence} on \gm. The ratios in these models are generally
$\gtrsim 10$ even for $\alpha = 0$. This is also true for extreme mechanical
heating rates $\alpha = 1$ (except for [OI] 63\mum/[CII] 158\mum~in M1).
Unlike the low density models, models M1 to M4 exhibit a jump in the line ratios
only {when} $\alpha \gtrsim 0.5$.

In summary, what we have found in this section that [CII] 158\mum/[CI]
369\mum, [OI] 63\mum/[CI] 369\mum~and [CI] 369\mum/[CI] 609\mum~are good 
diagnostic line ratios for low density clouds.  One can use those lines 
to constrain \gm~if the density, \go~and the $A_V$ of the object are known.  Those 
line ratios show a stronger dependence on \gm~at higher or lower metallicities as 
well, except in models at Z $=$ 0.1\zsun~(see Figure-\ref{fig:bar-plot-lineRatios-atomic-z-other}). However, 
further investigation is needed in-order to see if atomic line ratios are good diagnostics
of {mechanical feedback} for the whole range in density, \go~ and $A_V$.

In addition to the reference models, the grids of the line ratios from
which those models were picked are presented in Appendix-A
({see} Figure-\ref{fig:atomicLineRatios}).

\subsection{Molecular species \label{subsub:molspec}}
{The molecular emission lines were computed with LVG models}
, unlike those of the atomic species {which were computed within the 
discretized PDR.  We utilized the LVG code RADEX \citep{radex} in computing all the
emission intensity grids}. In this paper, we have limited ourselves to the 
rotational line emission from CO, $^{13}$CO, HCO$^{+}$, HCN, HNC, CS, and CN.

We first present an analysis that is common to most of the molecular
lines considered.  {In pure PDRs ($\alpha = 0$), the gas temperatures in
the molecular region are very low ($10 \sim 15$K) and vary little within
the grid (see the grids corresponding to $\alpha = 0$ in Appendix B for all 
the molecular species).  Thus whenever the gas density of a cloud is below 
the critical density ($n_{cr}$) of the line considered, the chemistry determines 
the shape of the contours in the emission grid.  There the contour lines are 
independent of \go~ and appear as almost vertical lines. Also in the 
absence of \gm~,  the emission contours follow the shape of the temperature contours of
 the molecular region.  This is particularly valid in the case of the high-$J$
transitions}.  However, when we introduce \gm, the chemistry is altered significantly 
along with the {gas temperature.  This} causes the  emission to increase by orders of 
magnitude (cf. the figures for CO with \gm~in the Appendix).

On the other hand, for densities above $n_{cr}$, the contours in the
emission grids depend mainly on temperature.  This is not
surprising, as we may demonstrate by considering a simple two level
system.  The ratio of the upper population density ($x_2$) to the
lower one ($x_1$) at equilibrium is \citep{drainBook}:

\begin{equation} \label{eq:1}
  \frac{x_2}{x_1} = \frac{1}{1 + n_{cr}/n_{col}} \frac{g_2}{g_1} \exp{\left(-\frac{T_{12}}{T_{kin}}\right)}
\end{equation}

\noindent where $n_{col}$ is the density of the colliding species
(mainly H$_2$). $g_1$ and $g_2$ are the degeneracies of the lower and upper
levels respectively {and $T_{12}$ is the energy difference of the two levels in
Kelvins $(E_{12}/k_b)$}.  For simplicity, we assume $n_{col} = n({\rm H}_2)$, and
also ignore a second term on the right involving the radiation
temperature $(T_{rad}$) that introduces only minor changes in the LVG
treatment (since $T_{rad} = T_{cmb} \ll T_{kin}$ in general).  

{When} $n_{col} \gg n_{\rm cr}$, this equation reduces to :

\begin{equation} \label{eq:x2x1eq}
  \frac{x_2}{x_1} = \frac{g_2}{g_1} \exp{\left(-\frac{T_{12}}{T_{kin}}\right)} 
\end{equation}
\noindent which is independent of the gas density, and depends only on
the kinetic gas temperature.  Also, when $n_{col} > n_{cr}$, all the
lines are thermalized, and the excitation temperature is effectively
equal to the gas temperature.  Although these equations are for a
two-level system, they provide a general idea of what might be
happening in a multi-level systems (which might include multiple
colliding species whenever the rates are available).

\begin{figure*}[!tbh]
  \centering
  \includegraphics[scale=0.65]{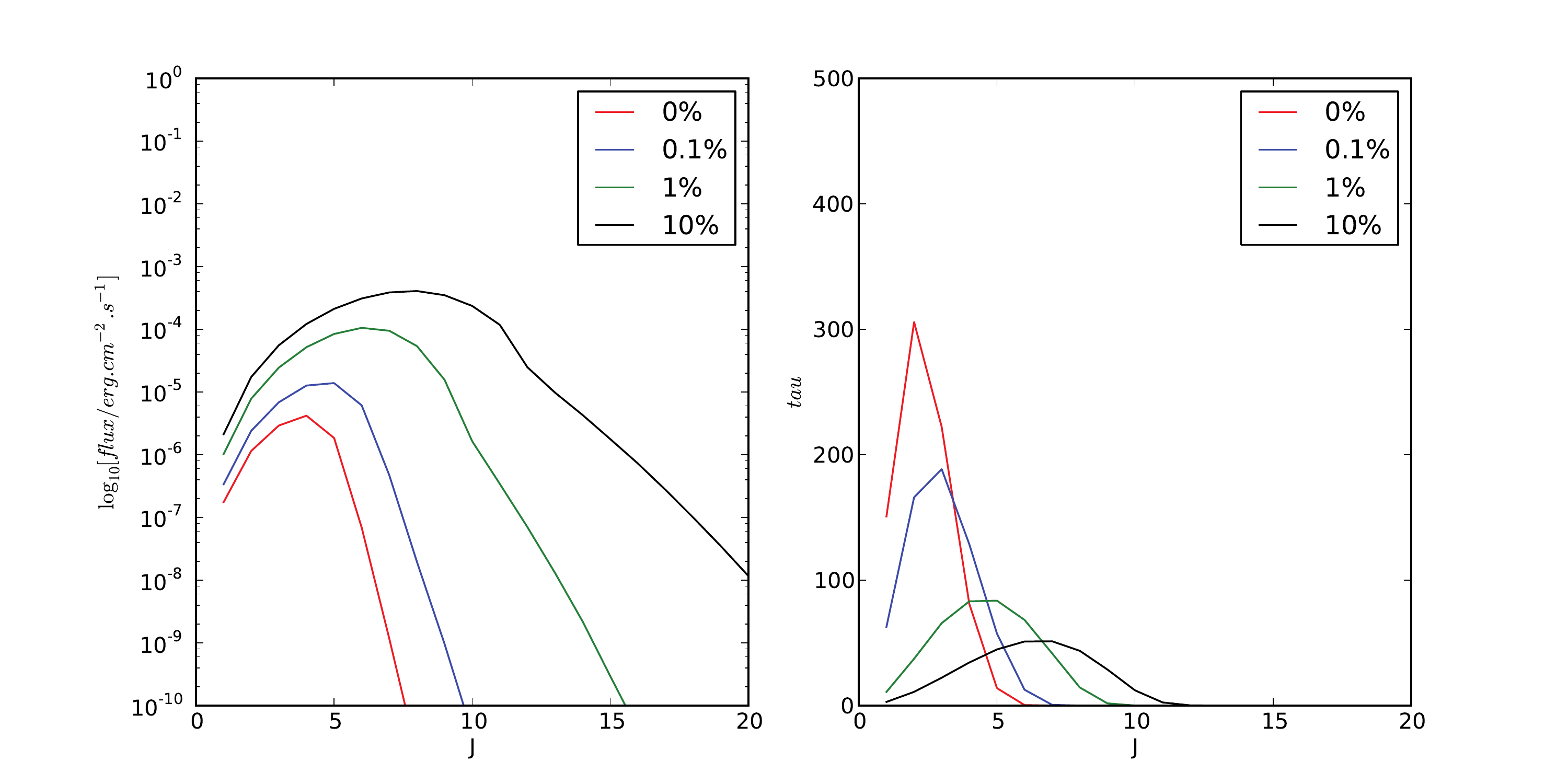}
  \caption{CO ladders for a PDR model with $n=10^3$~cm$^{-3}$ and
    $G_0=10^3$ for different values of mechanical heating. The numbers
    in the legend correspond to \gm~in terms of the surface
    heating. {\bf(left)} CO line intensities as a function of the
    rotational level $J$. Notice the boost in the intensities is much
    larger at high-$J$ compared to low-$J$. {\bf(right)} Optical depth
    at the line centres ($\tau_0$) for the same lines. The optical
    depth generally decreases as a function of increasing \gm~, but
    the most significant effect is that the low-$J$ CO lines become
    optically {thinner} with $\tau_0 \sim 1$. \label{fig:CO-ladder} }
\end{figure*}

As \gm~is introduced, the higher transitions gain prominence
since relative temperature increases are high.  This in turn leads to
increased high-level populations (see the left panel of
Figure-\ref{fig:CO-ladder}).  Another way to look at the pumping of
the higher levels is by looking at the collision rate coefficients
$K_{12}$ and $K_{21}$.  Ignoring degeneracies, those two are related
via $K_{12} \sim K_{21} \exp(-T_{12} / T_{kin})$. 

\begin{figure}[!tbh]
  \begin{minipage}[b]{1.2\linewidth} 
    \includegraphics[scale=0.75]{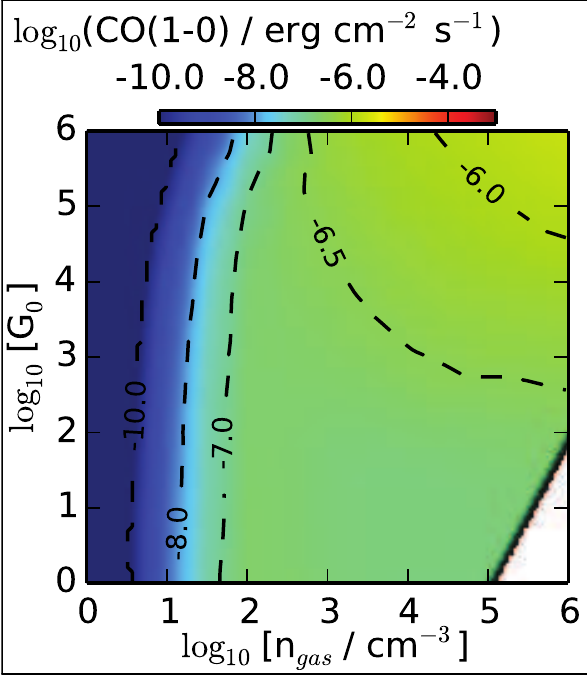}
    \includegraphics[scale=0.75]{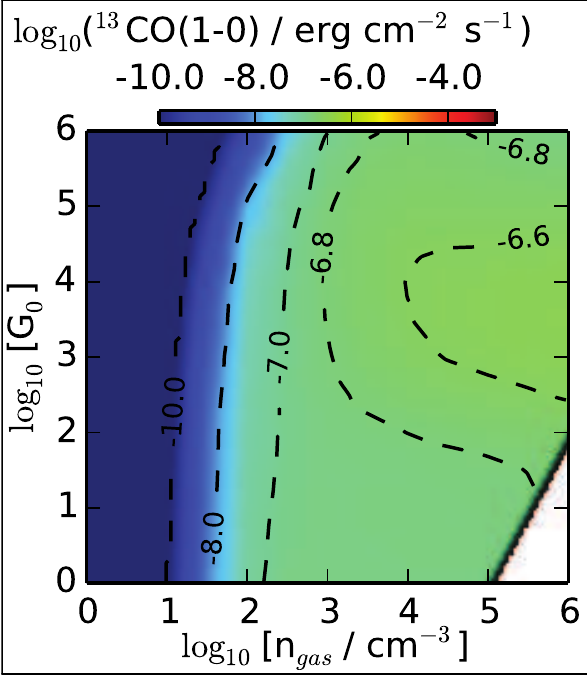}
  \end{minipage}
  \caption{{Emission intensity} grids of CO($J=1-0$) and $^{13}$CO($J=1-0$) {transitions} for 
  the 1D-PDR models.  The emission of the models in the grid correspond to an $A_V = 10$ mag,
    which amounts to a column density of H, $N_H \sim 10^{22}$~ \cms~ 
    at solar metallicity.  The intensities were computed using the 
    RADEX LVG code.  A line-width of 1 km s$^{-1}$ was used. No
    mechanical heating is added for the models in this grid.  This is
    used as a reference grid for the ones in which \gm~is introduced (see
    Fig-\ref{fig:CO-1-0-gmech}). \label{fig:CO-13CO-1-0-base} }
\end{figure}

\begin{figure}[!tbh]
  \centering
  \includegraphics[scale=1.0]{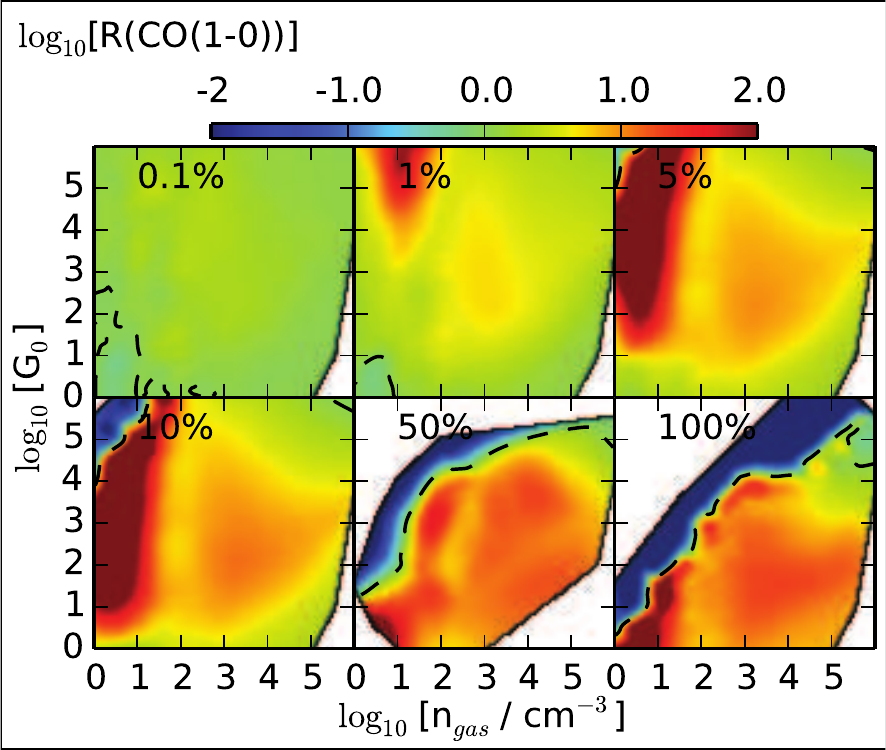}
  \caption{{Relative change in the emission} of the CO $J$=1-0 rotational line for
    different values of \gm. The colours in the panels correspond to
    the relative change of the {emission} with respect to the reference
    grid in the left panel of Figure-\ref{fig:CO-13CO-1-0-base}.  {See 
    also the caption of Fig.\ref{fig:C+-1-0-grids}}.\label{fig:CO-1-0-gmech} }
\end{figure}

Another interesting common feature is the optical depth at the line
centre, $\tau_0$.  We observe that it decreases as a function of
increasing \gm.  
%
%
%
%
As temperatures increase, more levels get populated leading to smaller
optical depths.  These are shown in Figure-\ref{fig:CO-ladder}, which
shows the CO ladder along with curves for $\tau_0$ for {a range in} \gm.  In the {right panel}, 
the trend in $\tau_0$ is clearly visible.
\vspace{0.7cm}
\newline
\indent {\bf CO and $^{13}$CO }\\
CO is the second most abundance molecule in the ISM after H$_2$.
Since the critical density of the CO(1-0) rotational line is {moderate} 
(see Table-\ref{tbl:atomSpecsProp}), it is a good tracer of molecular 
gas with {average} densities.  The
critical densities increase gradually as we go higher up the CO ladder,
reaching $\sim 10^{6.5}$~\cmt~for the 16-15 transition\footnote{The
  critical densities depend only lightly on temperature.  For instance
  $n_{\rm cr}$ for the 16-15 transition changes only by 5\% as the
  gas temperature changes from 10K to 1000K}.  In
Figure-\ref{fig:CO-13CO-1-0-base} we show the grids for CO $J$=1-0
emission.  In Figure-\ref{fig:CO-1-0-gmech}, we show the relative
increase/decrease in CO $J$=1-0 emission as a function of \gm.  The
analogous grids for the CO $J$=2-1, 3-2, 4-3, 6-5, 7-6, 
10-9 and 16-15 transitions are presented in
Figure-\ref{fig:CO-grids} (and a subset of those for the $^{13}$CO
lines) are displayed in Figure-\ref{fig:13CO-grids}. {We note that we 
will be using $^{12}$CO and CO interchangeably in referring to carbon monoxide.}

At low densities ($n < 10$~\cmt), {the emission of CO} is very weak
($\sim 10^{-10}$~ \inten) compared to emission at mid- to high
densities ($n > 10^3$~\cmt). This is obvious in
Fig.-\ref{fig:13CO-grids}.  {As was mentioned also earlier}, we notice
that at {such low densities} the emission contours show almost no
dependence on \go. This is simply explained by the fact that there are
few collisions to excite the upper rotational levels, and that the
background dust emission (which is only weakly dependent on $G_0$) is
dominating.  Another contributing factor is the low gas temperature
($\sim 10$~ K) in the molecular zone.  On the other hand at mid- and
high-densities, where $n \gtrsim n_{cr}$ of CO($J=$ 1-0), we start
seeing a strong dependence of these emission on \go.

In general the emission intensity is positively correlated with \gm, i.e. it increases 
with increasing \gm~(see Figs.-\ref{fig:CO-1-0-gmech}, \ref{fig:CO-grids},\ref{fig:13CO-grids}).  The 
only exception occurs at some of the high density regions in the CO(1-0) and $^{13}$CO grids.
{We can see such a behaviour in} the upper right corner of the CO grids {when} $\alpha = 0.5$ and 1.  In the
$^{13}$CO grids the emission decrease is clearest (and covers a larger part
of the grid), for example see the $J$=1-0, $J$=2-1 and $J$=3-2 grids
in Fig.-\ref{fig:13CO-grids}. 

In well-irradiated, low-density regions ($n < n_{\rm cr}$ and \go~
$>10$), the emission increases up to two orders of magnitude (see red
regions in Figure-\ref{fig:CO-1-0-gmech}). This is caused by (1) a
rise in the temperature induced by \gm, and (2) a higher abundance of
CO (and $^{13}$CO), causing a double increase.  The abundance of CO is
boosted because its is accelerated via the reaction H + CO$^+$
$\rightarrow$ CO + H$^+$.

At the higher densities ($n > n_{\rm cr}$ and $\gg n_{\rm cr}$) the
response to \gm~becomes weaker.  More than one factor contributes to
this.  From the thermal perspective the temperatures in the molecular
regions are already high ($\sim 50$ to 100 K).  {This is due to} the
tight coupling between dust and gas.  Hence the relative increase in
the gas temperature is small, in contrast to what happens in the
low-density models.  Another contributing factor is the decrease in
the abundance of CO.  In the absence of \gm, most C atoms are locked
in CO molecules. For extreme \gm~ ($\alpha \ge 0.5$) the abundance of
most molecules, including H$_2$, {decreases drastically \citep{mvk12};
  leading to} \xco~lower by almost 3 orders of magnitude (from $\sim
10^{-4}$ to $\sim 10^{-7}$).

{Although we would suspect} that the higher temperatures ({due to} \gm) {would lead to enhanced 
emission, the reduced abundance and column density of CO counteracts that enhancement.  These
two effects combined lead to a small relative increase in emission.  As an example, for low
densities an increases in the emission by one order of magnitude is easily attained for 
$\alpha = 0.01$.  On the other hand an $\alpha = 0.1$ is required 
to enhance the CO emission by the same factor in the high density region of the 
grids (cf. bottom-left panel of Figure-\ref{fig:CO-1-0-gmech}).  The only exception 
where a decrease in the emission is observed occurs in the CO(1-0) line.
This is not surprising, since in looking at the left panel of
Figure-\ref{fig:CO-ladder} we see that the 1-0 transition is weakly
effected by \gm~(compared to the higher transitions).  So a reduced
$N$(CO) leads to lower emission of the 1-0 line.}

A similar behaviour is observed for the $^{13}$CO {emission} which is,
however, more sensitive to \gm~than $^{12}$CO.  In particular, as \gm~{increases, the
mid- and high}-density regions show a stronger decrease in the 
emission {of the first three $J$ transitions.  This decrease is due to a reduced $N$(\thco) and 
an already low optical depth.  $N$(\thco) is about five times lower 
than $N$($^{12}$CO).  Moreover the Einstein} $A$ coefficients of \thco~are about 
20\% lower than those of $^{12}$CO.  Hence the upper levels, {which are mainly excited 
collisionally, are} de-populated less frequently, and a higher upper-level population
density is maintained.  
We see that {these two factors lead to the reduced optical depth of \thco.  
At mid- and high-densities $N$(\thco) decreases as $\alpha$ increases. 
Since the optical depth is already low, it neither plays a significant role in 
blocking the emission nor in enhancing the emission by allowing 
``trapped'' radiation to escape from the cloud; this is also true when the cloud becomes 
more transparent as $\alpha$ increases.
Consequently, since intensity is proportional to the column density of the emitting 
species, a decrease in $N$(\thco) results in lower emission.  On the other hand, 
the $J > 3 - 2$ transitions show enhanced emission all over the grid.  This occurs
despite the reduced column densities of \thco. High kinetic temperature due to 
\gm~enhances the strong pumping of the associated level populations; eventually, this 
counteracts the effect of the reduced $N$(\thco) on the emission.}

Although the generic behaviour of the CO and $^{13}$CO grids are similar,
the differences among them are interesting enough to have a closer look 
at their line ratios. Especially for ones involving the low-$J$ lines 
where the \thco~ lines are optically thin.

\begin{figure*}[!tbh]
  \centering
  \includegraphics[scale=0.39]{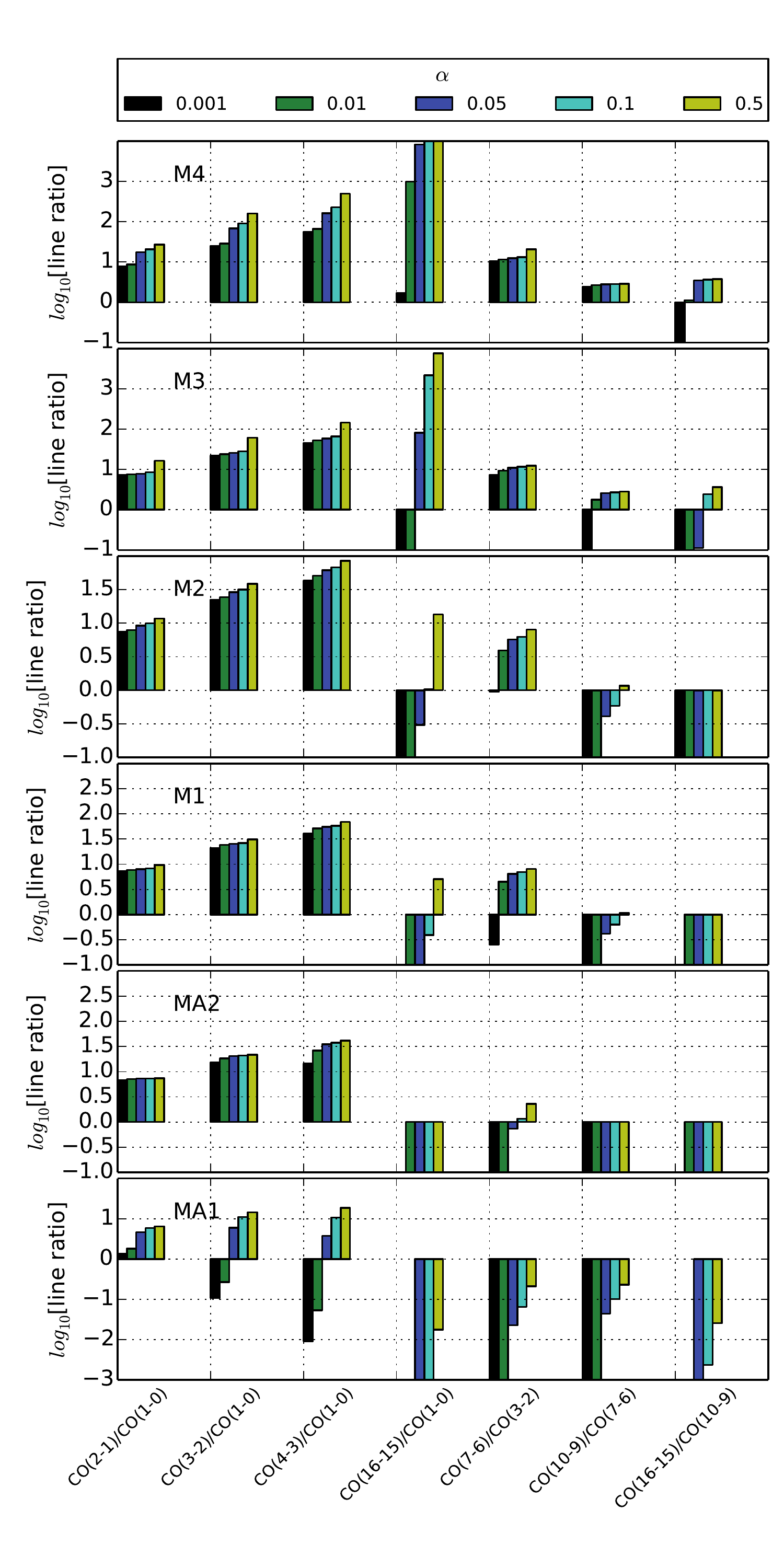}
  \includegraphics[scale=0.39]{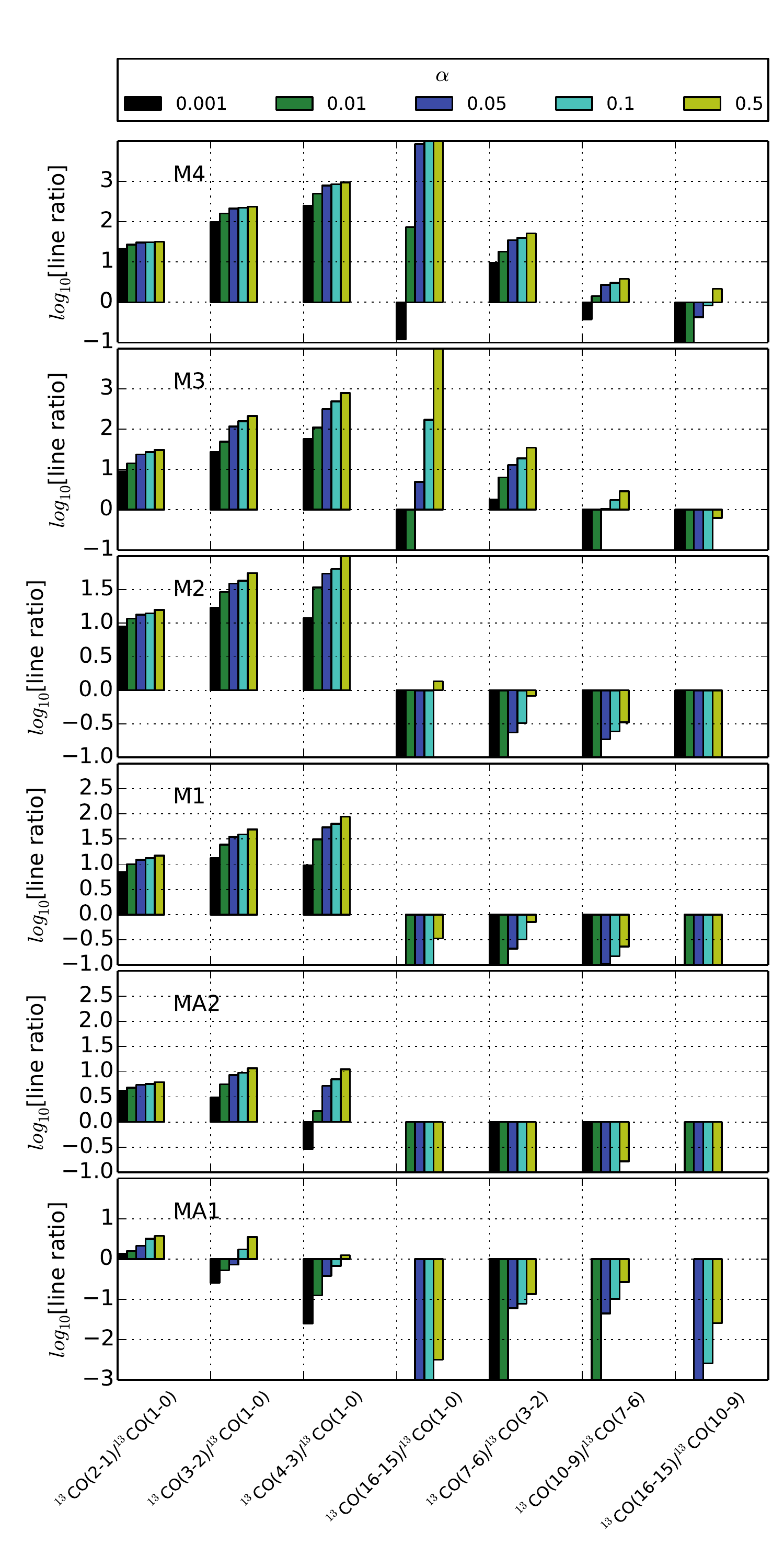}
  \includegraphics[scale=0.39]{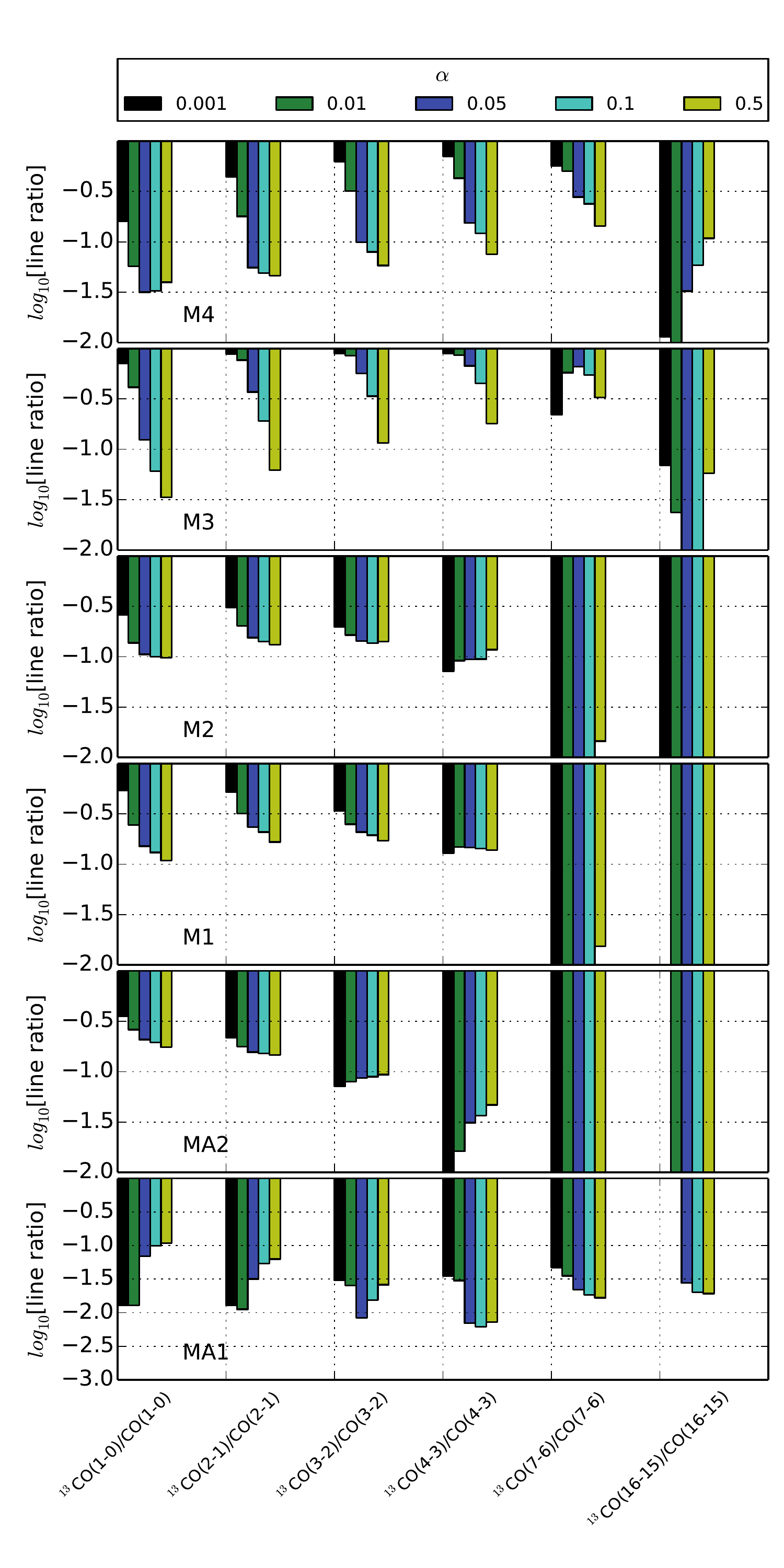}
  \caption{line ratios of CO and $^{13}$CO for different amounts of \gm~($Z=$ \zsun)
     for the reference models.
\label{fig:bar-plot-lineRatios-CO-13CO-13CO-CO-z-1.0}}
\end{figure*}

In Fig.\,\ref{fig:bar-plot-lineRatios-CO-13CO-13CO-CO-z-1.0}, we show
the line ratios for our reference models.  These ratios include
transitions for lines within the CO ladder (left panel), the $^{13}$CO
ladder (middle panel), and the $^{13}$CO to $^{12}$CO line ratios {(right panel)}.  The
ratios between high-$J$ and low-$J$ transitions show a strong
dependence on \gm.  For example, in M3 the CO(16-15)/CO(1-0) ratio
changes abruptly from $\lesssim 0.1$ to $>100$ when $\alpha > 0.05$.  A similar 
behaviour is observed for the corresponding $^{13}$CO ratio. In
Fig.\,\ref{fig:CO-ladder}, we see that the optical depth of $J$ = 16-15
CO is almost unaffected as \gm~increases, whereas that of the
$J=$ 1-0 line decreases rapidly from $\sim 100$ to $\sim 1$.  The
opposite is observed when it comes to the intensity, where
the $J$ = 16-15 CO line emission increases by $\sim$ four orders of
magnitude.  This explains the huge increase in the ratio of those
lines. This line ratio along with the ratio CO(16-15)/CO(10-9) are the only ones 
(among the ratios we looked at) that show {a significant change for the} 
high-density model M5 (see {the} top row of the middle panel in
Figure-\ref{fig:bar-plot-lineRatios-CO-13CO-13CO-CO-z-1.0}). 

In medium- to high-density models, ratios involving low-$J$
transitions are less sensitive to \gm. Those ratios are almost
constant in the high density models, since the lines are thermalized
and the population densities do not change relative to each other.
However in the low- and medium-density models MA1, MA2, and M1 the
CO(4-3)/CO(1-0) ratio might be a good diagnostic for \gm.  In MA1 we
see that this line ratio increases by a factor of $\sim$ 10 (as well
as the corresponding $^{13}$CO ratio).  Because these are ratios
within the same species, the increase is a pure measure of the
radiative properties of the specie.  As in the cases of MA1 and MA2,
we are in the non-LTE phase.  As the temperature increases, the
upper levels are populated faster leading to stronger emission in e.g. the
CO(4-3) line, which drives up the ratio.

The most interesting and useful behaviour of the $^{13}$CO/$^{12}$CO (right 
panel of Fig.-\ref{fig:bar-plot-lineRatios-CO-13CO-13CO-CO-z-1.0}) occurs 
in the high density models M3 and M4.  The ratios decrease
monotonously from $\sim 0.5$ (for $\alpha = 0$) to $\sim 0.1$ (for
$\alpha = 1$).

We {showed previously that the optical depth of CO decreases with
increasing $\alpha$.  Also \thco~ is optically thin in general.  So for
column densities corresponding to $A_V > 10$ mag, we expect to have a
steeper dependence of the line ratios of CO and \thco~ on \gm.}

{We} summarize the main results of this section by emphasizing that in
{low density PDRs, the} ratios of high-$J$ to
low-$J$ {emission lines of} CO and $^{13}$CO might be useful diagnostics for \gm.  The
{line ratio} of \thco~{with its isotopologue CO} are good diagnostics for \gm~in
high-density PDRs. They show a strong and clear trend.
\vspace{0.7cm}
\newline
\indent {\bf HCN and HNC}\\
HNC and HCN are linear molecules with very similar radiative
properties in the infra-red regime.  Both have (a) large dipole moments
(3.05 and 2.98 respectively \citep{Botschwina93}). This allows both of them to be
easily observed (b) both have high critical densities $> 10^5$~\cmt~ 
for the 1-0 transition (see Table-\ref{tbl:atomSpecsProp}). {This renders them 
as good tracers for high-density molecular gas}.  In this paper,
we consider the rotational transitions from $J = 1-0$ up to $J = 4-3$,
which are commonly observed.  {Since the emission of these lines is} very sensitive  
to temperature changes, {they} might be useful {in identifying molecular} clouds  
dominated by mechanical feedback.  In Paper I we studied the column
density ratios of the two species. We showed that HCN becomes more
abundant than HNC as \gm~increases.  Here we {study the effect of 
\gm~on the emission of these two species and their ratios.}

\begin{figure}[!tbh]
  \begin{minipage}[b]{1.2\linewidth} 
  \includegraphics[scale=0.7]{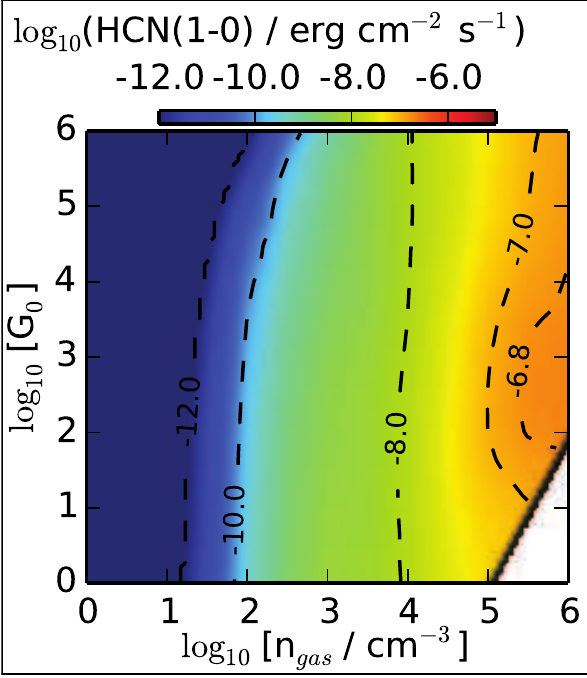}
  \includegraphics[scale=0.7]{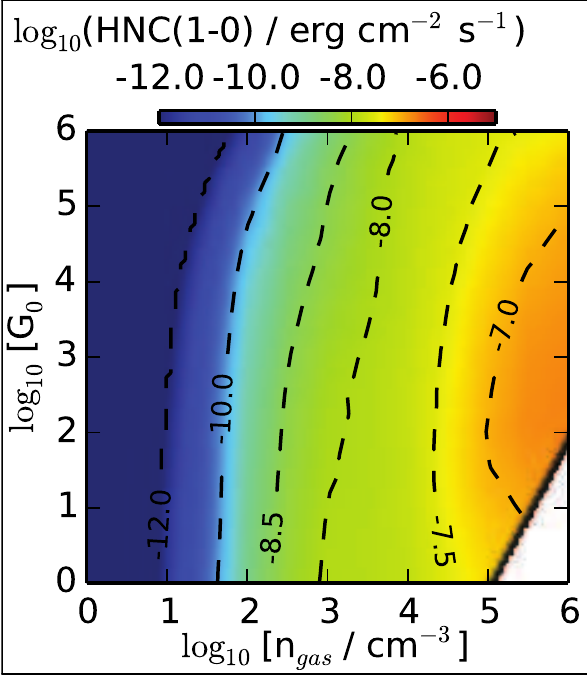}
  \end{minipage}
  \caption{HCN and HNC emission of the 1-0 line in the absence of \gm~($Z=$~ \zsun). 
    \label{fig:HNC-HNC-base} }
\end{figure}

In the absence of any \gm~the emission grids show a very weak
dependence on \go, as compared to their dependence on $n$. This 
dependence is illustrated in Figure-\ref{fig:HNC-HNC-base} .  
{This kind of dependence is generic to cases where the gas density
is (much) lower than the critical density of the line considered; which 
is also the case here.}

In contrast to the weak dependence on \go, the dependence on \gm~is
quite strong.  The enhancements in the emission of HCN are stronger
than those of HNC the bottom two rows of Figure-\ref{fig:HNC-HNC}).
This is {simply} because HCN becomes more abundant than HNC.  {The
  main channel through which the conversion occurs is} via the
reaction H + HNC $\rightarrow$ H + HCN \citep[see][for a more
elaborate discussion on the chemistry]{meijerink11}.  {This process
  becomes dominant for $T_{kin} > 150$K in the molecular region
  \citep{schilke92}, which is achieved when $\alpha > 0.1$. Below that
  threshold in $\alpha$, HNC is equally destroyed via ion-neutral
  reactions with H$_{3}$O$^+$, especially at low densities}. Another
contributing factor to the increase in the abundance of HCN is its
efficient formation via the neutral-neutral reactions with H$_2$.  For
{completeness's} sake, we note that the relative change in the
emission of the $J = 4 - 3$ line is stronger than that of the 1-0 line
for reasons discussed in section-\ref{subsub:molspec}.

\begin{figure}[!tbh]
  \centering
  \includegraphics[scale=0.55]{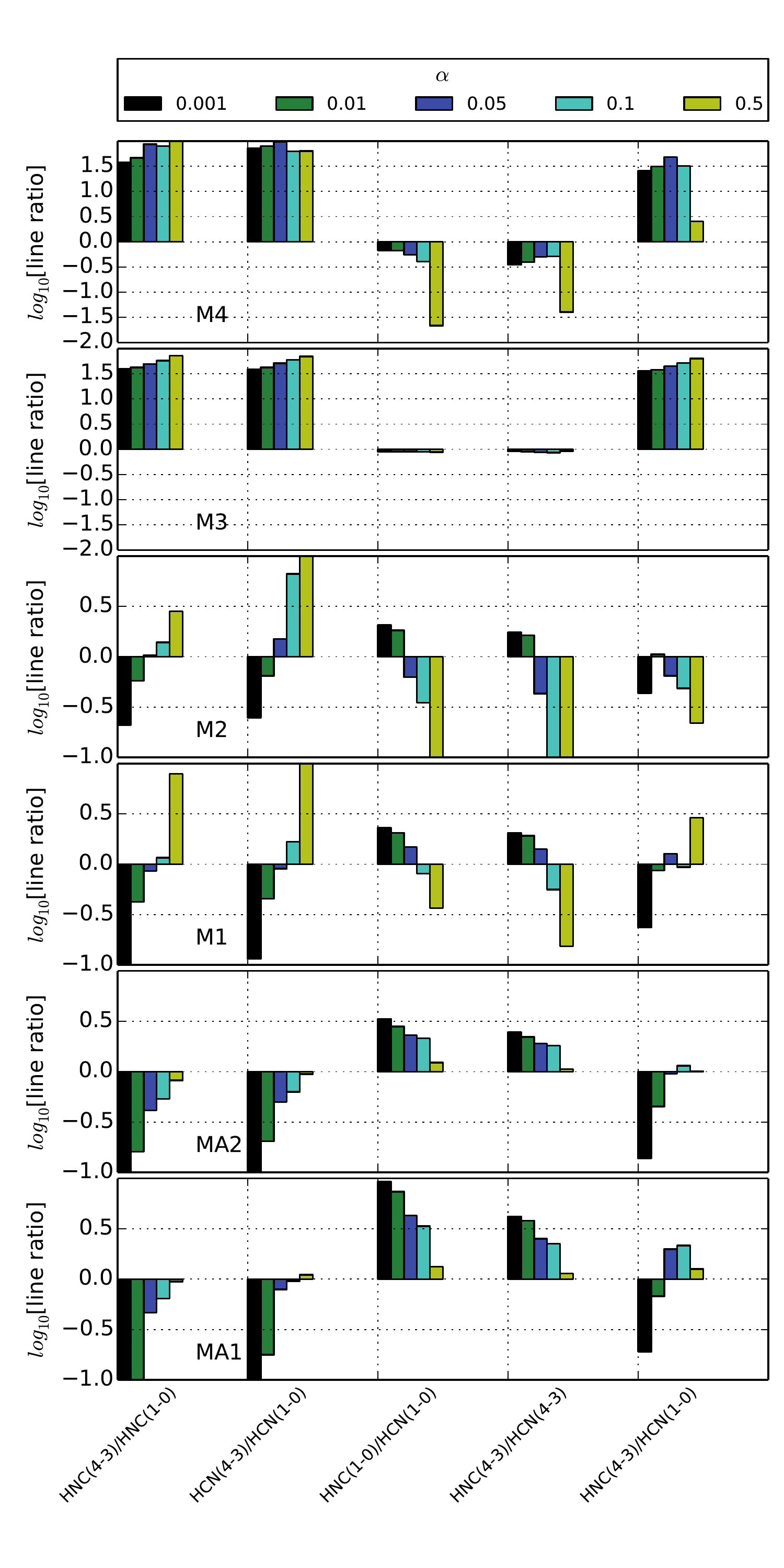}
  \caption{Various line ratios of HNC and HCN as a function of \gm~($Z=$\zsun)
    for the reference models.
\label{fig:bar-plot-lineRatios-HNC-HCN-z-1.0}}
\end{figure}

In Fig.\,-\ref{fig:bar-plot-lineRatios-HNC-HCN-z-1.0} we show the line
ratios considered for HCN and HNC.  Transition ratios such as
HNC(4-3)/HNC(1-0) and HCN(4-3)/HCN(1-0) behave as expected (first two
columns in the figure): with increasing $\alpha$, they increase as well
(since the higher levels are populated more easily at higher
temperatures).  In M1 and M2, the ratio increases quite fast linearly in log scale.  
Unfortunately, in M3 and M4 the ratios are almost constant and are thus of little 
use as a diagnostic for high-density PDRs.

Again in M1 and M2, the {inter-species} ratios HNC(1-0)/HCN(1-0),
HNC(4-3)/HCN(4-3) and HNC(4-3)/HCN(1-0) are strongly dependent on
\gm. We see that for $\alpha = 0.5$, the ratio HNC(1-0)/HCN(1-0) drops
from $\sim 2$ to $\sim 0.3$.  This is caused primarily by the
difference in column densities caused by the {chemical effects} discussed
above.  Since those ratios depends monotonously on $\alpha$ (see M2 in
Figure-\ref{fig:bar-plot-lineRatios-HNC-HCN-z-1.0}), we may consider
them as as a good diagnostic for such PDRs. This is not the case for
M3 and M4.  The line ratios have a weaker dependence on mechanical 
heating whenever $\alpha < 0.1$. However an abrupt decrease (from 
$\sim 0.3$ to $< 0.01$) is observed for $\alpha \gtrsim 0.5$.

Interestingly enough at metallicities typical for galactic {centre
regions ($Z = 2$~\zsun), clouds such as M4 show a significant
dependence on $\alpha$} (see Figure-\ref{fig:bar-plot-lineRatios-HNC-HCN-z-other}).  This 
is simply
a result of the fact that a higher metallicity implies a higher column
density of the gas and the species in question.  In such situations,
fluctuations of the abundances in the radical region play a minor
role.

\begin{figure}[!tbh]
  \begin{minipage}[b]{1.2\linewidth} 
  \includegraphics[scale=0.7]{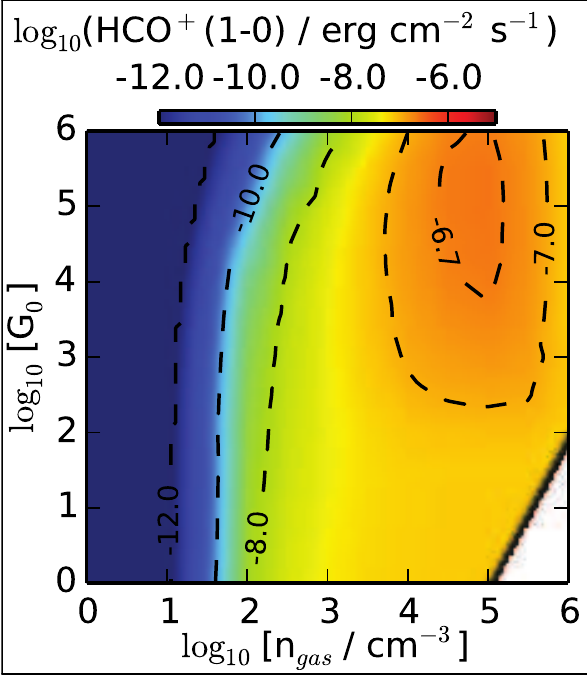}
  \includegraphics[scale=0.7]{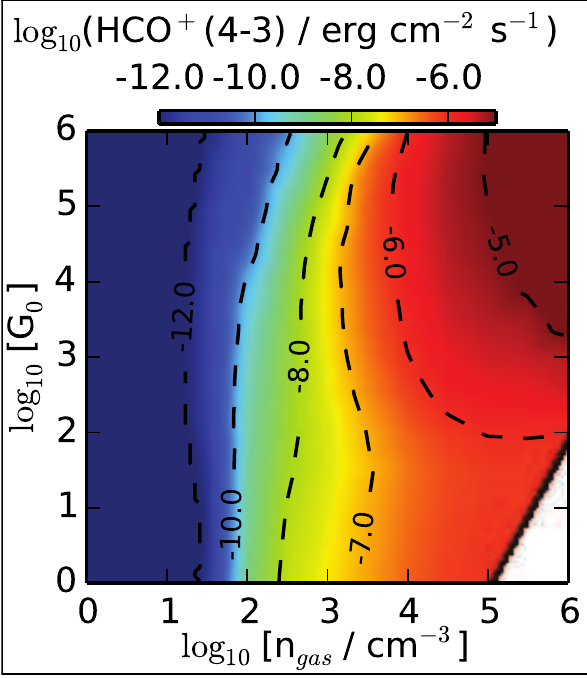}
  \end{minipage}
  \caption{HCO$^+$ $J= 1-0$ and $J= 4-3$ emission in the absence of \gm~for solar 
  metallicity ($Z =$~\zsun).
    \label{fig:HCOP-base} }
\end{figure}

In summary, line ratios such as HNC(1-0)/HCN(1-0) and
HNC(4-3)/HCN(4-3) are good diagnostics for \gm~in PDRs in the
following cases : (a) at gas densities less than the critical
densities of the lines mentioned, and  (b) in {high-density PDR} 
environments such as galaxy centres with super-solar metallicities
{and star-burst regions.}
\vspace{0.7cm}
\newline
\indent {\bf HCO+}\\
HCO$^+$ is another high density tracer.  The critical densities (at
50K) of its $J =1-0$ and $J =4-3$ emission lines are $\sim 2 \times 10^5$~\cmt~and 
$\sim 10^7$~\cmt~respectively.  The emission grids for both lines are
shown in Figs.\,\ref{fig:HCOP-base}.  The corresponding grids as a function of 
$\alpha$ are presented in Fig.\,\ref{fig:HCOP-base-and-gmech} in the Appendix.

{We will discuss only mid- and high-density models ($n > 10^3$ \cmt~), since
at lower densities the lines would be too weak to be observed.  With increasing \gm~,
the emission of the $J$=1-0 line in the high density region (and \go~ $>10^5$) decreases 
by a factor of $\sim 2$. At such high densities, \xhcop~drops by three orders of magnitude, 
leading to the reduced emission.  We trace the source of the reduced abundance of 
\xhcop~to it slow production rate; which is reduced  by an order of magnitude as $\alpha$ increases.
This slowing down is mainly due to the reaction of the ionic 
species HOC$^+$ and CO$^+$ with H$_2$ through which HCO$^+$ is formed.  The abundance of these 
two ionic reactants drops by factors of two and two hundred respectively as $\alpha$ increases, hence 
the slow production rate of HCO$^+$.  As long as the HCO$^+$($J = 1-0$) grid is concerned, the emission 
is enhanced everywhere else throughout the grid. See the dashed line 
in Figure-\ref{fig:HCOP-base-and-gmech}.}

For PDRs with moderate densities i.e in the non-LTE phase of the rotational line of HCO$^+$, the 
coupling between dust and gas is weak compared to that in high-density clouds.  This weaker coupling results 
in a weaker dependence of the abundance on \gm~(since the physical conditions do
not change much).  However, the temperature does increase in the
molecular zone where HCO$^+$ is present. This increase in the
temperature enhances the emission of the $J =1-0$ line for densities $n
< 10^4$~ \cmt~.

The $J=4-3$ line responds in similar way to changes in \gm, but the
emission decrease only for $\alpha=0.05$ and 0.1.  This emission is
more sensitive to temperature changes because of the ease of
populating upper levels.  Thus the emission is boosted again for
$\alpha > 0.1$, even though \xhcop~ decreases for $n > 10^5$\cmt.

\begin{figure}[!tbh]
  \centering
  \includegraphics[scale=1.0]{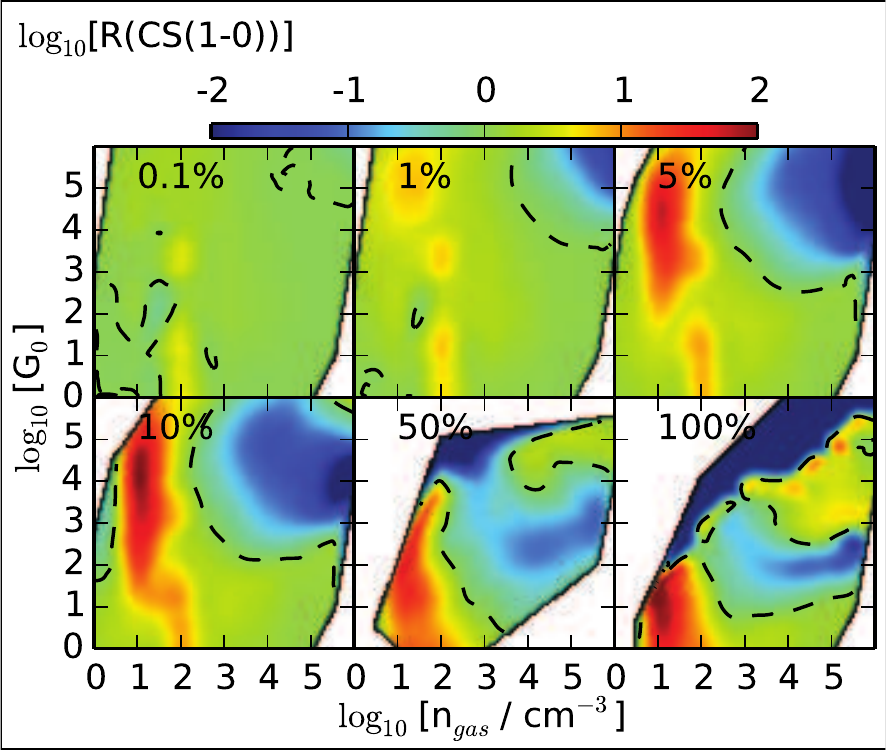}
  \caption{Grids of the relative changes in the emission of the CS(1-0)
    line for different values of $\alpha$ ($Z=$\zsun). See also 
    Figure-\ref{fig:C+-1-0-grids} for a description of the colours.
    \label{fig:CS-1-0-gmech} }
\end{figure}

HCO$^+$ is interesting because it behaves quite differently in the
{mid- and high-density} regimes. It can be used as a diagnostic
for both regimes in combination with other species (as we discuss below). 
\vspace{0.7cm}
\newline
\indent {\bf CN}\\
{
The critical densities of the CN($1_{1/2}$-$0_{1/2}$) and CN($2_{3/2}$-$1_{3/2}$) 
is in the high density part of the parameter space (see Table-\ref{tbl:atomSpecsProp}). Similar 
to HCO$^+$, the emission grid of those lines also exhibits a peculiar dependence on 
increasing amounts of mechanical feedback.  In looking
at the abundance of CN, we see that at high densities $x_{\rm CN}$ correlates 
negatively with $\alpha$.  The reduction in the abundance is caused by the 
high temperatures in the molecular zone of the PDR.  The high temperatures leads to 
the destruction of CN, at a rate which is an order of magnitude higher compared to pure PDRs,
through the reaction H$_2$ + CN $\rightarrow$ HCN + C. The reduced abundance 
of CN results in the dimming emission as \gm~is introduced, which is evident 
in (Fig.\,-\ref{fig:CS-CN-base-and-gmech}).  Beyond $\alpha = 0.1$ 
$N$(CN) becomes too low, where the intensities (in both lines considered) decrease 
by a factor of 10. 

On the other end of the parameter space (at mid- and low densities), 
N + CN $\rightarrow$ N$_2$ + C is the dominant reaction for the full range in
$\alpha$.  This reaction maintains a high $x_{\rm CN}$ in that part of the 
parameter space, i.e the region bounded by the dashed contour line 
in the bottom row (where $\alpha = 0.1, 0.5, 1.0$) of the panel corresponding 
to CN in Figure-\ref{fig:CS-CN-base-and-gmech}.}
\vspace{0.7cm}
\newline
\indent {\bf CS}\\
This species has a very distinctive dependence on \gm~(compared to the
other molecular species we have so far considered).  In
Fig.\,\ref{fig:CS-1-0-gmech} we see that a region of suppressed
emission sweeps across the grid from high- to low- density as $\alpha$
increases.  This non-trivial behaviour is difficult to explain in
detail.

{The $n_{cr}$ for CS is $\sim 5 \times 10^4$~ \cmt~ and $\sim 3 \times
10^6$~ \cmt~ for the $J =1-0$ and $J =4-3$ lines respectively (at 50K).
Up to $\alpha = 0.05$ both grids indicate a strong decrease in the emission for
$n>10^3$\cmt~and \go$>10^3$.  The reduction in the emission is as 
low as a factor of 50 for high density PDRs. As $\alpha$ increases further 
the emission of those PDRs starts to increase again relative to the 
case of $\alpha = 0$.  This increase} reaches a factor of 50 for $J =4-3$ 
transition. Meanwhile, the region where the emission are suppressed is 
pushed to lower densities and lower \go. This is a consequence of the 
chemistry.  At the high densities, the drop in the emission is due to 
reactions with cosmic rays.  Those reactions become dominant in 
destroying CS instead of the neutral-neutral reaction O + CS $\rightarrow$ S + CO; 
{which otherwise} is the dominant reaction in {pure PDRs.

The strong dependence of the CS lines on $\alpha$ makes it a useful
candidate for mechanical feedback.}

\begin{figure}[!tbh]
  \centering
  \includegraphics[scale=0.55]{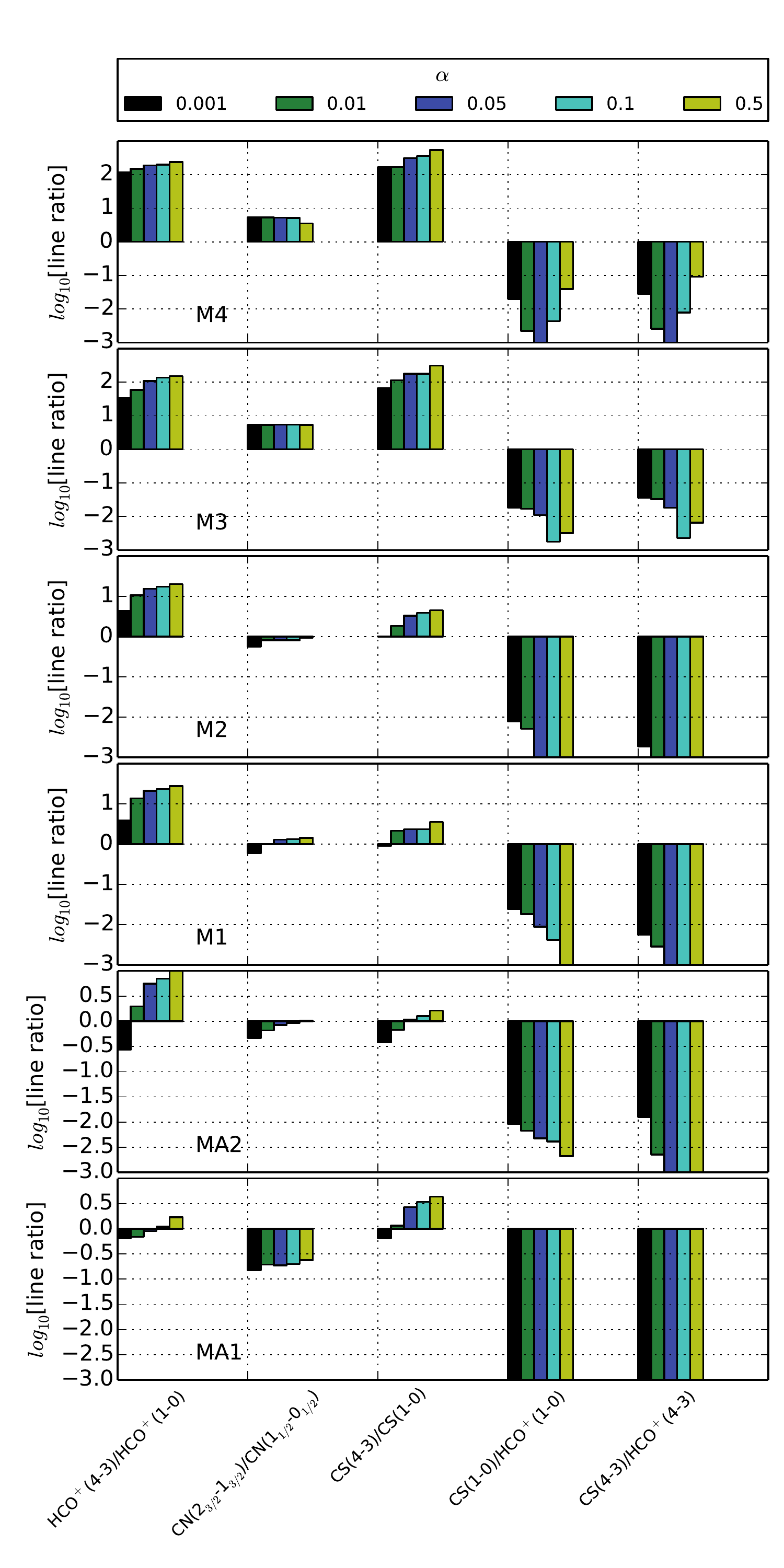}
  \caption{Line ratios of HCO$^+$, CN and CS for $Z=1$\zsun~and $A_V = 10$~ mag
    for the reference models. \label{fig:bar-plots-HCO+-CN-CS-z-1.0} }
\end{figure}

In Fig.\,\ref{fig:bar-plots-HCO+-CN-CS-z-1.0} we show the line ratio
for CS, CN, and HCO$^+$.  In all the reference models the line ratio of 
HCO$^+$(4-3)/HCO$^+$(1-0) and particularly the interspecies CS/HCO$^+$ ratios 
vary over more than one order of magnitude as a function of $\alpha$.
The CS(4-3)/CS(1-0) ratio shows some dependence on \gm, but
again {those variations} are small compared to the ones mentioned before.  
{The} line ratios CS(1-0)/HCO$^+$(1-0) and CS(4-3)/HCO$^+$(4-3) {are more interesting}.  In 
looking at the CS grids in Figs.\,\ref{fig:CS-1-0-gmech}, we see that
variations are well {described} in the CS/HCO$^+$ ratios, which
allows them to be used diagnostically to constrain \gm~in 
{extreme star-bursts}.  For instance a ratio less than
0.01 would {imply} an $\alpha \sim 0.05$; whereas a ratio around 0.1 
implies $\alpha > 0.1$ (see last column in Fig.\,\ref{fig:bar-plots-HCO+-CN-CS-z-1.0}).

In M1 and M2, the line ratio behaviour is the same at higher or lower
metallicities (see Fig.\,\ref{fig:bar-plot-lineRatios-HCO+-CN-CS-z-other}).  In M3 
and M4, t{his ratio's} response to changes in $\alpha$ is slightly weaker at
Z~$= 0.5$ \zsun.  In the lowest metallicity case, CN($2_{3/2}$-$1_{3/2}$)/CN($1_{1/2}$-$0_{1/2}$) 
decreases to unity as $\alpha$ increases. This might be useful in probing
\gm~in e.g. dwarf galaxies.

\subsection{Other line ratios \label{subsec:miscRatio}}
\begin{figure}[!tbh]
  \centering
  \includegraphics[scale=0.55]{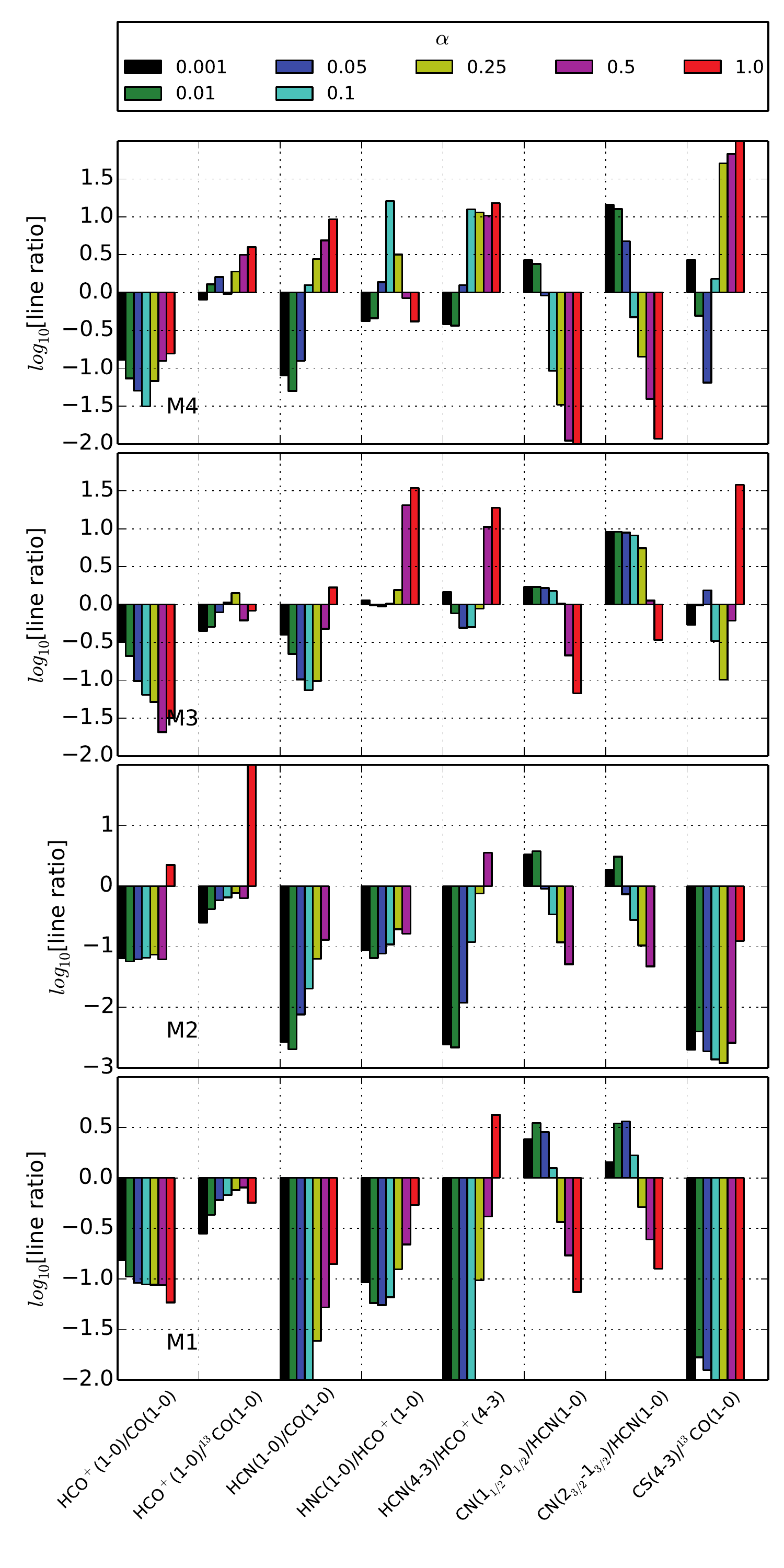}
  \caption{Line ratios with strong dependence on \gm~for
    $Z=1$\zsun for the reference models. \label{fig:bar-plots-msic1-z-1.0} 
  }
\end{figure}

Fig.\,\ref{fig:bar-plots-msic1-z-1.0} shows some other molecular line
ratios, selected to illustrate {their importance as a diagnostic for} \gm~in PDRs.

Most of the line ratios exhibit an order of magnitude change for
$\alpha \lesssim 0.25$.  Typical examples are HCO$^+$(1-0)/CO(1-0)
and HCN(1-0)/CO(1-0) in the {high-density} models such as M3
and M4.  On the other hand, HCO$^+$(1-0)/$^{13}$CO(1-0) shows an irregular increase
as a function of $\alpha$ in these models, but a stricktly monotonous increase 
(from $\sim$ 0.01 to 1) {is observed} in the lower density models M1 and M2.

{Ratios involving lines} of HCN with CO and HCO$^+$ are excellent
candidates for constraining \gm.  This is also the case at lower and
higher metallicities (see Fig.\,\ref{fig:bar-plot-lineRatios-misc1})
for all the representative models. In some cases, such as
HCN(4-3)/HCO$^+$(4-3) in M4, the ratio increase from $\sim 0.3$ to 10
for $\alpha \sim 0.1$.  One drawback in the HCN(1-0)/CO(1-0) ratio is
the degeneracy in its dependence on \gm~in M3.  For example in the 
absence of \gm, this ratio has a value of 0.3. It reaches a 
minimum of 0.01 (for $\alpha = 0.1$) and increases back to $\sim 0.3$ for 
extreme {cases where} $\alpha = 0.5$. Cases of such degeneracies can be 
resolved by simultaneously considering other line ratios, as we will 
demonstrate at the end of this section.

We draw special attention to the CN/HCN interspecies ratios. They
have a very strong dependence on \gm, showing a decrease by an order 
magnitude for the lowest transition ratios as $\alpha$ increases from 0
to 1. 

In summary, we found that line ratios between CN, HCN and HCO$^+$ are
quite useful in constraining \gm.  This is true in particular for CN
since \gm~seems to drive ratios to values well below unity for most
clouds when $\alpha > 0.1 $.  In high-density clouds whose heating budget is 
dominated by \gm, line ratios of HNC/HCO$^+$ tend to exceed unity.

\section{Application}\label{sec:app}
We have presented model predictions for line intensities and line ratios of many molecular species. 
However, we have not yet reflected on any observational data to 
which these models can be applied. In this section we use actual data
and demonstrate a) the importance of molecular line ratios as a diagnostic 
for mechanical heating and b) their usefulness as a tool for constraining it.

In Table-\ref{tbl:loenen-data}, we present the range of line ratios involving HCN, 
HNC, CO and HCO$^+$ for infra-red luminous galaxies taken from Table-B.2 in
\cite{baan2008}.

\begin{table}[h]  
\centering
\begin{tabular}{c c c}
  \hline
  ratio & range & reference\\
  \hline
  \hline
  HCN(1-0)/HNC(1-0)     & 1.5 -- 4.0 &  \tablefootmark{a}\\
  HCN(1-0)/HCO$^+$(1-0) & 0.6 -- 3.2 &  \tablefootmark{a}\\
  HNC(1-0)/HCO$^+$(1-0) & 0.3 -- 1.0 &  \tablefootmark{a}\\
  \hline
  HCO$^+$(4-3)/CO(1-0) &  0.1 -- 0.5 &   \tablefootmark{b,c}\\
  \hline
\end{tabular}
\tablefoot{\\
  \tablefoottext{a}{From \cite{baan2008}}\\
  \tablefoottext{b}{From \cite{israel95} }\\
  \tablefoottext{c}{From \cite{kundsen07} }\\
}
\caption{Observed ranges of molecular line ratios for some starburst galaxies.
  \label{tbl:loenen-data}}
\end{table}

We follow a simplistic approach to an ambitious goal. Our aim is to constrain 
($n$,\go, $A_V$, $\alpha$) using the data at our disposal. The only major 
assumption we make is the metallicity of the 
source. Here we will assume solar metallicity.

\begin{figure*}[!tbh]
  \begin{minipage}[b]{1.0\linewidth} 
    \centering
    \includegraphics[scale=1.05]{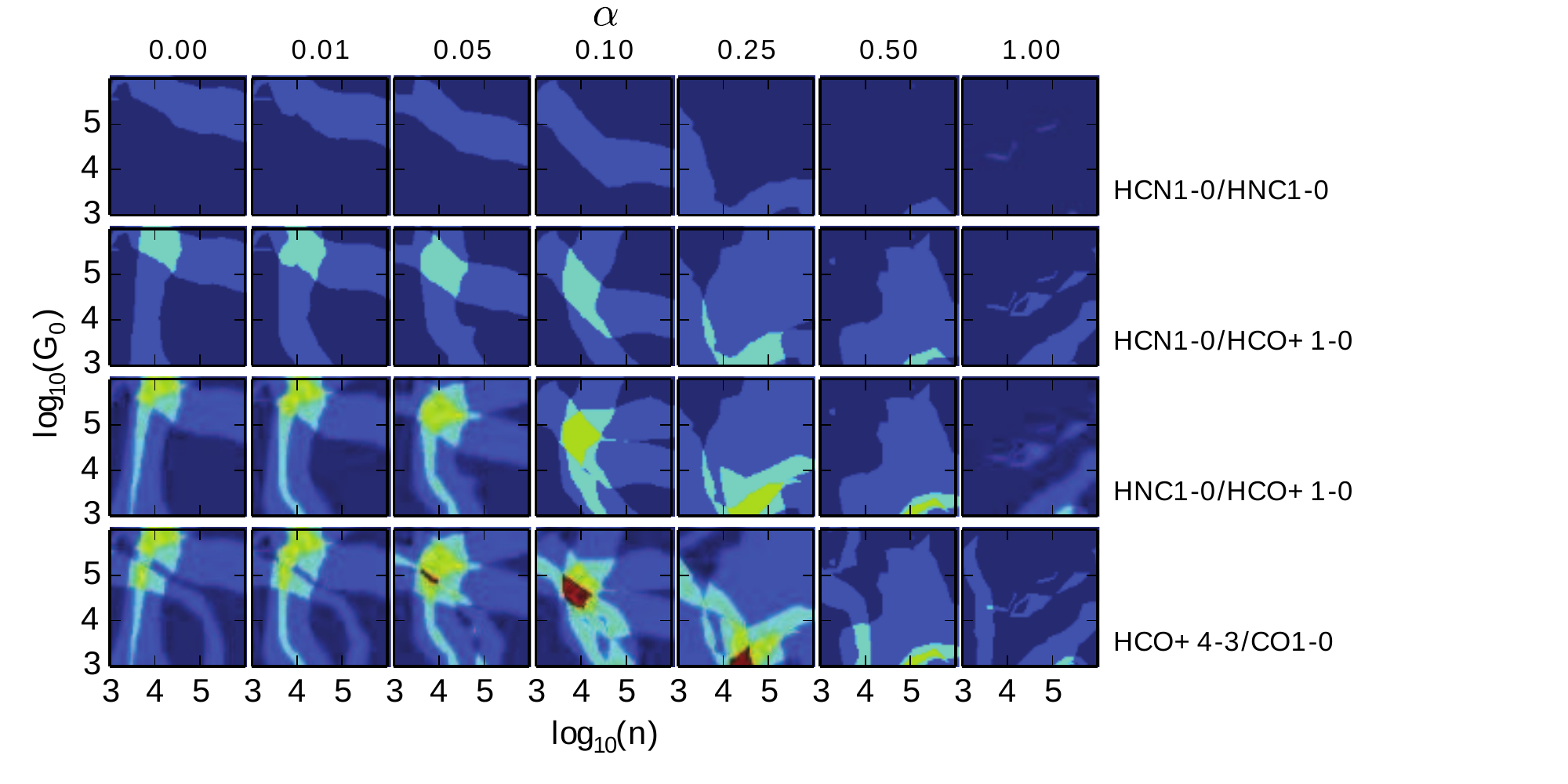} 
  \end{minipage}
  \caption{Constraining the \gm~, $A_V$, $n$ and \go~ for starburst galaxies. In this figure we illustrate the procedure
    used to constrain those parameters for $A_V = 5$~mag, the figures for the remaining $A_V$ are in Figures-\ref{fig:constraining10},
    \ref{fig:constraining30}. Each row corresponds to a certain line ratio, and each column corresponds to a certain $\alpha$. The 
    colors correspond to regions where the observed line ratio in each grid is within the observed range of 
    Table-\ref{tbl:loenen-data}. In the first row regions where HCN(1-0)/HNC(1-0) is between 1.5 and 4.0 is shaded in light blue. In 
    the second row we include HCN(1-0)/HCO$^+$(1-0). The regions which are within the observed line ratios for both 
    HCN(1-0)/HCO$^+$(1-0) and HCN(1-0)/HNC(1-0) are shaded in cyan.  In the third row, we introduce HNC(1-0)/HCO$^+$(1-0), the 
    region satisfying all three line ratios is shaded in green. We notice that green regions exist for all values of $\alpha$ 
    in the third row. Hence we can not constrain $\alpha$ so far. It is only when HCO$^+$(4-3)/CO(1-0) is included, $\alpha$
    is constrained to 0.1 (red region in the $\alpha = 0.1$ column) in the last row. Although much smaller red regions
    are also visible for  $\alpha = 0.05$ and $\alpha = 0.25$, the global minimum would be most likely around $\alpha = 0.1$ 
    region. In applying a similar procedure to grids corresponding to $A_V =$~10 and 30 mag, we do not observe any red region.
    \label{fig:constraining}}
\end{figure*}

In Figure-\ref{fig:constraining} we present a step-by-step procedure to 
constrain ($n$,\go, $A_V$, $\alpha$) using line ratio grids 
{parametrized} with those four quantities. Each small square represents 
a grid as a function of $n$ (horizontal axis) and \go (vertical axis), like all the 
previous grids that we have shown so far. Each collection of grids illustrates the 
constraining procedure for a certain $A_V$.  The collection of grids in Figure-\ref{fig:constraining}
corresponds to $A_V = 5$~mag.  The ones for $A_V =$ 10 and 30 mag can be found in the 
appendix in Figures-\ref{fig:constraining10} and \ref{fig:constraining30}.  In each column, grids for 
different $\alpha$ are presented.

In the top row of each collection, regions where the line ratio of HCN(1-0)/HNC(1-0) 
is within the observed range (see Table-\ref{tbl:loenen-data}), are delineated in light blue.
Clearly, the HCN(1-0)/HNC(1-0) ratio does not constrain all four parameters. However if $A_V$ 
and $\alpha$ are known, one can constrain the range in $n$ and \go~ for the source. For 
example, if we know a priori that $A_V = 5$~mag and $\alpha=0$ (pure PDR), then the UV flux 
is constrained to \go~ $>10^5$ (see the grid with $\alpha = 0$ in the first row of 
Figures-\ref{fig:constraining}) while the gas density is not constrained.  On the other hand, if 
$\alpha$ is known to be 0.5 (which is quite extreme), then $n$ and \go~ are constrained to a 
much narrower region.  Including the information about HCN(1-0)/HCO$^+$(1-0) from Table-\ref{tbl:loenen-data}, 
helps us better constrain all four parameters (see the second row in the mentioned Figures).  
Although {now} $n$ and \go~ are better constrained (cyan regions), $\alpha$ is still degenerate. 
Similarly, HNC(1-0)/HCO$^+$(1-0) fails in achieving our goal [yellow zones in the third row]. 

Based on our observation in the results section, that \gm~has a strong signature on high-$J$ 
transitions (which was more or less ubiquitous for all species), we attempt adding ratios of 
observed lines involving a high-$J$ and a low-$J$ transition. We use 
the $J=4-3$ transition of HCO$^+$ of NGC253 as a guide. It is clear that this ratio 
manages to constrain all four parameters, with moderate certainty to,  $A_V \sim 5$~mag, 
$ 10^{3.5} < n < 10^4$, $ 10^{4} < G_0 < 10^{4.5}$ and $\alpha \sim 0.1$. This may not be a 
unique find; however, we expect the $\chi^2$ value (or the minimum for a 
proper statistical fit) to be close to the range constrained by this proof of concept 
simple demonstration.

\section{Conclusion and discussion}

We have studied the effect of \gm~on a wide range of parameter space
in $n$ and \go~ covering six order of magnitude in both ( $1 < n <
10^6$~ \cmt~ and $1 < G_0 < 10^6$).  Throughout this parameter space
we investigated the the most important and commonly
observed molecular emission and atomic fine-structure lines and their ratios.
The explored range in mechanical heating (\gm) covers {quiescent 
regions}, with almost no star-formation, as well as violently turbulent 
star-bursts.  The star-formation rates for those range from 0.001
\modotyr~ to $\sim 100$ \modotyr~ respectively.

The two fundamental questions we try to answer in this paper are: {\bf(a)} is it
possibile to constrain the mechanical heating rate in a {star-forming region} by using molecular
line ratios as a diagnostic? {(b)} how important is \gm~in recovering the 
molecular H$_2$ mass of a {star-forming region} using observed molecular line emission such as 
those of CO?

Before discussing these questions, we {state the main 
characteristics of mechanically heated PDRs we observed in our grids: }

\begin{itemize}
\item The most significant contribution of \gm~to the atomic
  fine-structure line intensities results from enhanced temperatures
  in the molecular zone.  This is especially the case for the [CI]
  lines.  {For clouds whose density is below the critical density of 
  those lines}, half of the emission {intensity emanates from} 
  the molecular zone. FS line ratios, such as [CII] 158\mum/[CI] 369\mum, 
  [OI] 63\mum/[CI] 369\mum~ and [CI] 369\mum/[CI] 609\mum, are good 
  diagnostics for \gm~in low-density PDRs ($n <10^3$~\cmt).
\item High-$J$ to low-$J$ transitional ratios of $^{12}$CO and
  $^{13}$CO, such as CO(16-15)/CO(1-0), are good diagnostics for
  \gm~over the whole density range considered.  In cotrast
  low-$J$ CO line ratios, such as CO(2-1)/CO(1-0) or CO(4-3)/CO(1-0), 
  are useful as diagnostics\footnote{We refer the reader 
  to the end of this section for a small discussion on the difference 
  about regions dominated by cosmic-rays in comparison to ones dominated 
  by \gm.} only for clouds with $n < 10^3$ \cmt.
\item The line ratios of \thco/$^{12}$CO (in both low-$J$ and high-$J$
  transitions) have a strong dependence on \gm.  They decrease as \gm~
  increases. This complements the range in density where low-$J$
  transitions of $^{12}$CO (and $^{13}$CO) can be used as diagnostics for
  \gm.
\item At high metallicities ($Z = 2$~\zsun), HCN and HNC are very good diagnostics
  for \gm~{when \go~$\gtrsim 10^5$, such sources include} star-bursts in galaxy 
  centers.
\item {Line ratios such as} HCN(1-0)/CO(1-0),
  HCN(1-0)/HCO$^+$(1-0), CN(1$_{1/2}$-0$_{1/2}$)/HCN(1-0), CN(2$_{3/2}$-1$_{3/2}$)/HCN(1-0),
  CS(1-0)/HCO$^+$(1-0) {show a strong dependance of \gm, hence they 
  are a good diagnostic of it.}
\end{itemize}

The major conclusions of the paper, which we demonstrated in the
application section is: low-$J$~ transitions alone are not good enough
to constrain mechanical heating; ratios involving high-$J$~ to low-$J$
transitions do a much better at that.

Another major conclusion is the importance of \gm~in constraining
$A_V$ or, equivalently, the hydrogen column density $N_{\rm H}$, which can
be used to determine the molecular mass of the cloud.  In comparing
Figure-\ref{fig:constraining} to the ones of higher $A_V$ in the
appendix, one can see that if \gm~is ignored, it is easy to under- or
over-estimate the $A_V$ by a factor of five (or more).  Ignoring
\gm~also results in more than an order of magnitude error in
estimating the $n$ and \go.  For instance in looking at the last row
of Figure-\ref{fig:constraining}, one can see that when \gm~is
ignored, an error up to two orders of magnitude can be done in
constraining the ranges of $n$ and \go.

We emphasis that our approach in constraining the physical parameters of clouds using
the observed line ratios, is just a proof of concept demonstration. Ultimately
one must use more elaborate minimization methods to attempt to constrain the 
physical parameters.  However, it is most 
likely that the parameters {which best fit} the observations, will be very close to the ones obtained using 
the method adopted in the application section.  We leave it to the interested reader 
to make use of the grids which are published as well with this paper 
(see Figures-\ref{HCN-HCO+-grid-grids}, \ref{HNC-HCO+-grid-grids}, \ref{HCN-HNC-grid-grids} in the 
appendix). 

We finalize our discussion by touching on the effect of cosmic rays (CR). Although it is 
outside the scope of this paper, we explored the effect of enhanced CR rates.  Diagnostic 
line ratio grids for HCN, HNC and HCO$^+$ are fundamentally different from those which 
are dominated by \gm~(see Figure-\ref{CR-grid-grids}). Hence, we expect 
that in using diagnostics presented in this paper, {clouds which are embedded in environments where 
the CR rate is enhanced}, would not be mistaken with {clouds whose heating budget is 
dominated by \gm.}

\begin{acknowledgements}
MVK would like to thank Marissa Rosenberg for useful insight on the observations
and some aspects of the of the modelling and the fitting procedures.  The author is 
grateful also to Alessandra Candian, who provided some guidance on molecular 
properties.  The author would also like to thank F. S. van der Tak for some 
information on RADEX. {Finally, MVK would like to thank the anonymous referee 
whose comments and suggestions helped improve the paper significantly.}
\end{acknowledgements}

\bibliographystyle{aa}


\clearpage

\begin{appendix}{Appendix A: Atomic species} \label{app:a}

\setcounter{figure}{0}
\makeatletter
\renewcommand{\thefigure}{A\@arabic\c@figure} 

\begin{figure*}[!tbh]
  \begin{minipage}[b]{1.0\linewidth}   
  \centering 
  \includegraphics[scale=0.7]{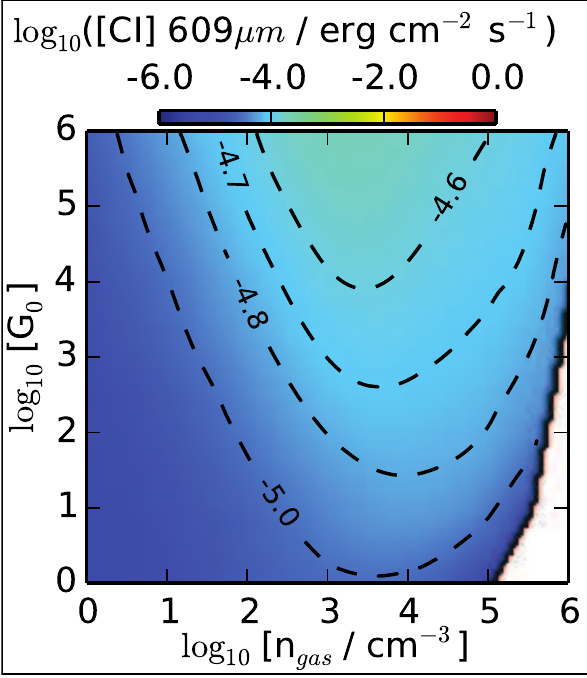}
  \includegraphics[scale=0.7]{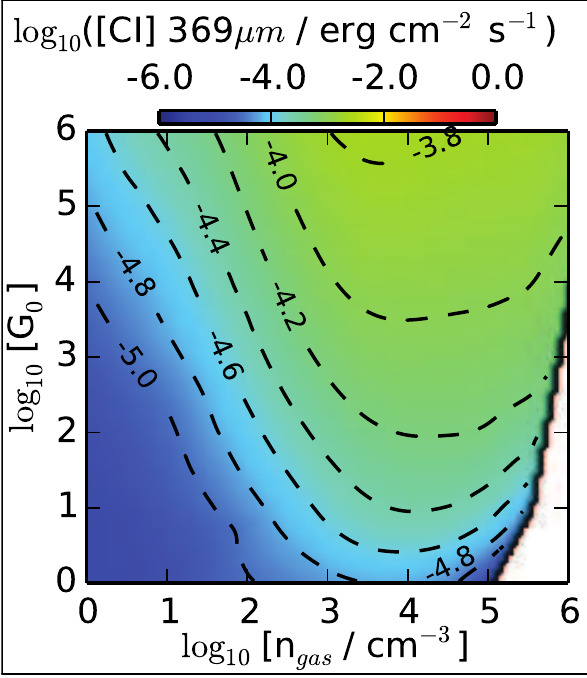}
  \includegraphics[scale=0.7]{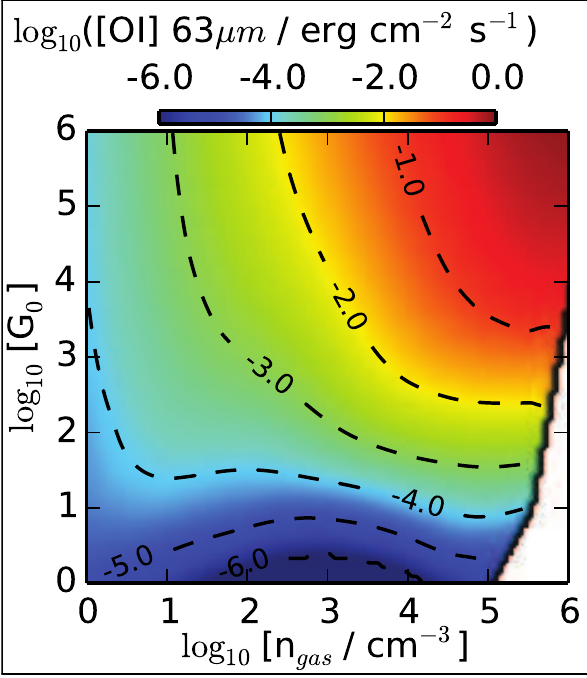}
  \end{minipage}
  \begin{minipage}[b]{1.3\linewidth}   
  \includegraphics[scale=0.7]{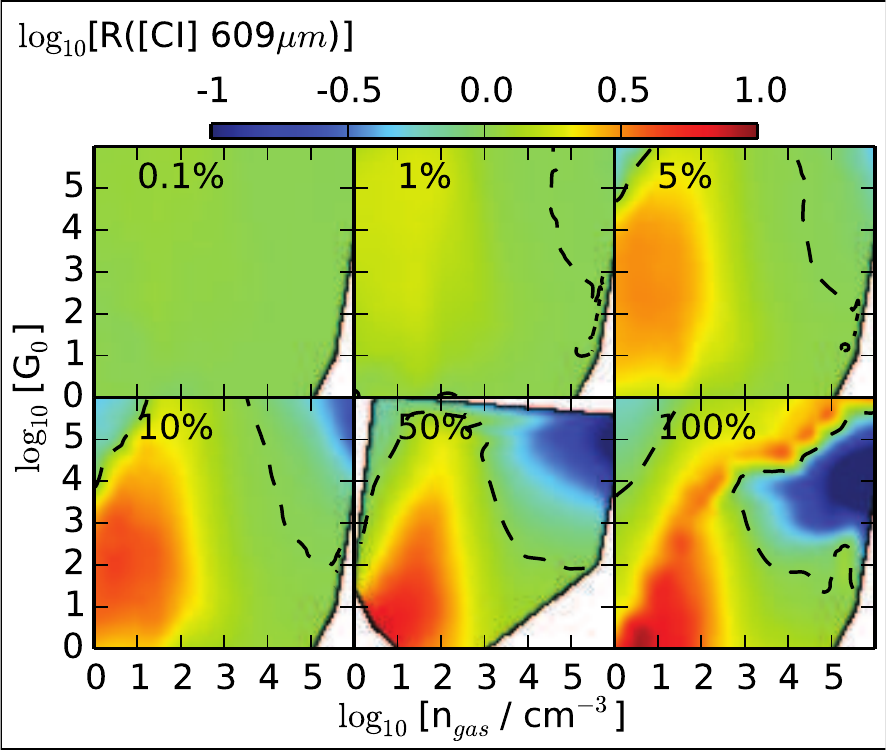}\label{fig:C-1-0}
  \includegraphics[scale=0.7]{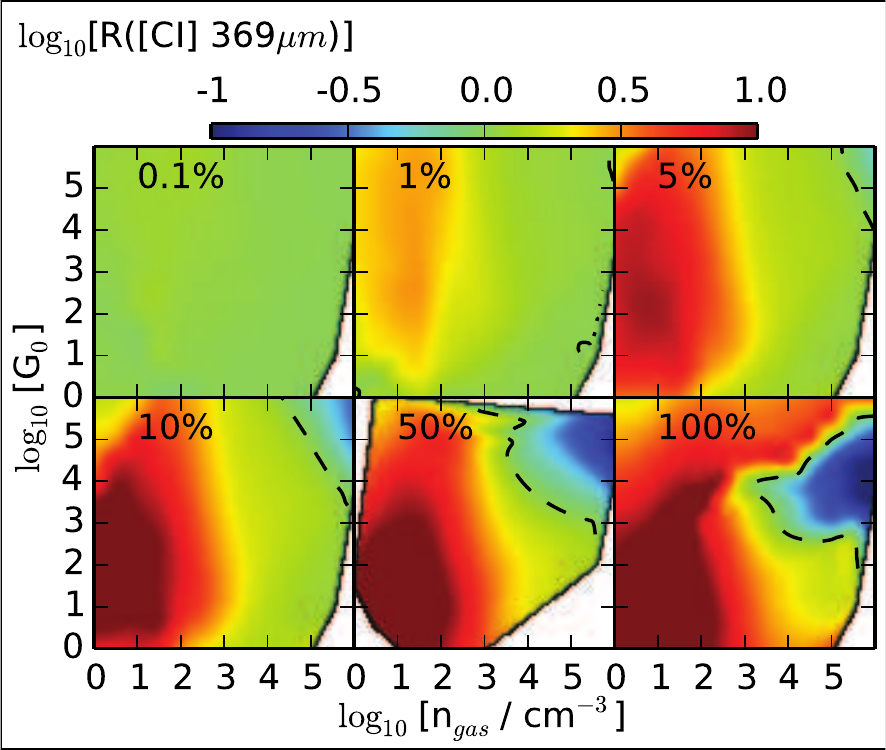}\label{fig:C-2-1}
  \includegraphics[scale=0.7]{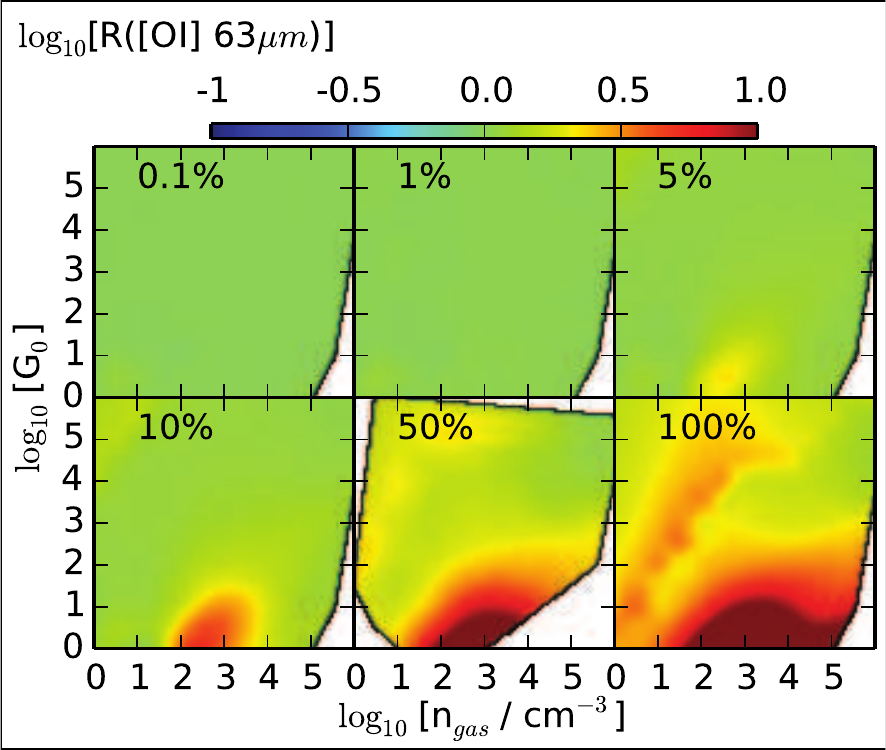}\label{fig:O-1-0}
  \end{minipage}

  \caption{Fine-structure emission grids of [CI] 609~$\mu$m, [CI] 369~$\mu$m, and [OI] 63~$\mu$m
    with and without \gm ($Z=$\zsun). {\bf [Top]} Emission grids. 
    {\bf [Bottom]} Relative changes in the emissions as a function of $\alpha$. The dashed contour
    traces the $R = 1$ line, where the emissions with and without extra heating are the same. (See caption
    of Figure-\ref{fig:C+-1-0-grids})
    \label{fig:fs-lines-app} }
\end{figure*}

\begin{figure*}[!tbh]
  \begin{minipage}[b]{1.5\linewidth} 
  \includegraphics[scale=0.39]{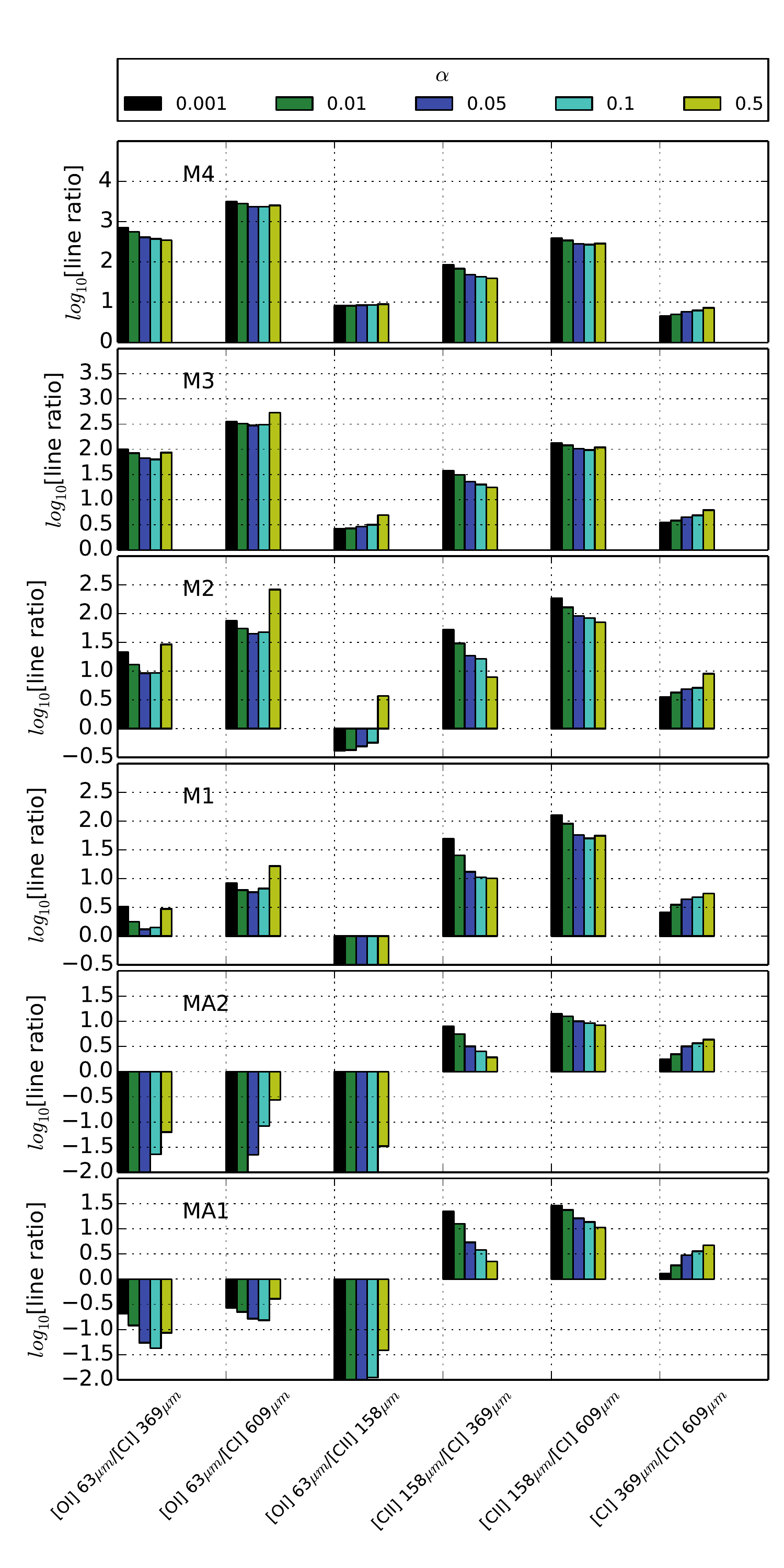}
  \includegraphics[scale=0.39]{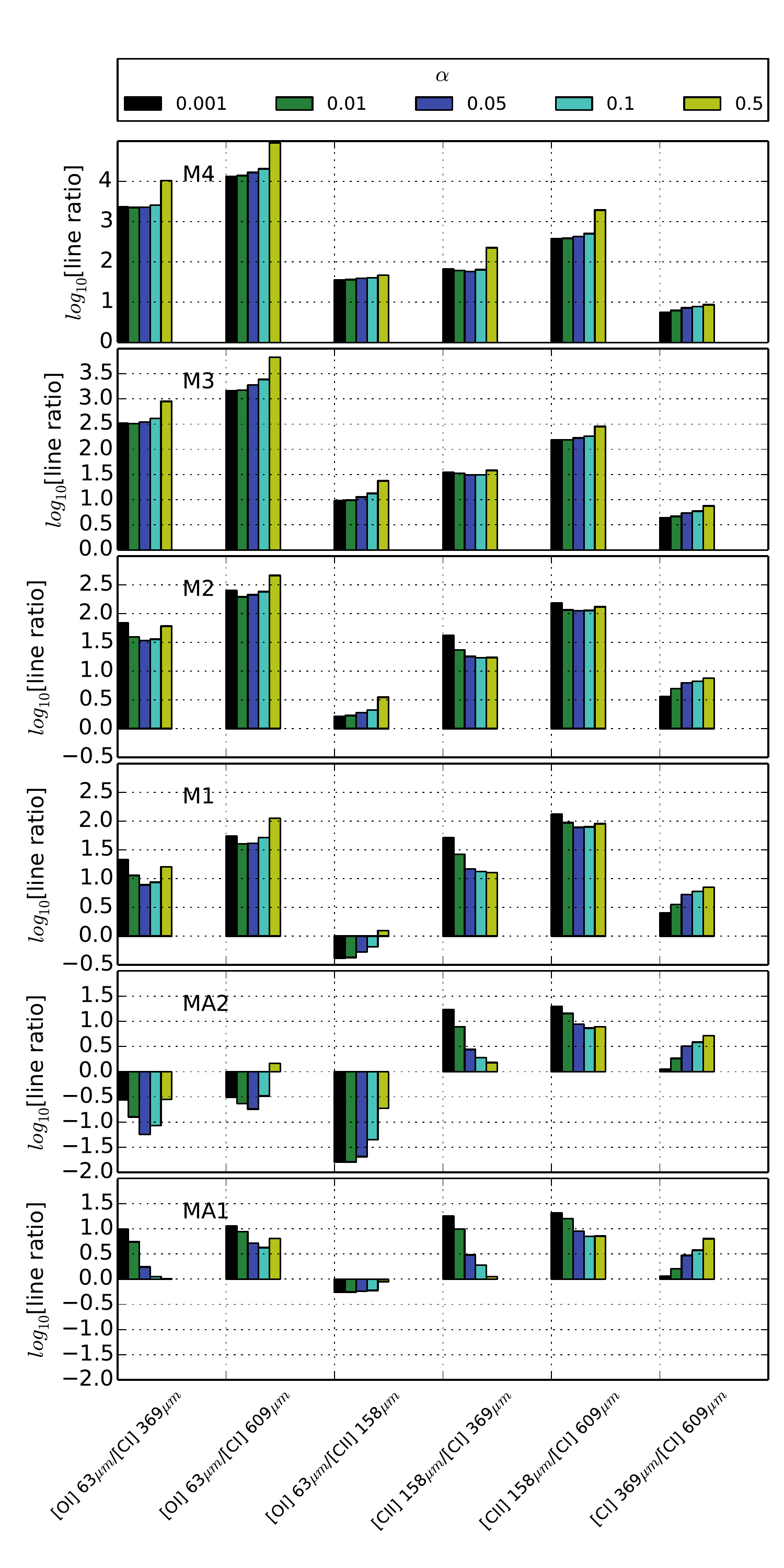}
  \includegraphics[scale=0.39]{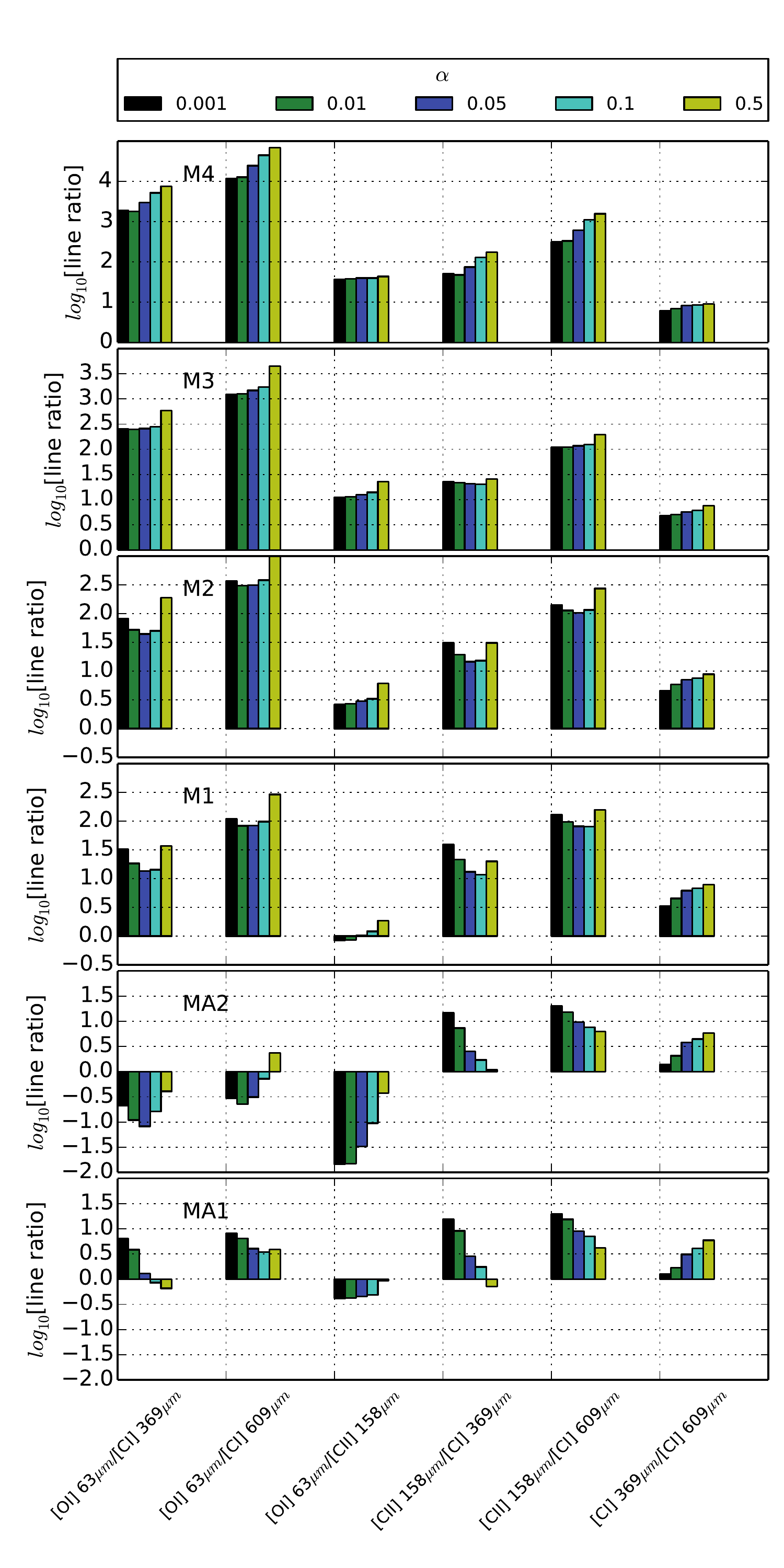}
  \end{minipage}
  \caption{Fine-structure line ratios of [OI] 63~$\mu$m, [CI]
    369~$\mu$m, [CI] 609~$\mu$m, and [CII] 158~$\mu$m, for the reference
    models (see Table-\ref{tbl:refModels}) as a function of \gm~, with
    $Z=$~ 0.1, 0.5, and 2\zsun~ in the left, middle and right panels
    respectively. \label{fig:bar-plot-lineRatios-atomic-z-other}}
\end{figure*}

\begin{figure*}[!tbh]
  \begin{minipage}[b]{1.5\linewidth}   
  \includegraphics[scale=0.5]{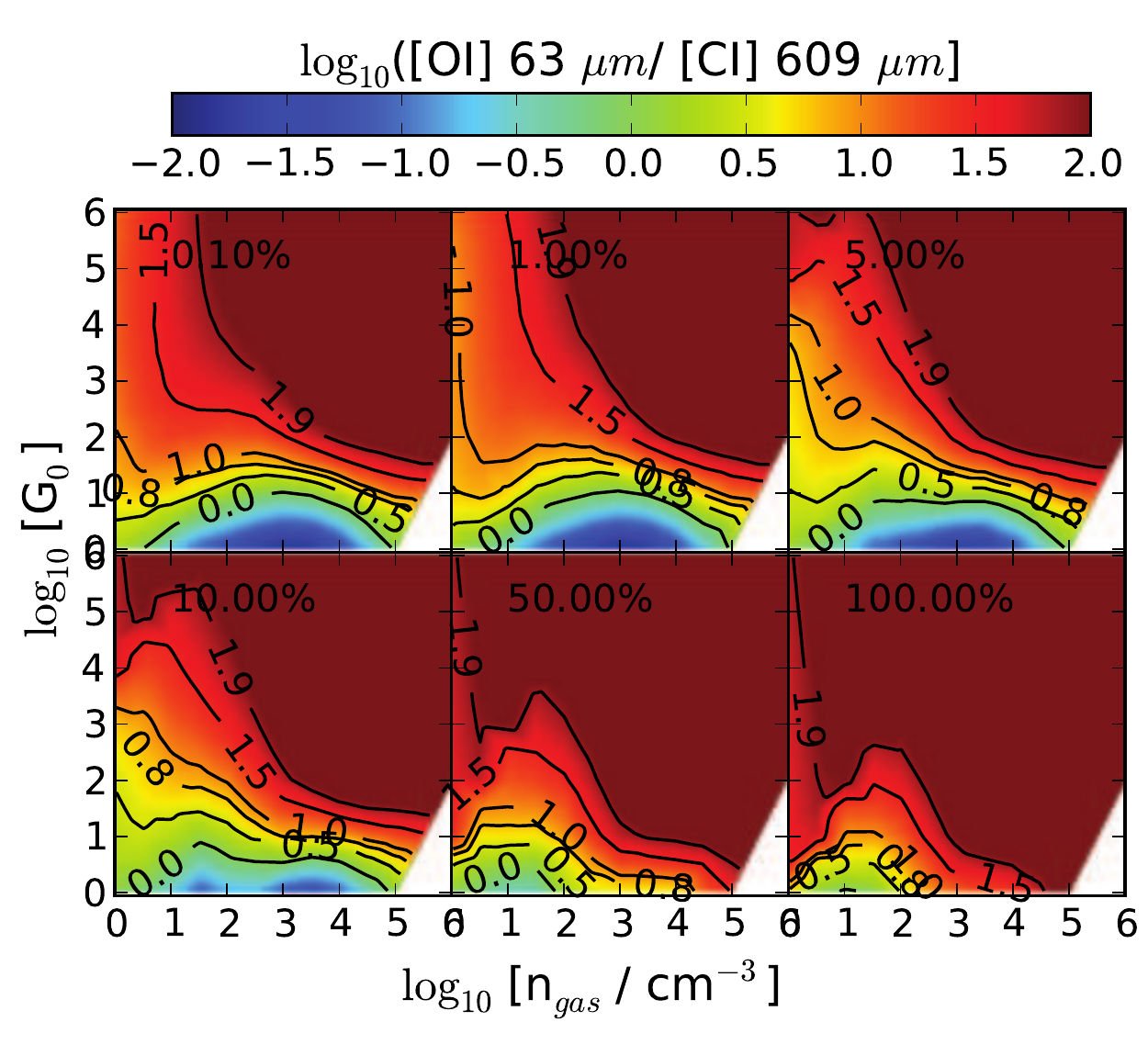} 
  \includegraphics[scale=0.5]{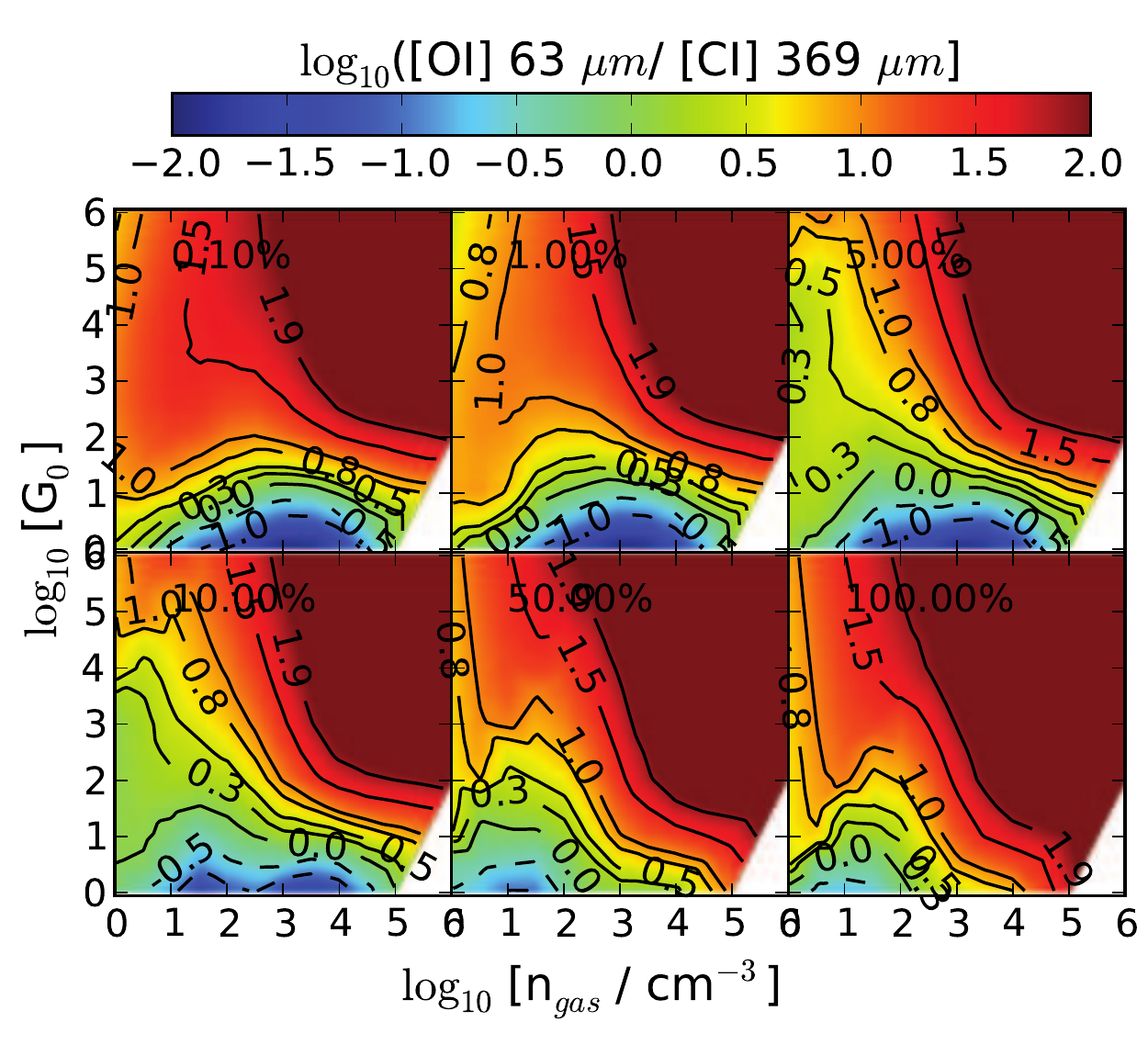} 
  \includegraphics[scale=0.5]{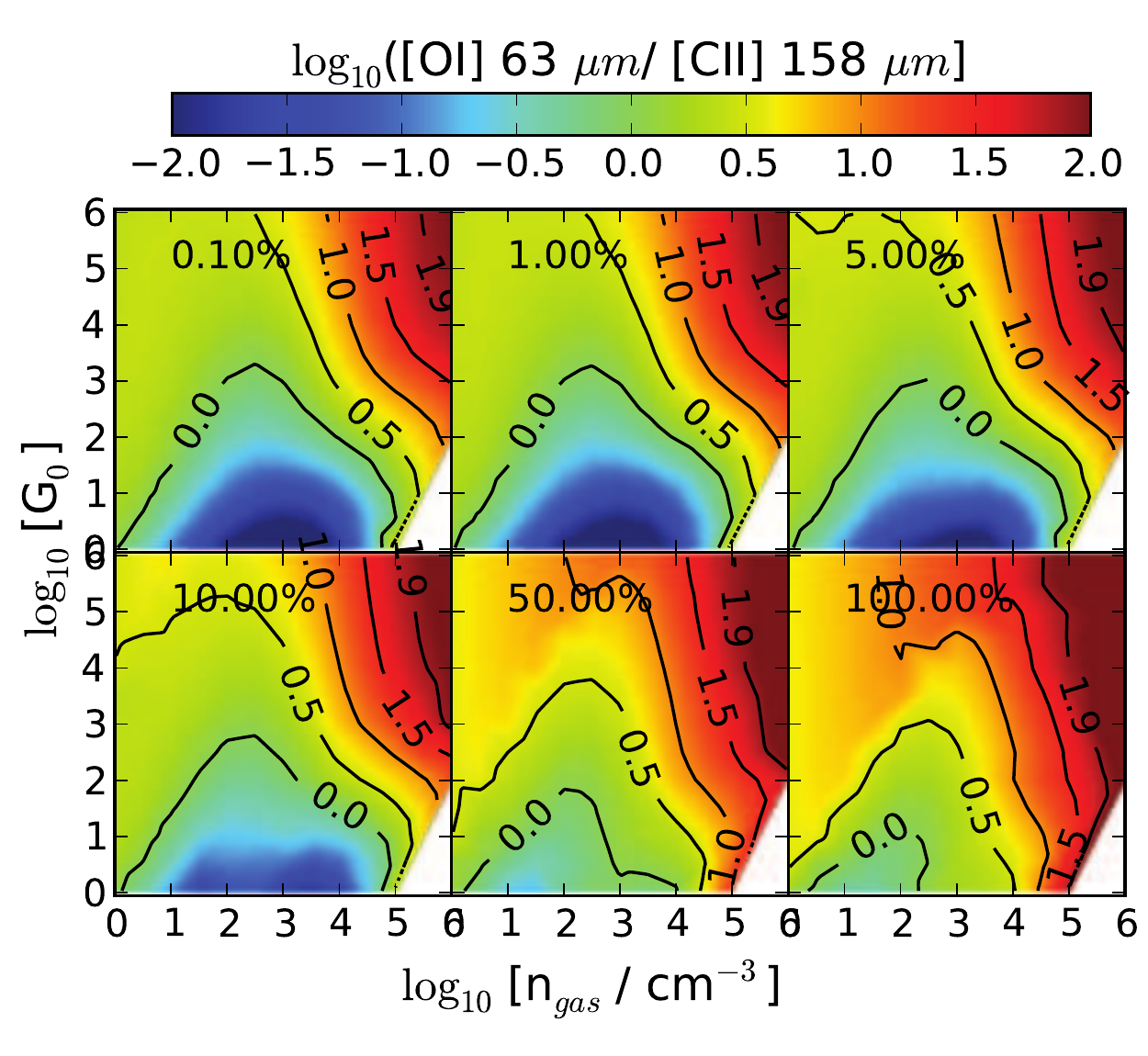}
  \end{minipage}
  \begin{minipage}[b]{1.5\linewidth}   
  \includegraphics[scale=0.5]{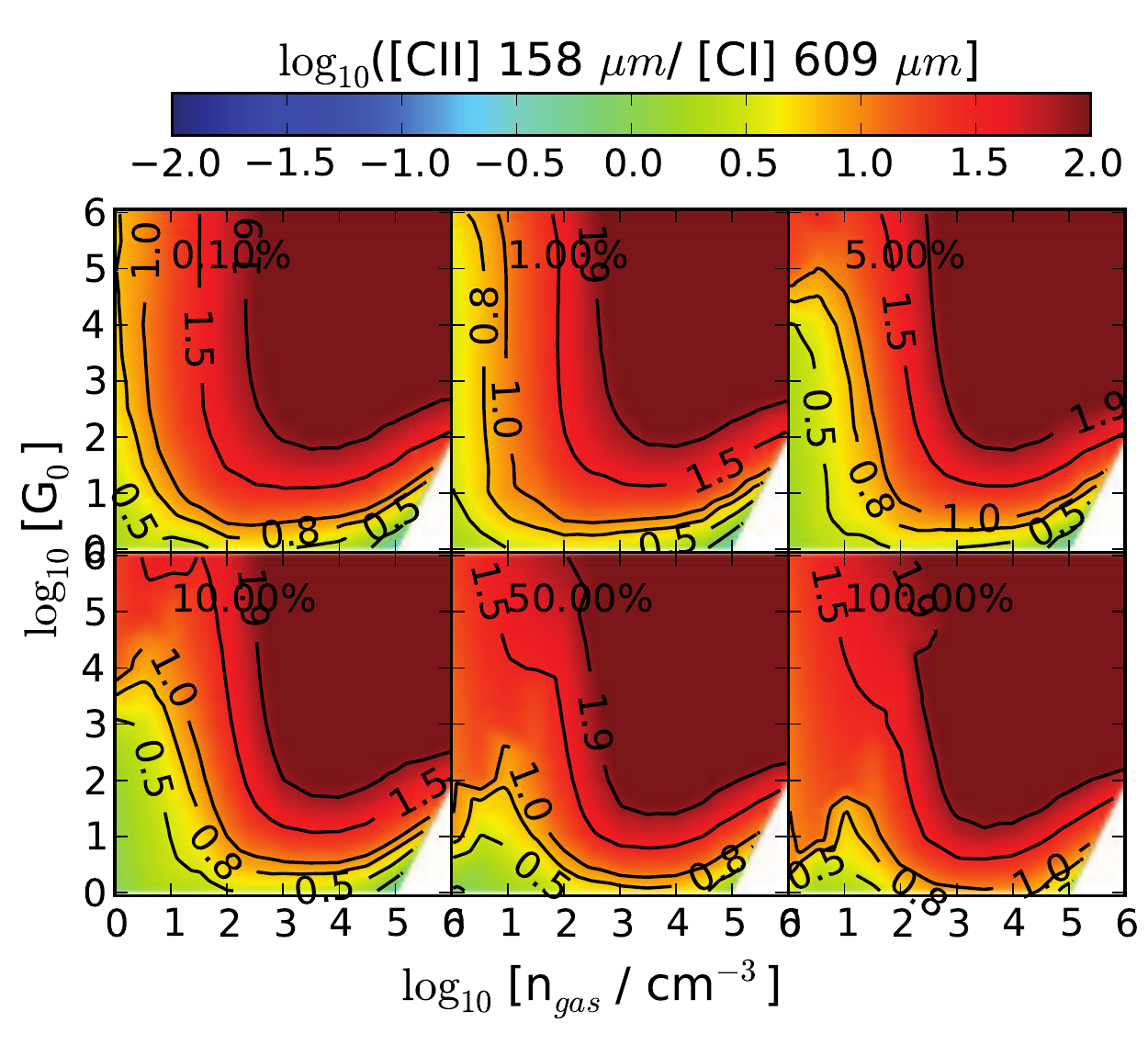}
  \includegraphics[scale=0.5]{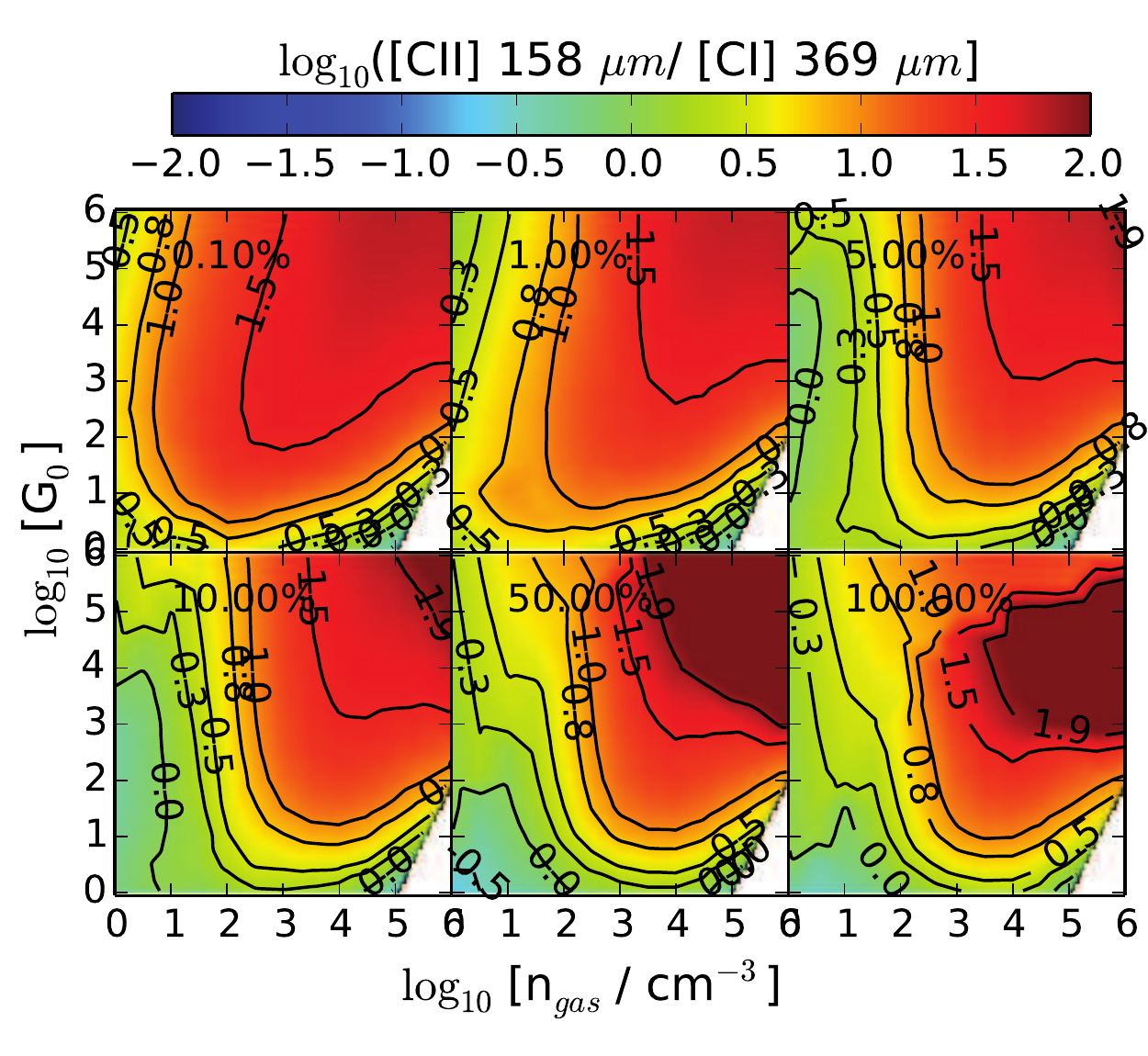}
  \includegraphics[scale=0.5]{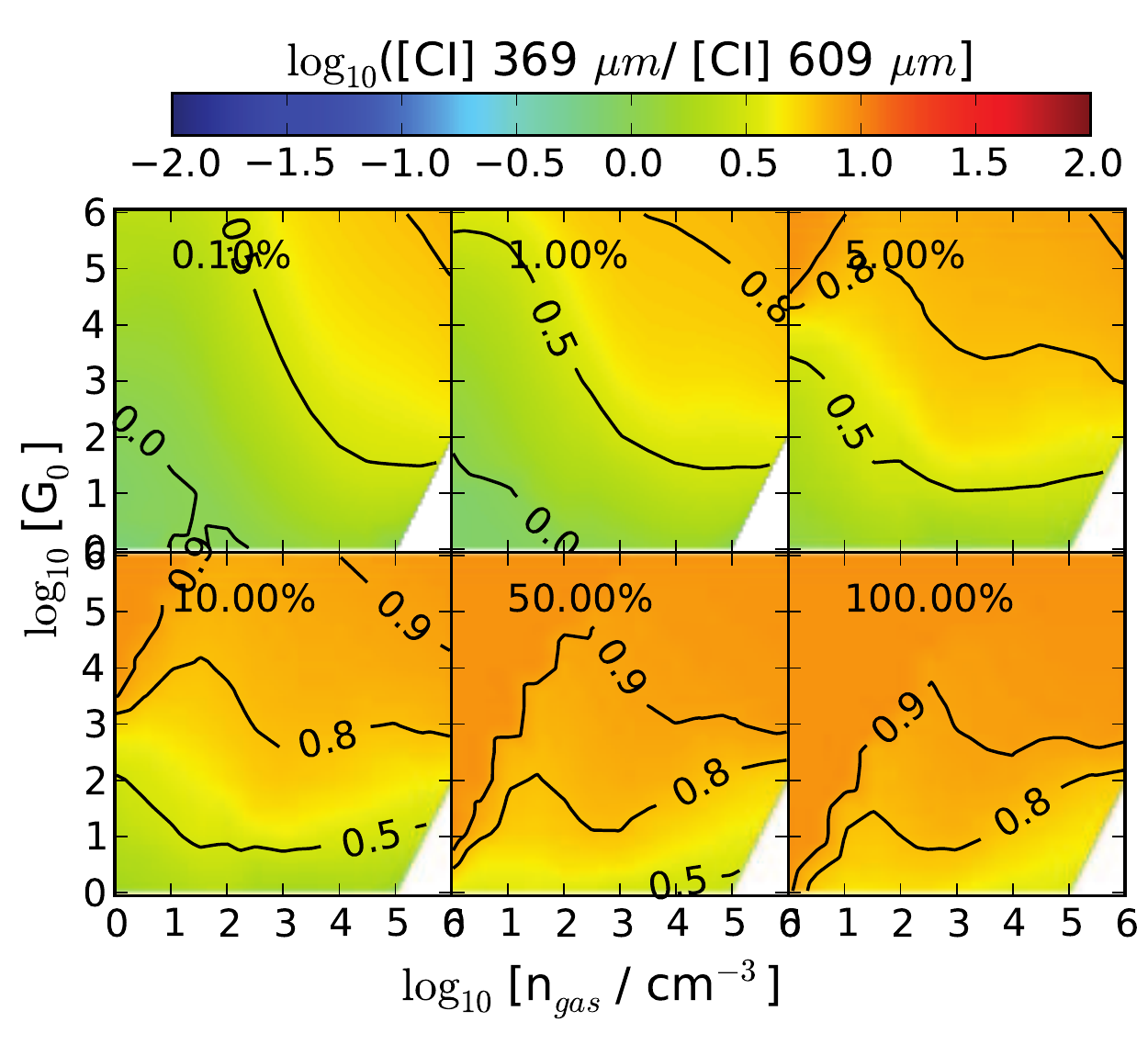} 
  \end{minipage}  
  \caption{Atomic fine-structure line ratio grids of [OI] 63~$\mu$m,
    [CI] 609~$\mu$m, [CI] 369~$\mu$m, and [CII] 158~$\mu$m as a function
    of \gm~($\alpha$ increases from 0.001 to 1), for $A_V =
    10$~ mag ($Z=$~\zsun).
    \label{fig:atomicLineRatios}
  }
\end{figure*}

\end{appendix}

\clearpage

\begin{appendix}{Appendix B: Molecular Species}

\setcounter{figure}{0}
\makeatletter
\renewcommand{\thefigure}{B\@arabic\c@figure} 

\begin{figure*}[!tbh]
  \begin{minipage}[b]{1.5\linewidth} 
  \includegraphics[scale=0.6]{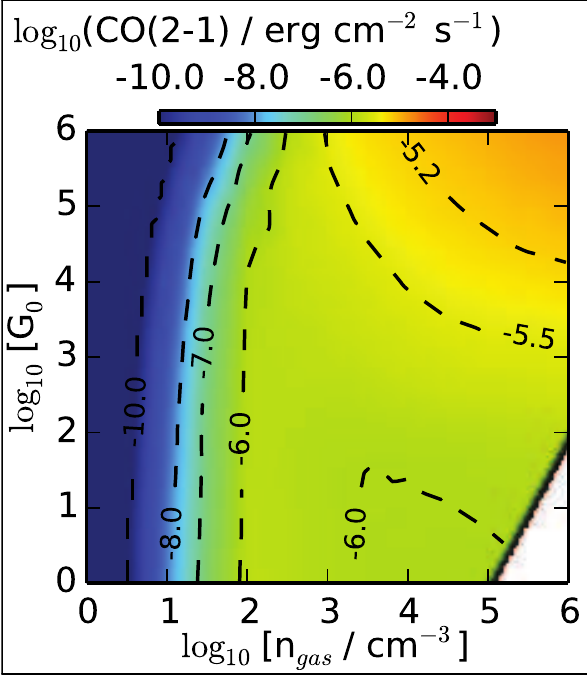}
  \includegraphics[scale=0.6]{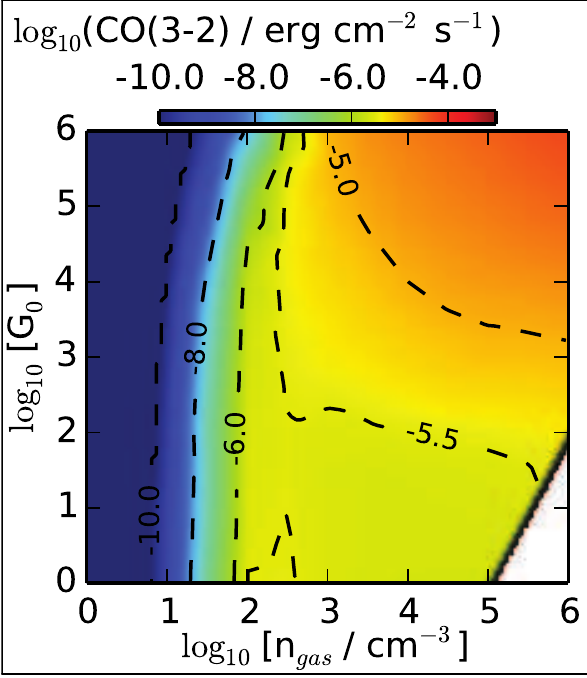}
  \includegraphics[scale=0.6]{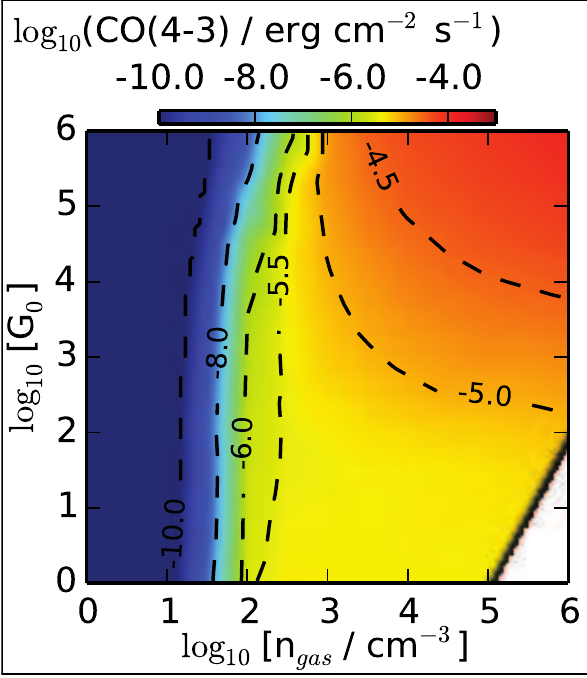}
  \includegraphics[scale=0.6]{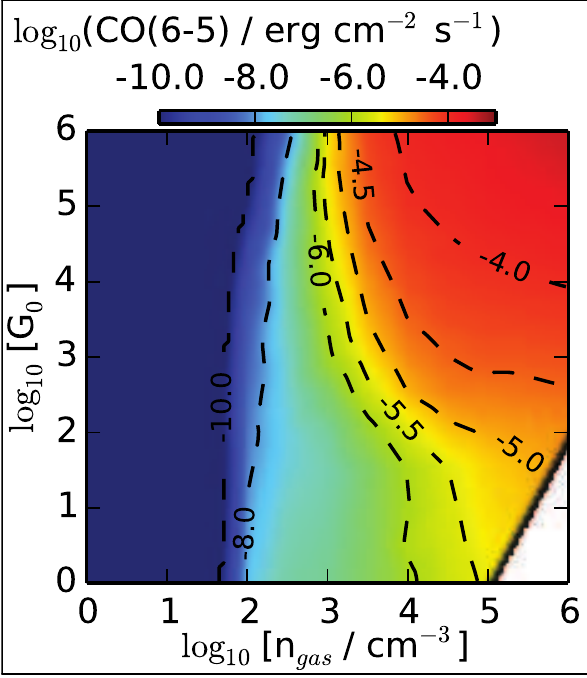}
  \end{minipage}
  \begin{minipage}[b]{1.5\linewidth} 
  \includegraphics[scale=0.6]{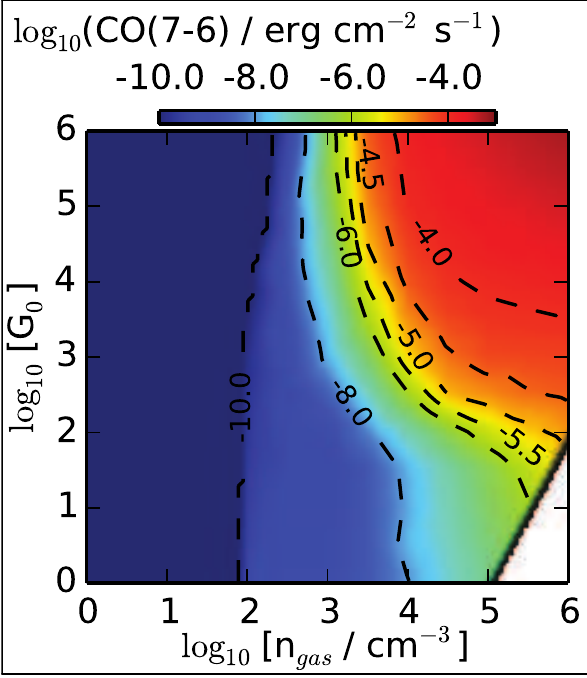}
  \includegraphics[scale=0.6]{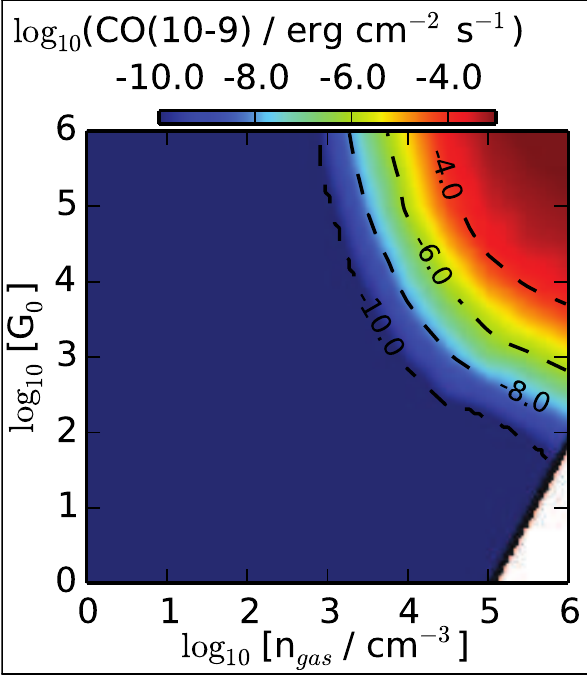}
  \includegraphics[scale=0.6]{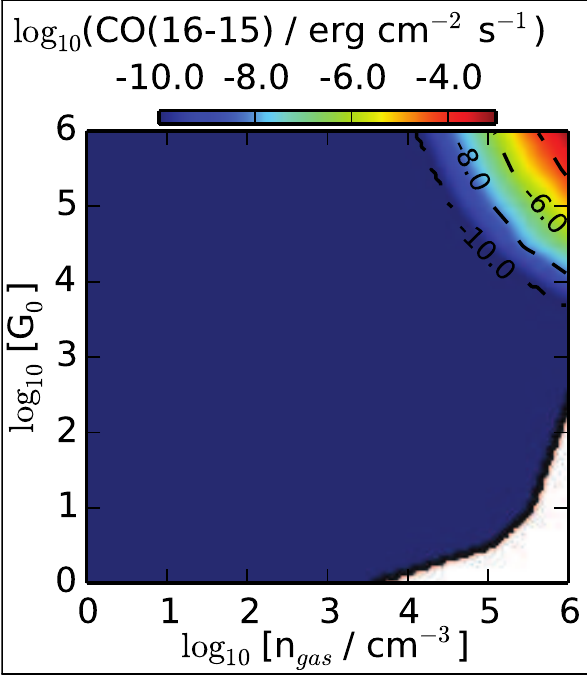}
  \end{minipage}
  \begin{minipage}[b]{1.5\linewidth} 
  \includegraphics[scale=0.65]{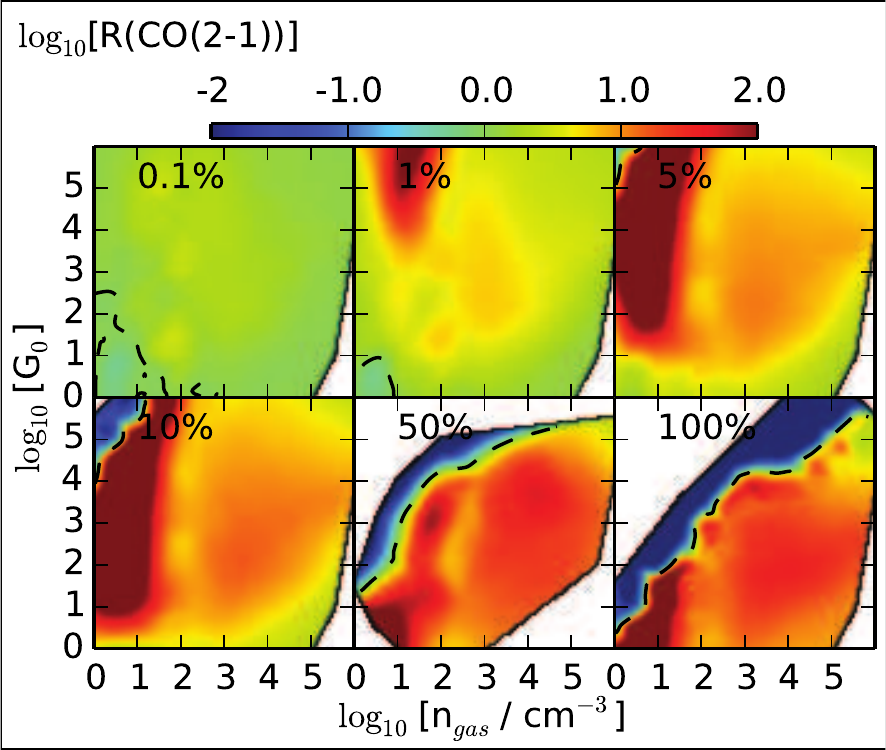}\label{fig:CO-2-1}
  \includegraphics[scale=0.65]{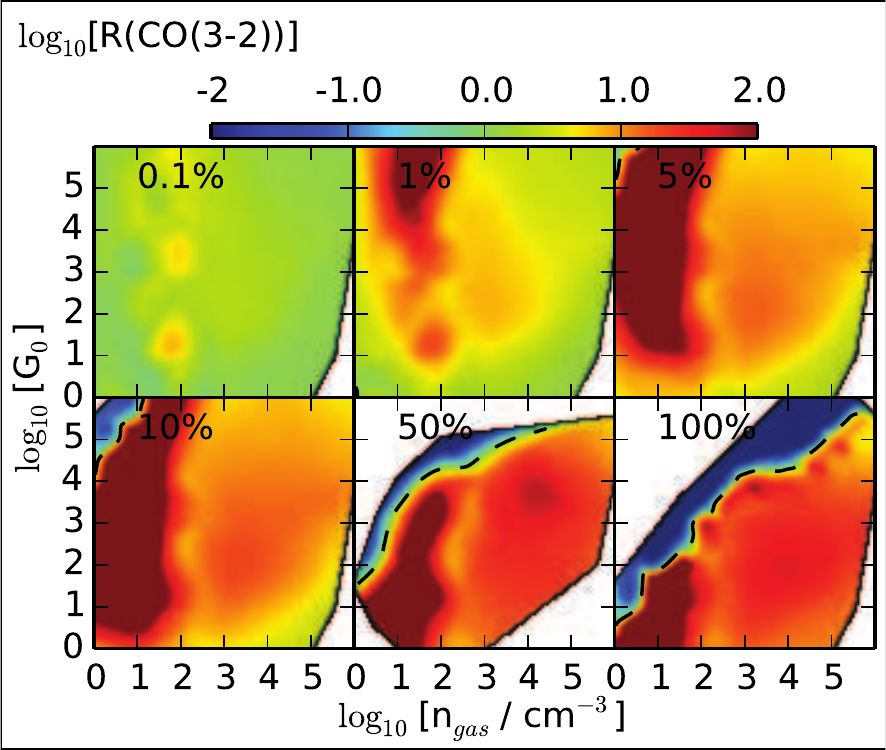} \label{fig:CO-3-2}
  \includegraphics[scale=0.65]{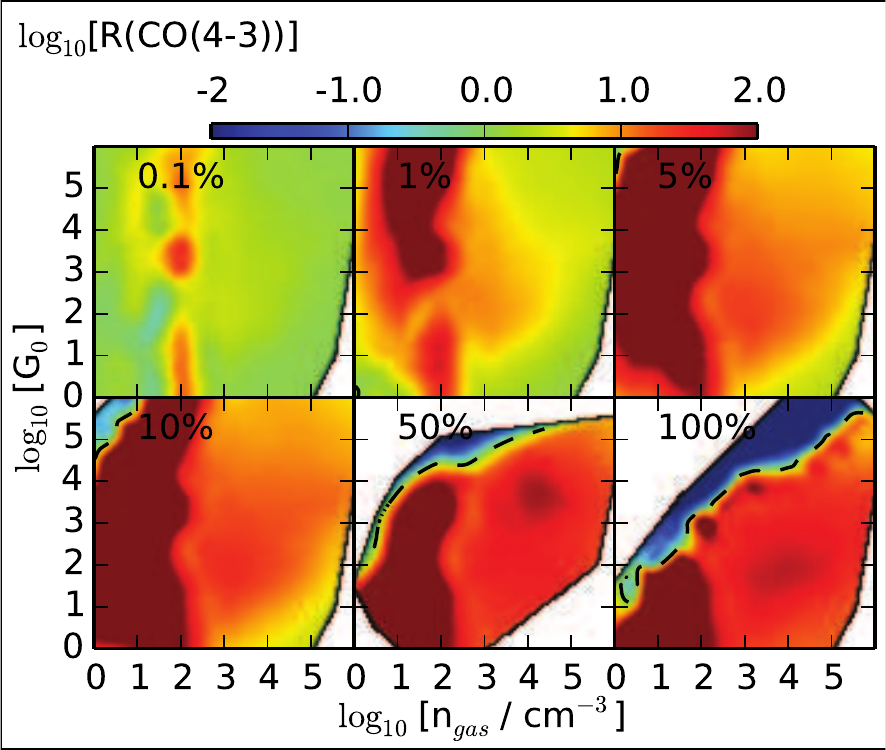}\label{fig:CO-4-3}
  \end{minipage}
  \begin{minipage}[b]{1.5\linewidth} 
  \includegraphics[scale=0.65]{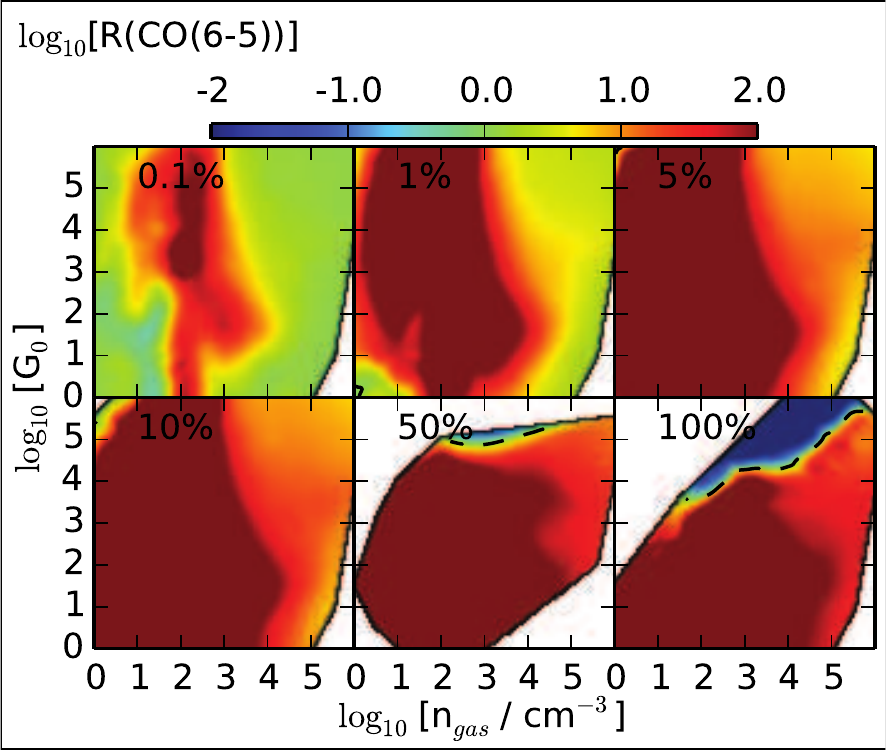}\label{fig:CO-6-5}
  \includegraphics[scale=0.65]{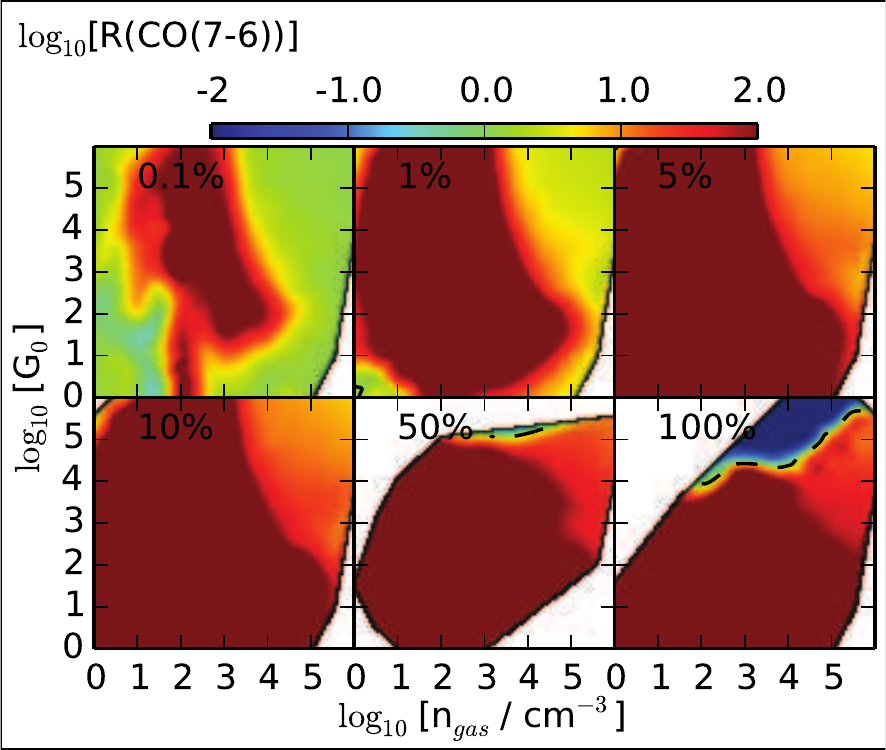}\label{fig:CO-7-6}
  \includegraphics[scale=0.65]{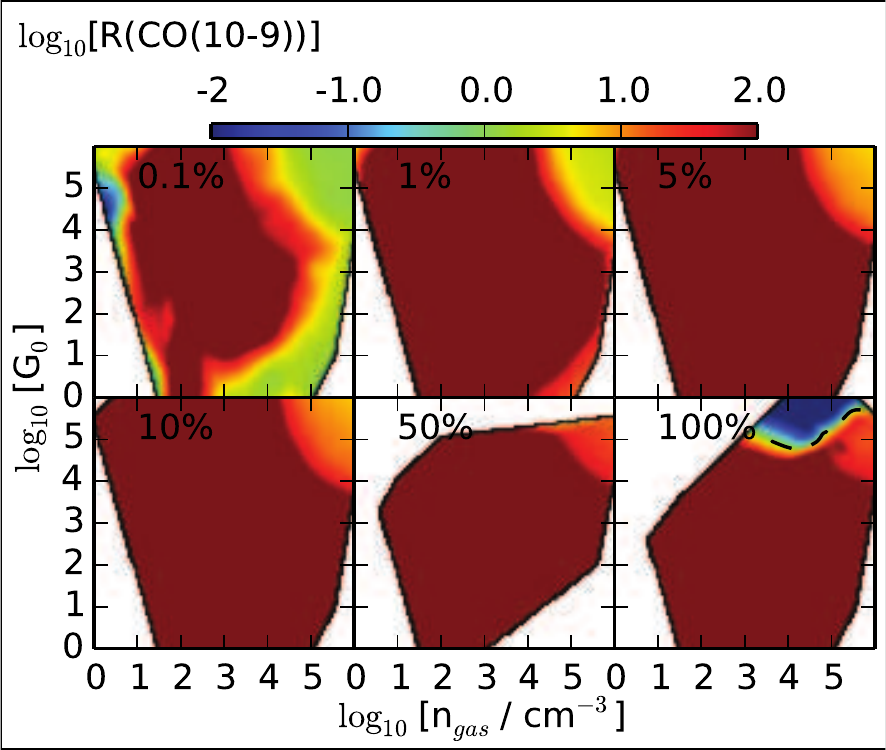}\label{fig:CO-10-9}
  \end{minipage}
  \begin{minipage}[b]{1.5\linewidth} 
  \includegraphics[scale=0.65]{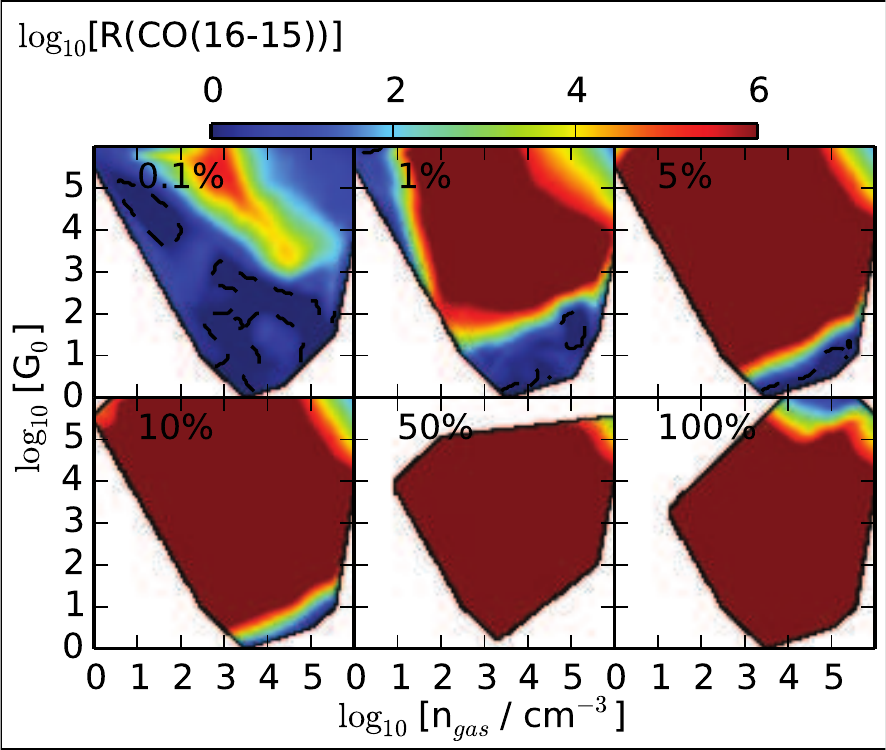}\label{fig:CO-16-15}
  \end{minipage}
  \caption{{\bf [Top]} Emission grids of PDR models without mechanical heating
    for a selection of CO transitions for $A_V = 10$~ mag
    ($Z=$~\zsun). {\bf [Bottom]} Relative changes in the emission as a
    function of $\alpha$. The dashed contour traces the $R = 1$ line,
    where the emission with and without extra heating are the
    same. (See caption of Figure-\ref{fig:C+-1-0-grids})
    \label{fig:CO-grids} }
\end{figure*}

\begin{figure*}[!tbh]
  \begin{minipage}[b]{1.5\linewidth} 
  \includegraphics[scale=0.6]{13CO1-0-base.pdf}
  \includegraphics[scale=0.6]{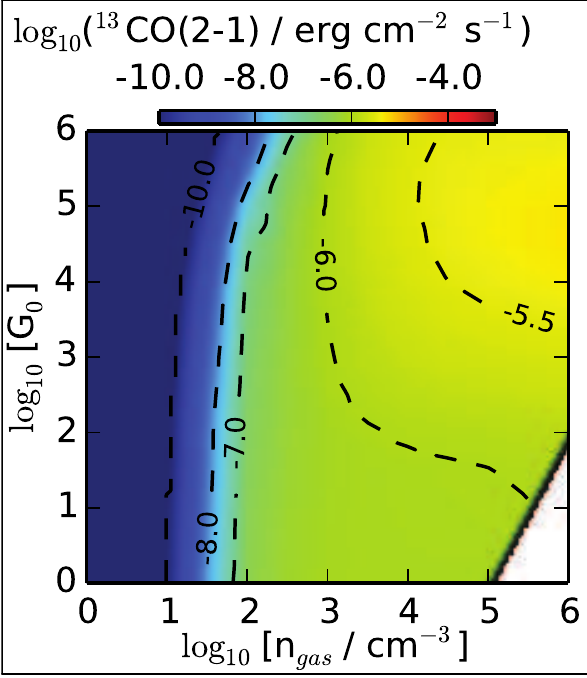}
  \includegraphics[scale=0.6]{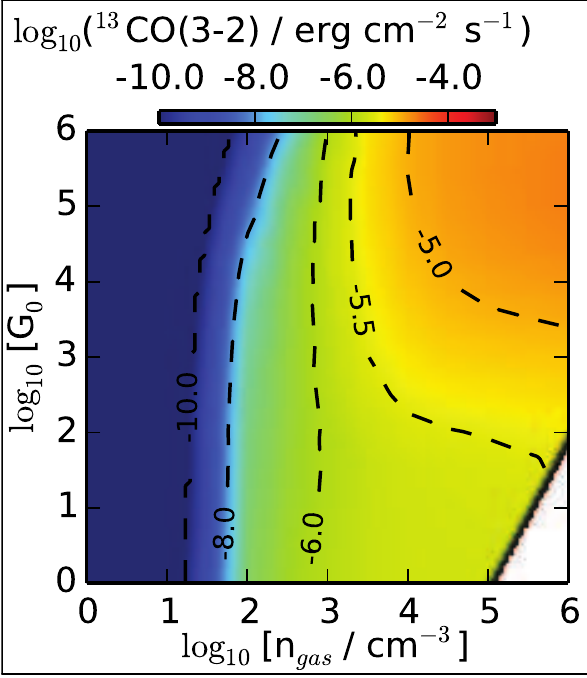}
  \includegraphics[scale=0.6]{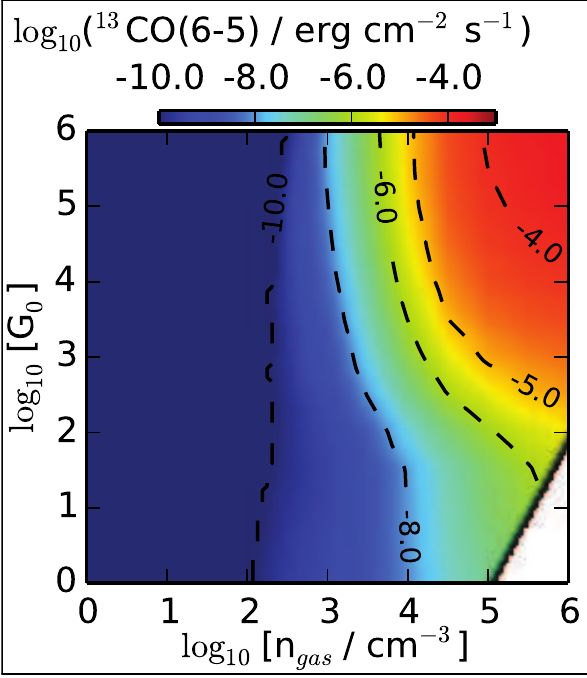}
  \end{minipage}
  \begin{minipage}[b]{1.5\linewidth} 
  \includegraphics[scale=0.68]{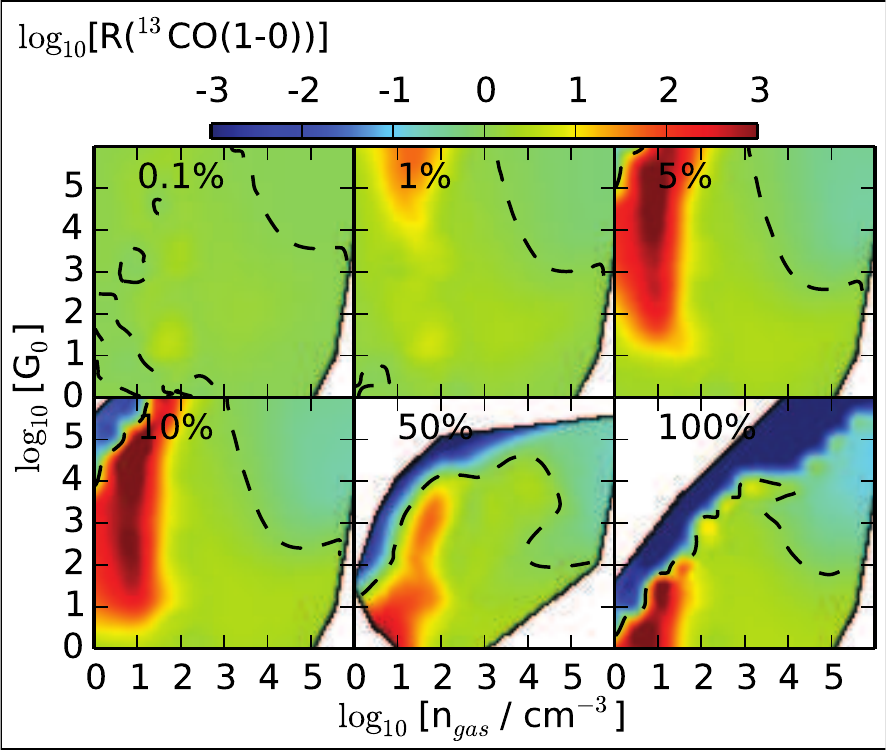}
  \includegraphics[scale=0.68]{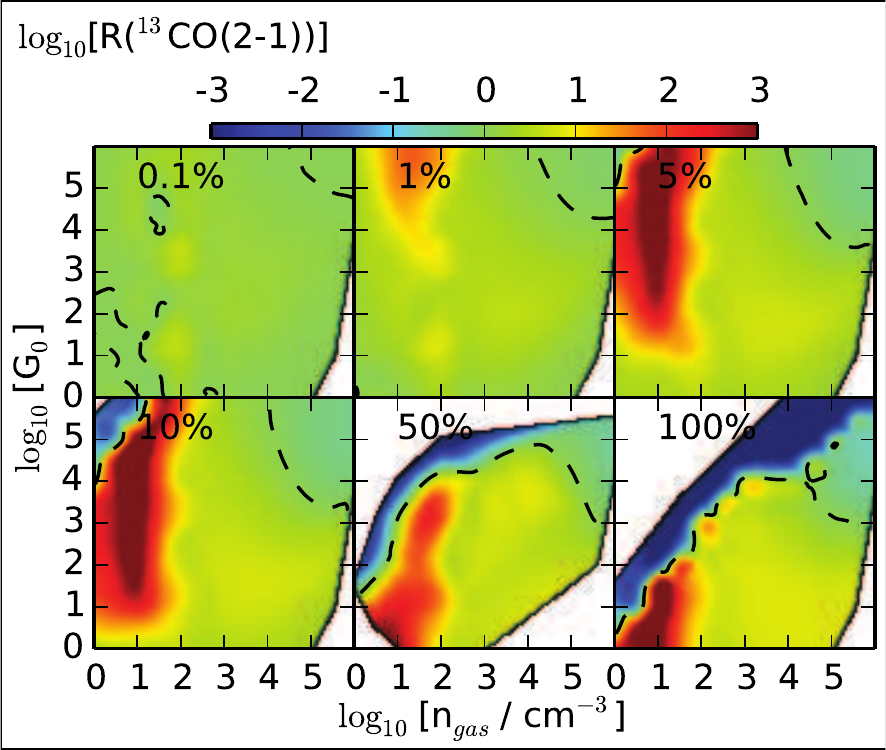}
  \end{minipage}
  \begin{minipage}[b]{1.5\linewidth} 
  \includegraphics[scale=0.68]{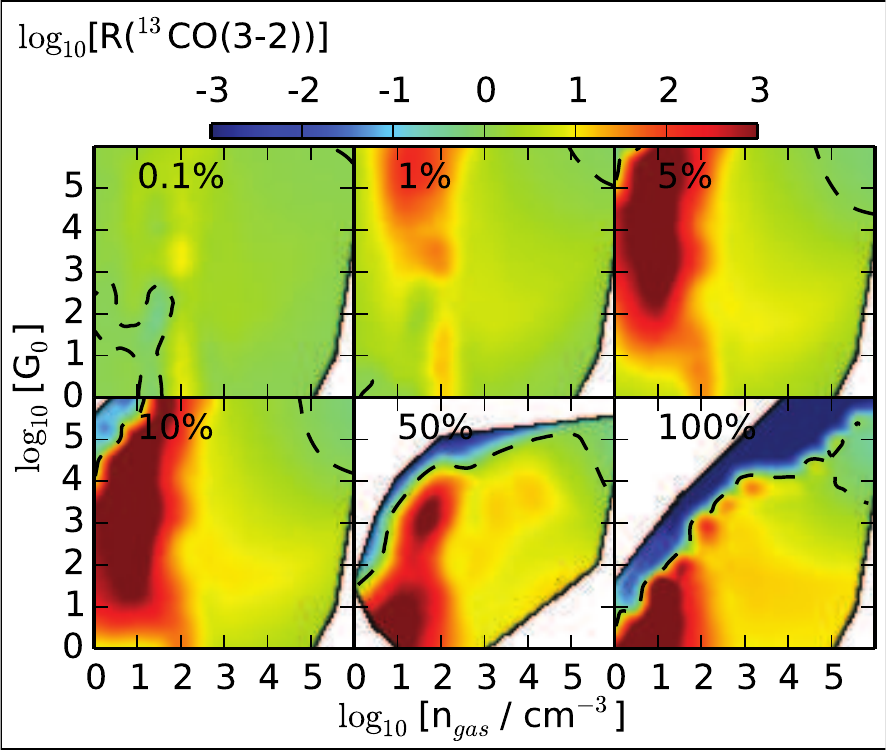}
  \includegraphics[scale=0.68]{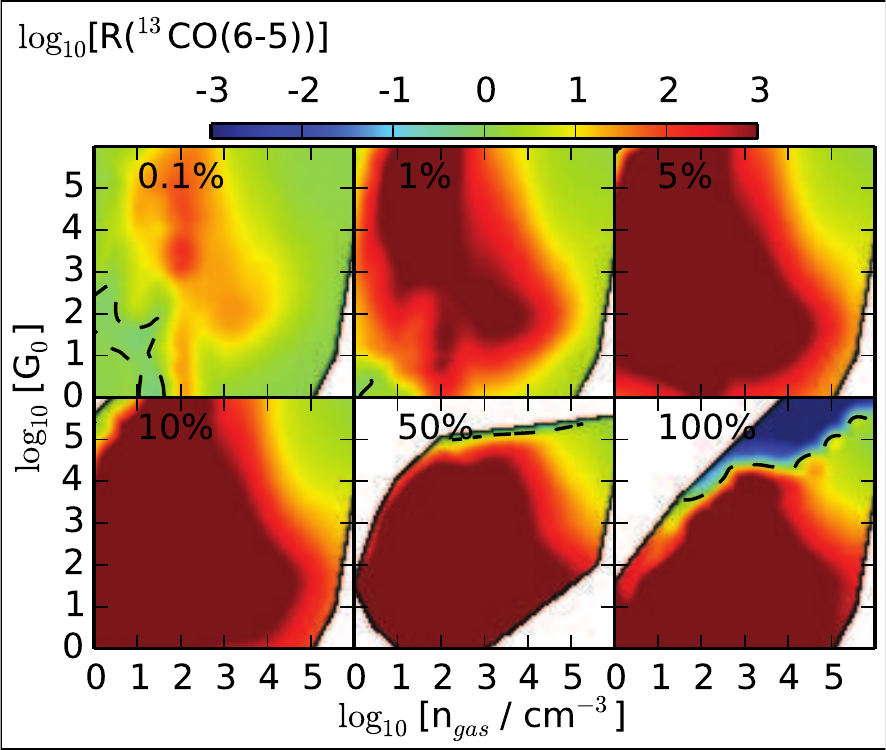}
  \end{minipage}
  \caption{{\bf [Top]} Emission grids of PDR models without mechanical heating for a 
    selection of $^{13}$CO transitions for $A_V = 10$~ mag ($Z=$~\zsun). {\bf [Bottom]} 
    Relative changes in the emission as a function of $\alpha$. The dashed contour traces the $R = 1$ line, 
    where the emission with and without extra heating are the same. (See caption of Figure-\ref{fig:C+-1-0-grids})
    \label{fig:13CO-grids} }
\end{figure*}

\begin{figure*}[!tbh]
  \begin{minipage}[b]{1.5\linewidth} 
  \includegraphics[scale=0.6]{HCN1-0-base.pdf}
  \includegraphics[scale=0.6]{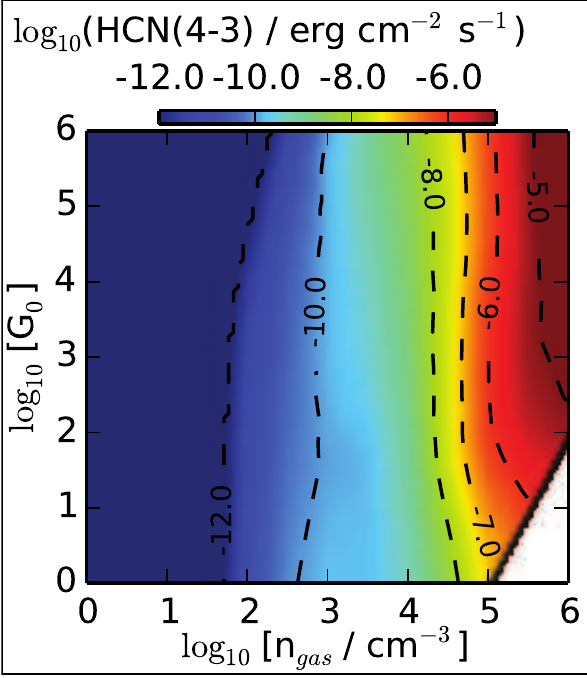}
  \includegraphics[scale=0.6]{HNC1-0-base.pdf}
  \includegraphics[scale=0.6]{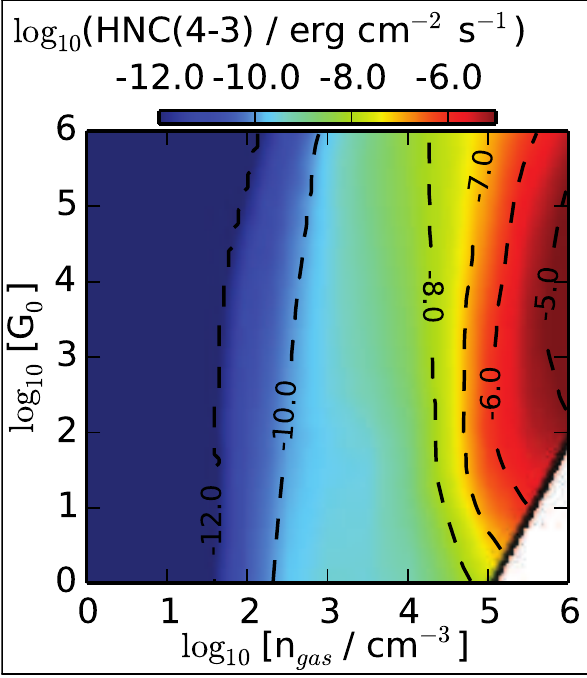}
  \end{minipage}
  \begin{minipage}[b]{1.5\linewidth} 
  \includegraphics[scale=0.68]{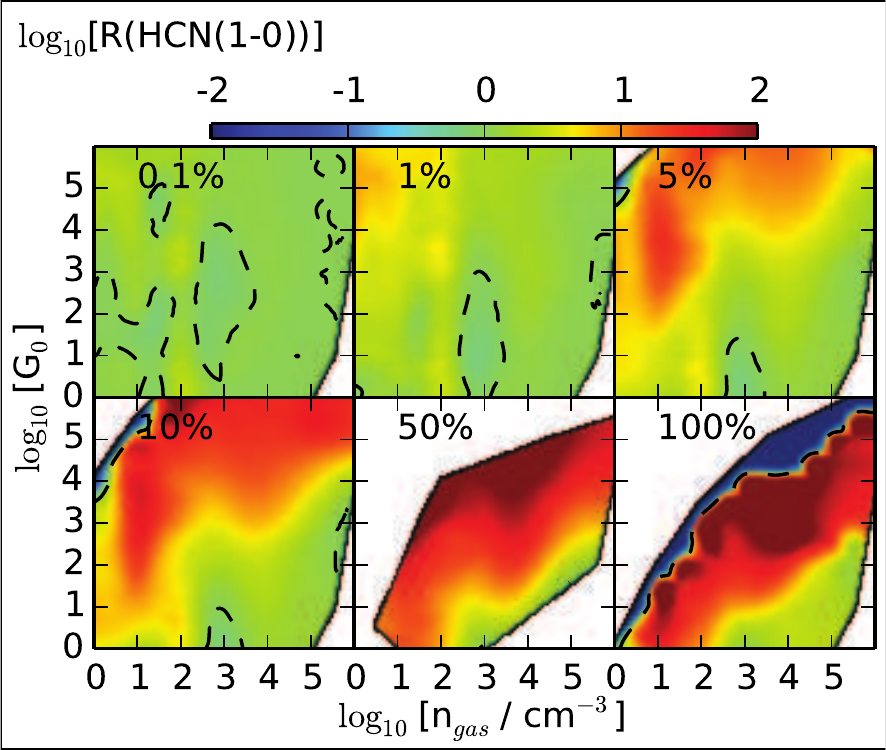}\label{fig:HCN-1-0}
  \includegraphics[scale=0.68]{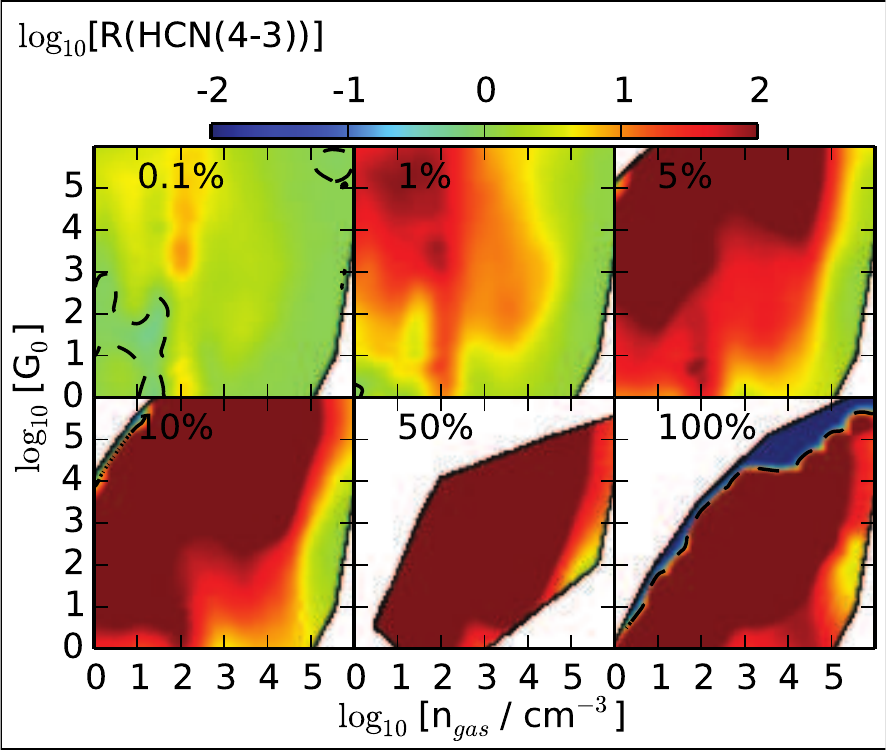}\label{fig:HCN-4-3}
  \end{minipage}
  \begin{minipage}[b]{1.5\linewidth} 
  \includegraphics[scale=0.68]{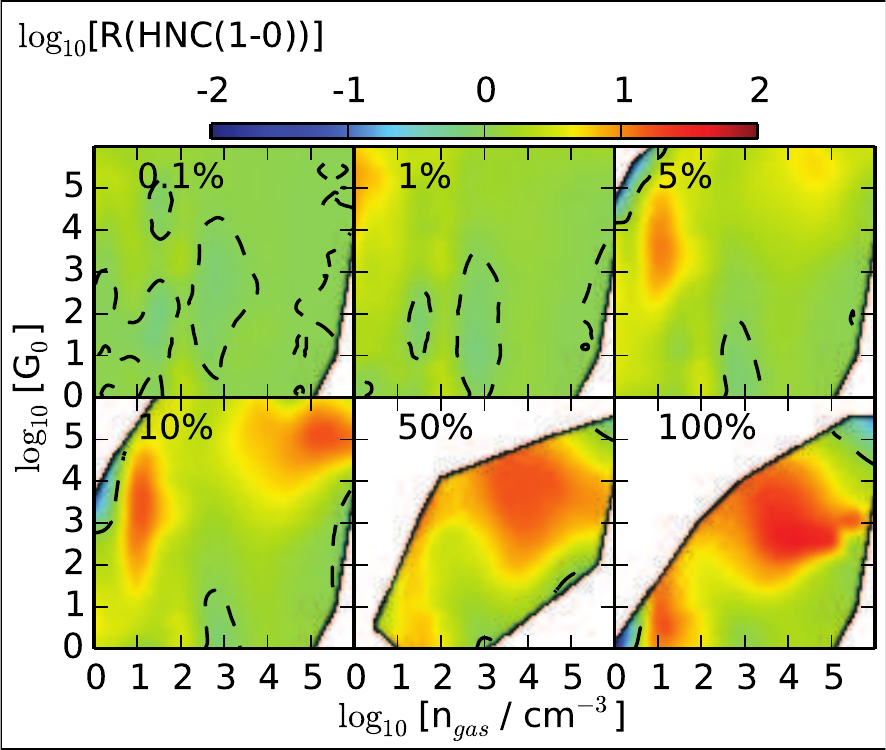}\label{fig:HNC-01-00}
  \includegraphics[scale=0.68]{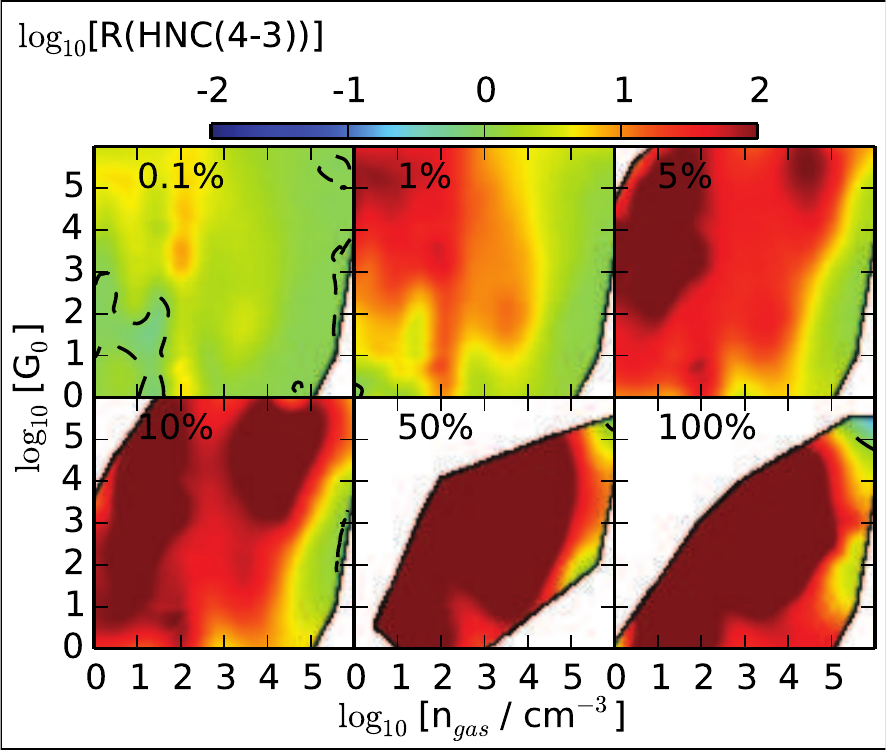}\label{fig:HNC-04-03}
  \end{minipage}
  \caption{{\bf [Top]} Emission grids of PDR models without mechanical heating for a selection of 
    HCN and HNC transitions for $A_V = 10$~ mag ($Z=$~\zsun). {\bf [Bottom]} 
    Relative changes in the emission as a function of $\alpha$. The dashed contour traces the $R = 1$ line, 
    where the emission with and without extra heating are the same. (See caption of Figure-\ref{fig:C+-1-0-grids})
    \label{fig:HNC-HNC} }
\end{figure*}

\begin{figure*}[!tbh]
  \begin{minipage}[b]{1.5\linewidth} 
  \includegraphics[scale=0.39]{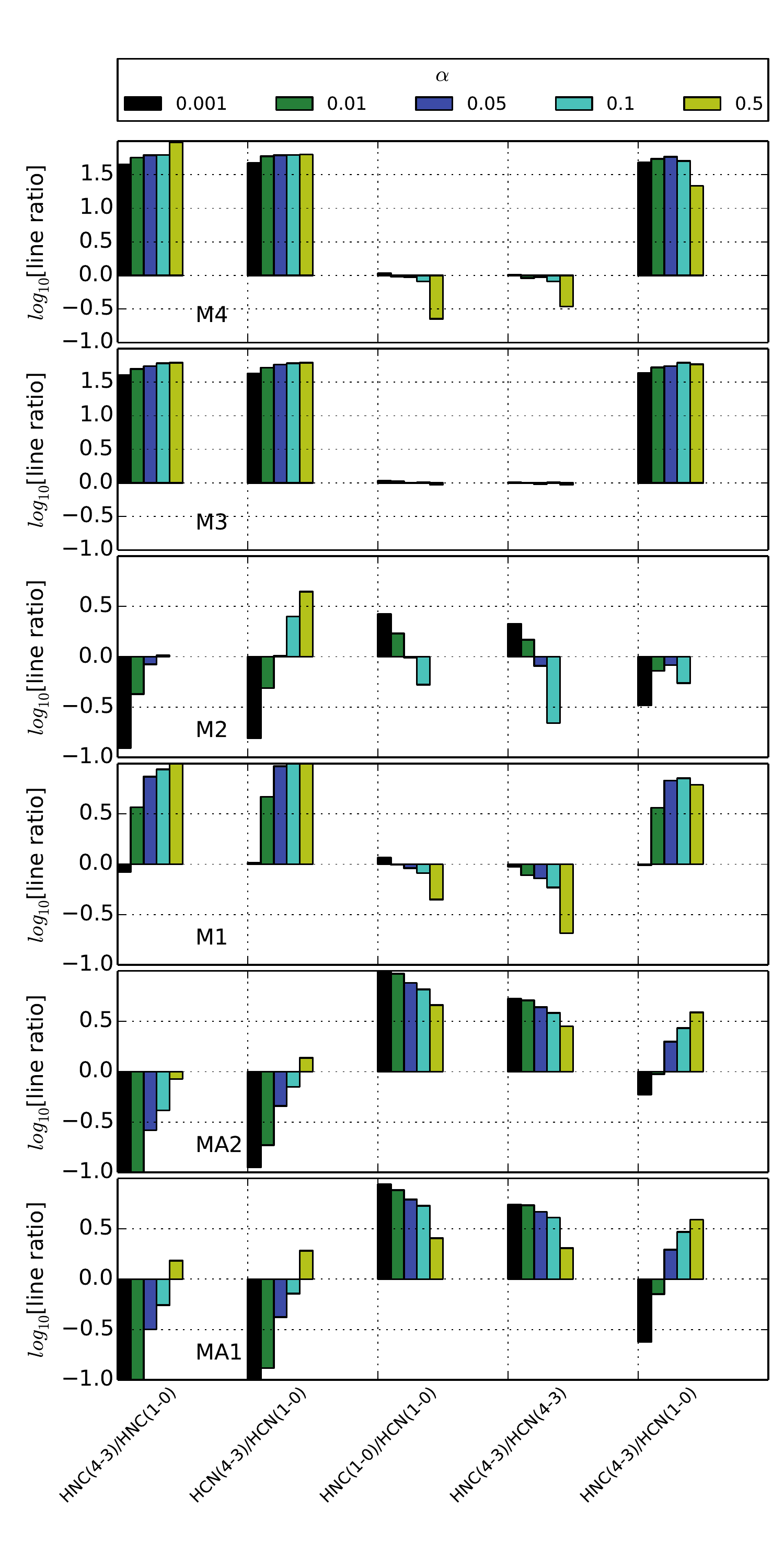}
  \includegraphics[scale=0.39]{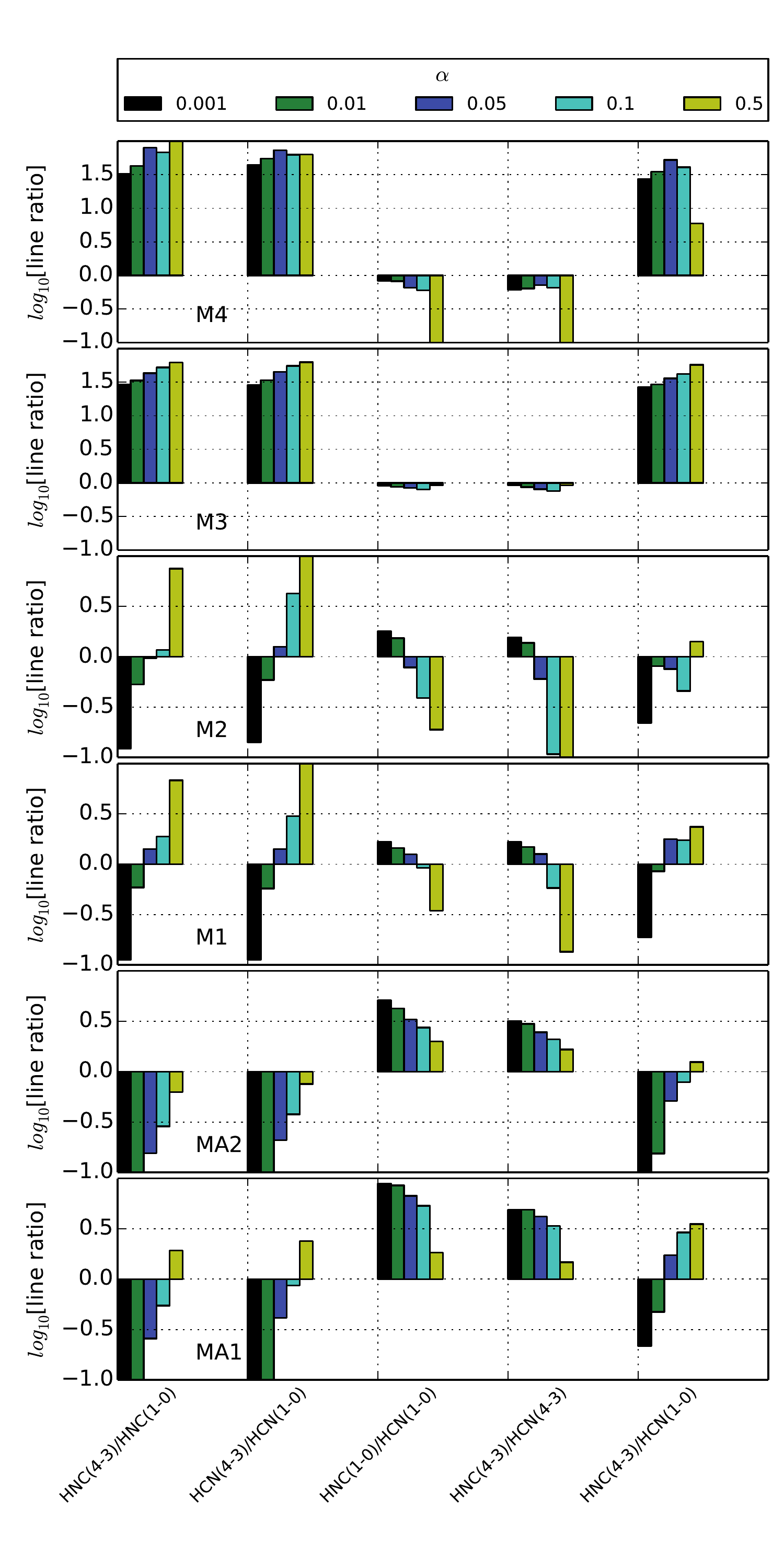}
  \includegraphics[scale=0.39]{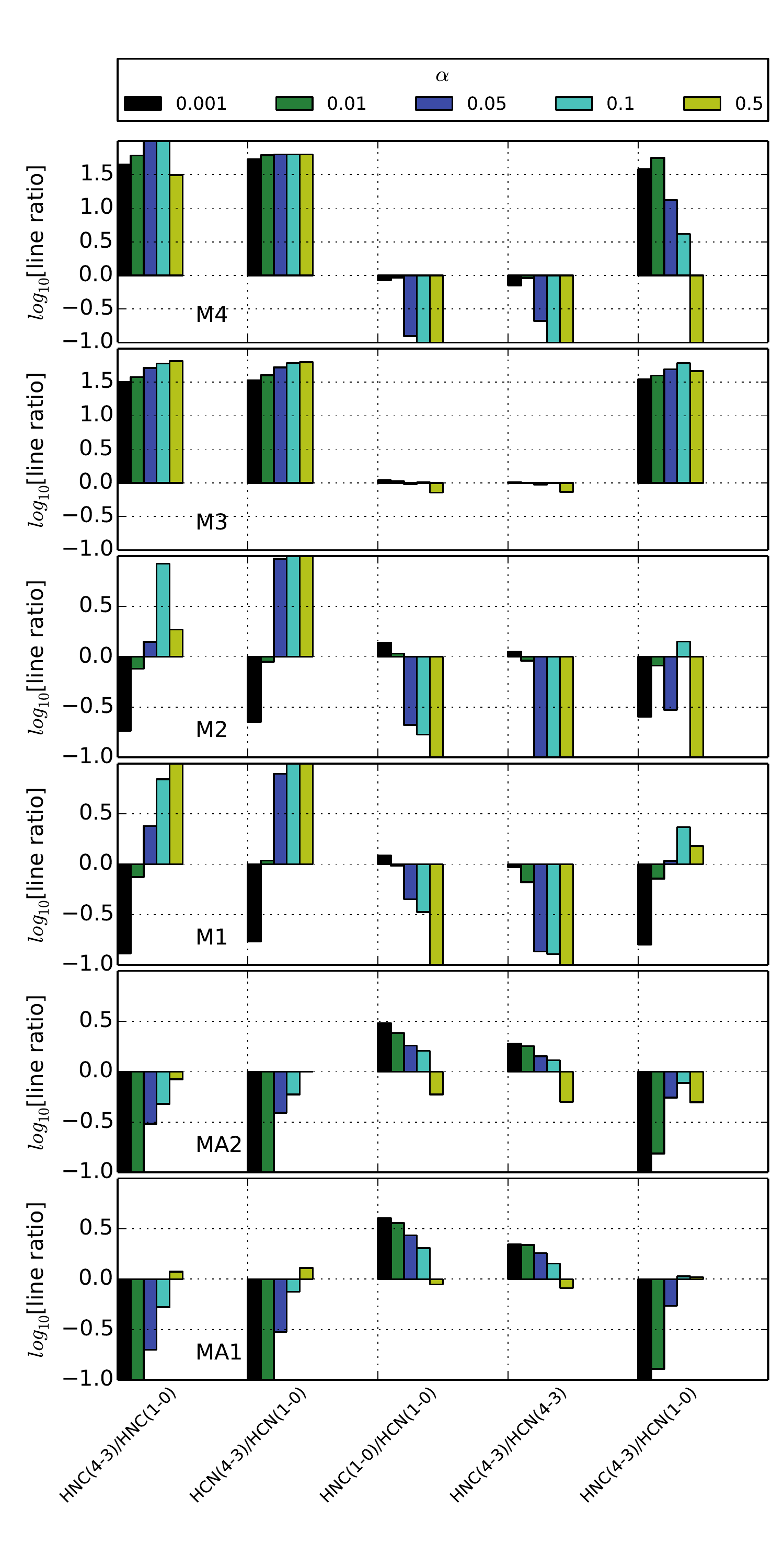}
  \end{minipage}
  \caption{Various line ratios of HNC and HCN for the reference models (see Table-\ref{tbl:refModels}) 
    as a function of \gm~ for $Z=$~0.1~\zsun~(left), 
    0.5~\zsun~(middle) and 2.0~\zsun~(right)
    \label{fig:bar-plot-lineRatios-HNC-HCN-z-other}}
\end{figure*}


\begin{figure*}[!tbh]
  \centering
  \begin{minipage}[b]{1.0\linewidth} 
  \centering
  \includegraphics[scale=0.6]{HCO+1-0-base.pdf}
  \includegraphics[scale=0.6]{HCO+1-0-base.pdf}
  \end{minipage}
  \newline
  \includegraphics[scale=0.7]{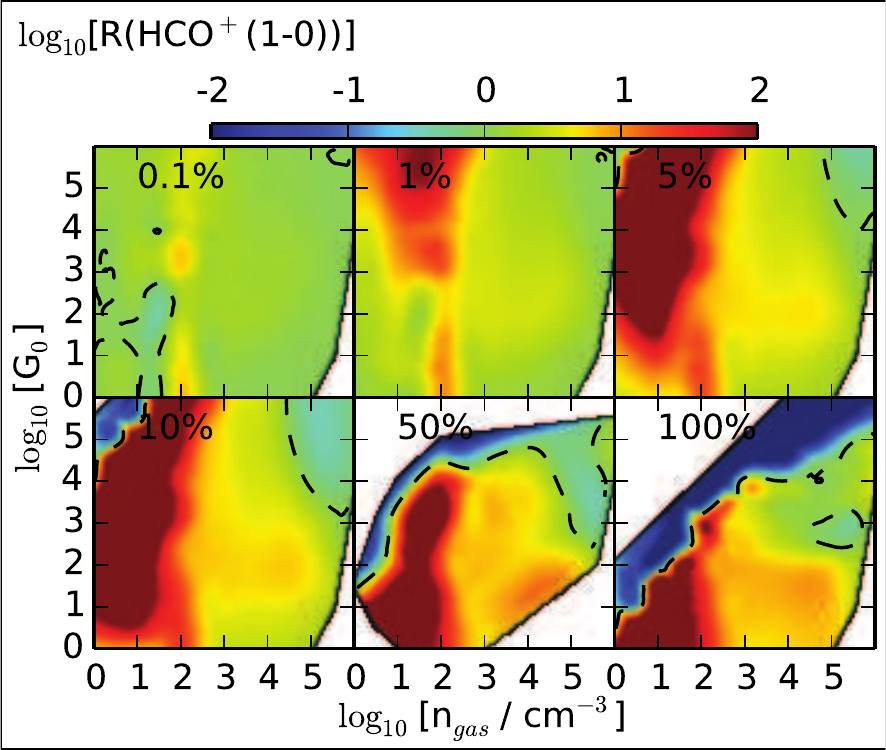}
  \includegraphics[scale=0.7]{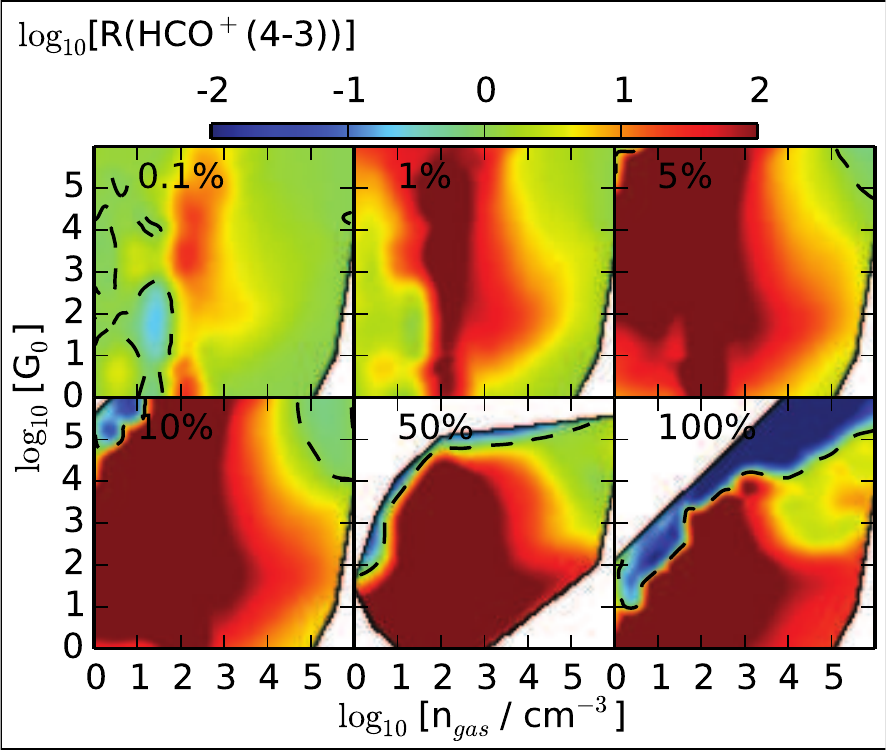}\label{fig:HCOP-4-3}
  \caption{{\bf [Top]} Emission grids of PDR models without mechanical heating for a selection of 
    HCO$^+$ transitions for $A_V = 10$~ mag ($Z=$~\zsun). {\bf [Bottom]} 
    Relative changes in the emission as a function of $\alpha$. The dashed contour traces the $R = 1$ line, 
    where the emission with and without extra heating are the same. (See caption of Figure-\ref{fig:C+-1-0-grids})
    \label{fig:HCOP-base-and-gmech} }
\end{figure*}

\begin{figure*}[!tbh]
  \begin{minipage}[b]{1.5\linewidth} 
  \includegraphics[scale=0.6]{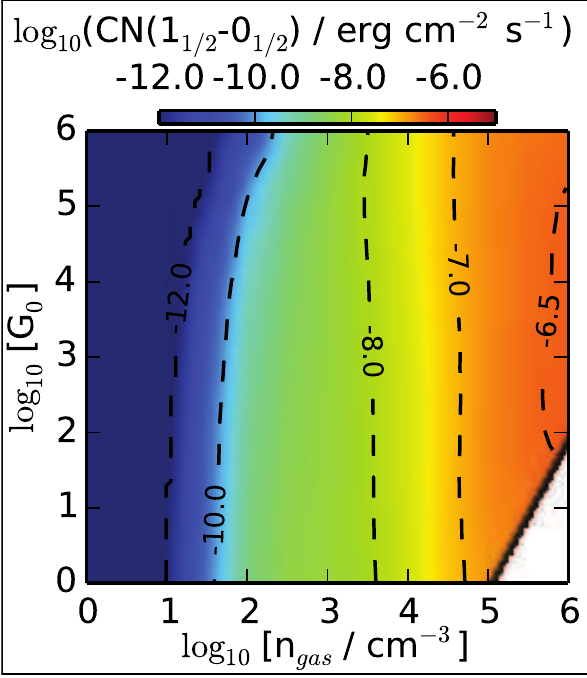}
  \includegraphics[scale=0.6]{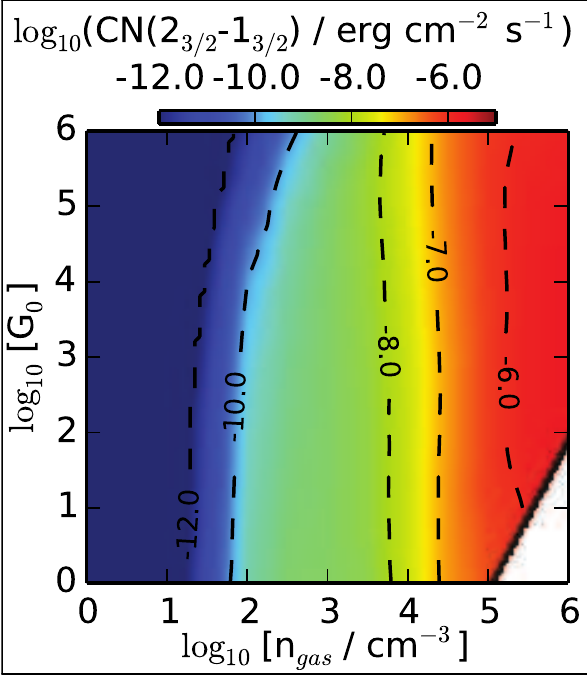}
  \includegraphics[scale=0.6]{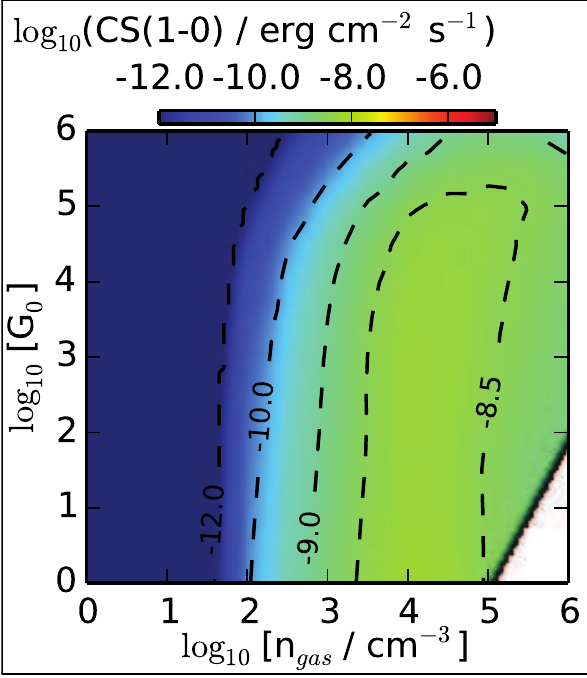}
  \includegraphics[scale=0.6]{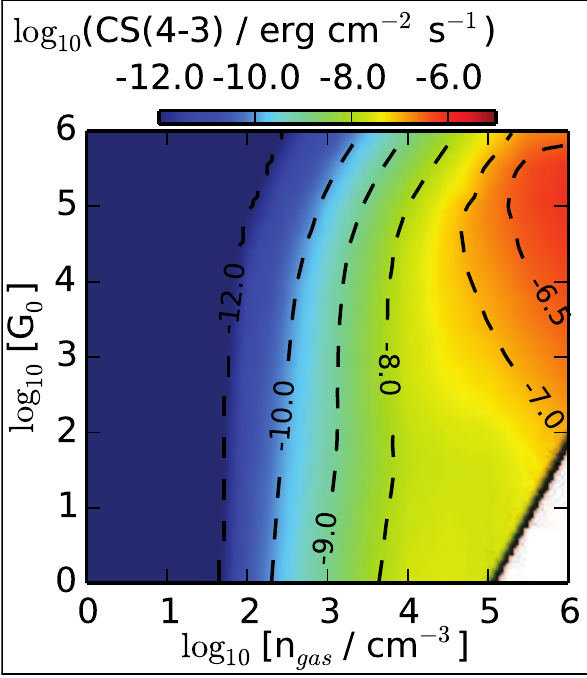}
  \end{minipage}
  \begin{minipage}[b]{1.5\linewidth} 
  \includegraphics[scale=0.68]{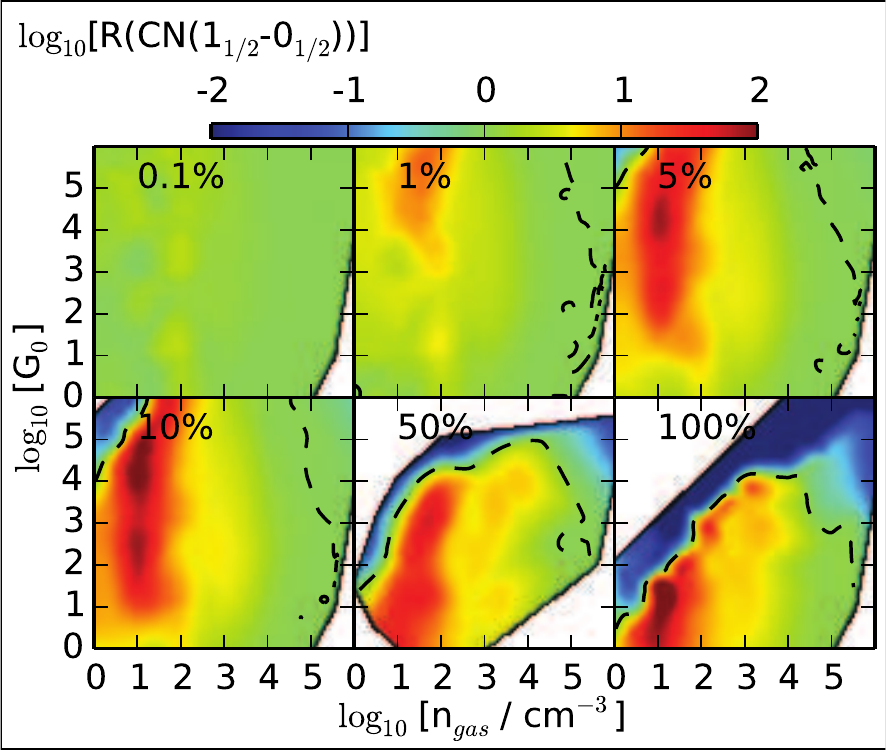}\label{fig:CN-1_0.5-0_0.5}
  \includegraphics[scale=0.68]{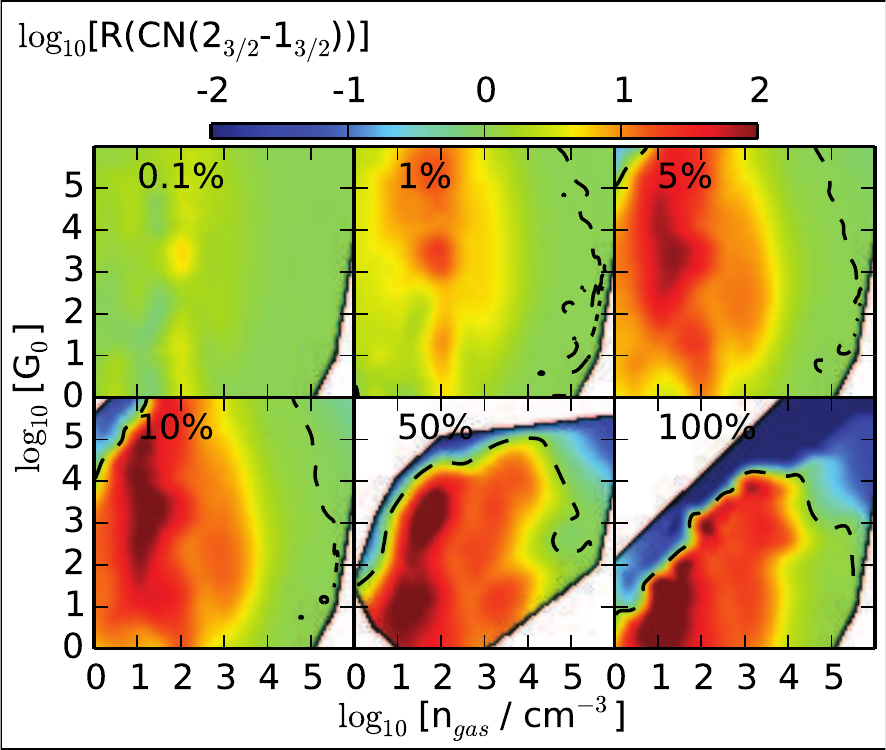}\label{fig:CN-2_1.5-1_1.5}
  \end{minipage}
  \begin{minipage}[b]{1.5\linewidth} 
  \includegraphics[scale=0.68]{CS1-0-gMech.pdf}\label{fig:CS-1-0}
  \includegraphics[scale=0.68]{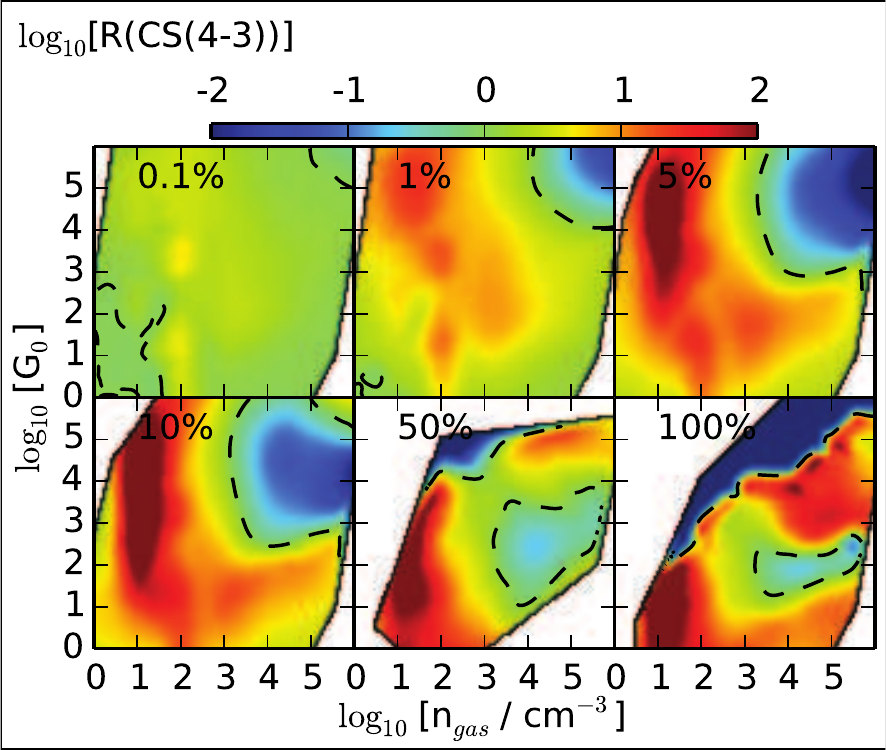}\label{fig:CS-4-3}
  \end{minipage}
  \caption{{\bf [Top]} Emission grids of PDR models without mechanical heating for a selection of 
    CN and CS transitions for $A_V = 10$~ mag ($Z=$~\zsun). {\bf [Bottom]} 
    Relative changes in the emission as a function of $\alpha$. The dashed contour traces the $R = 1$ line, 
    where the emission with and without extra heating are the same. (See caption of Figure-\ref{fig:C+-1-0-grids})
    \label{fig:CS-CN-base-and-gmech} }
\end{figure*}

\begin{figure*}[!tbh]
  \begin{minipage}[b]{1.5\linewidth} 
  \includegraphics[scale=0.39]{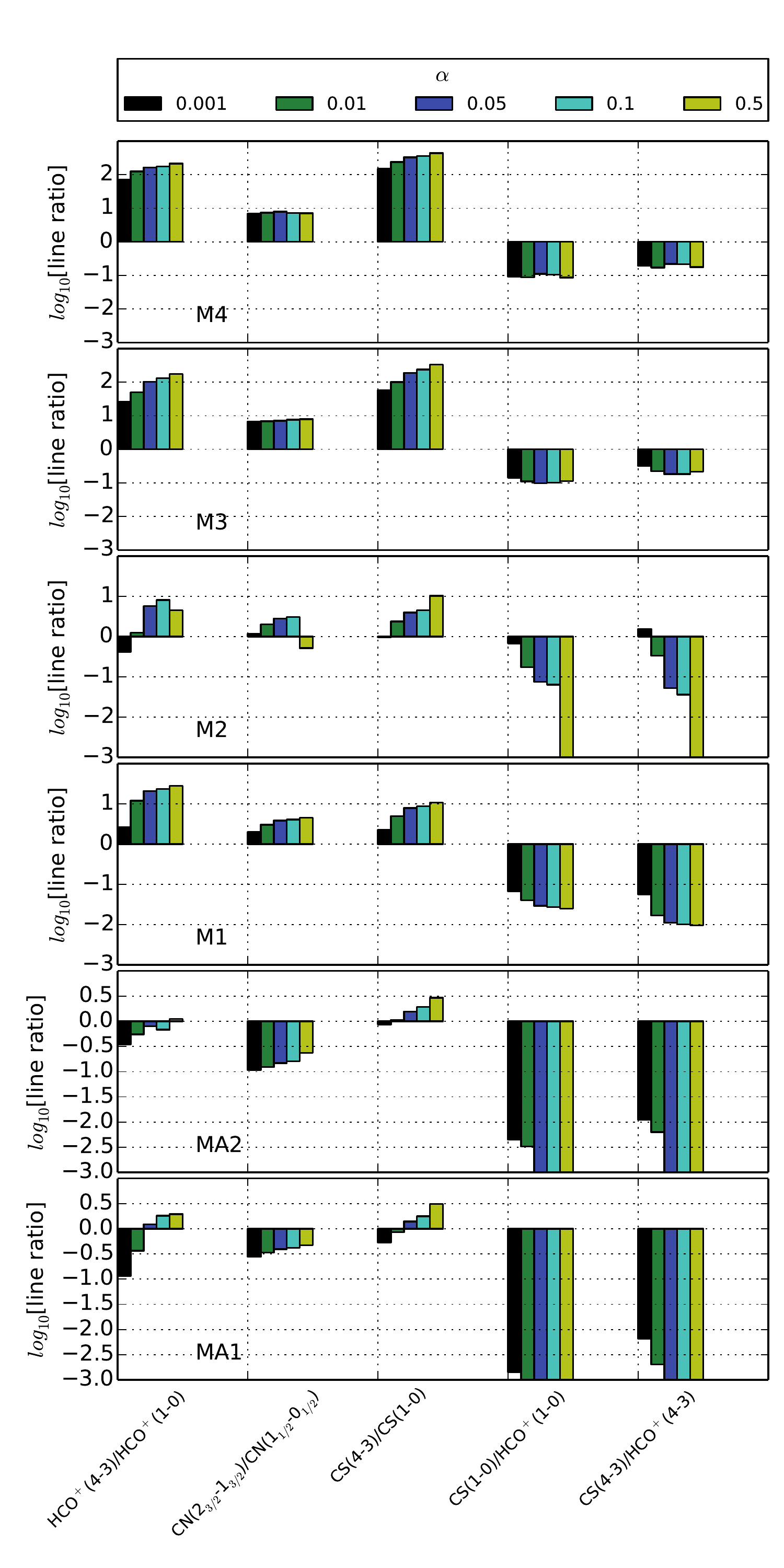}
  \includegraphics[scale=0.39]{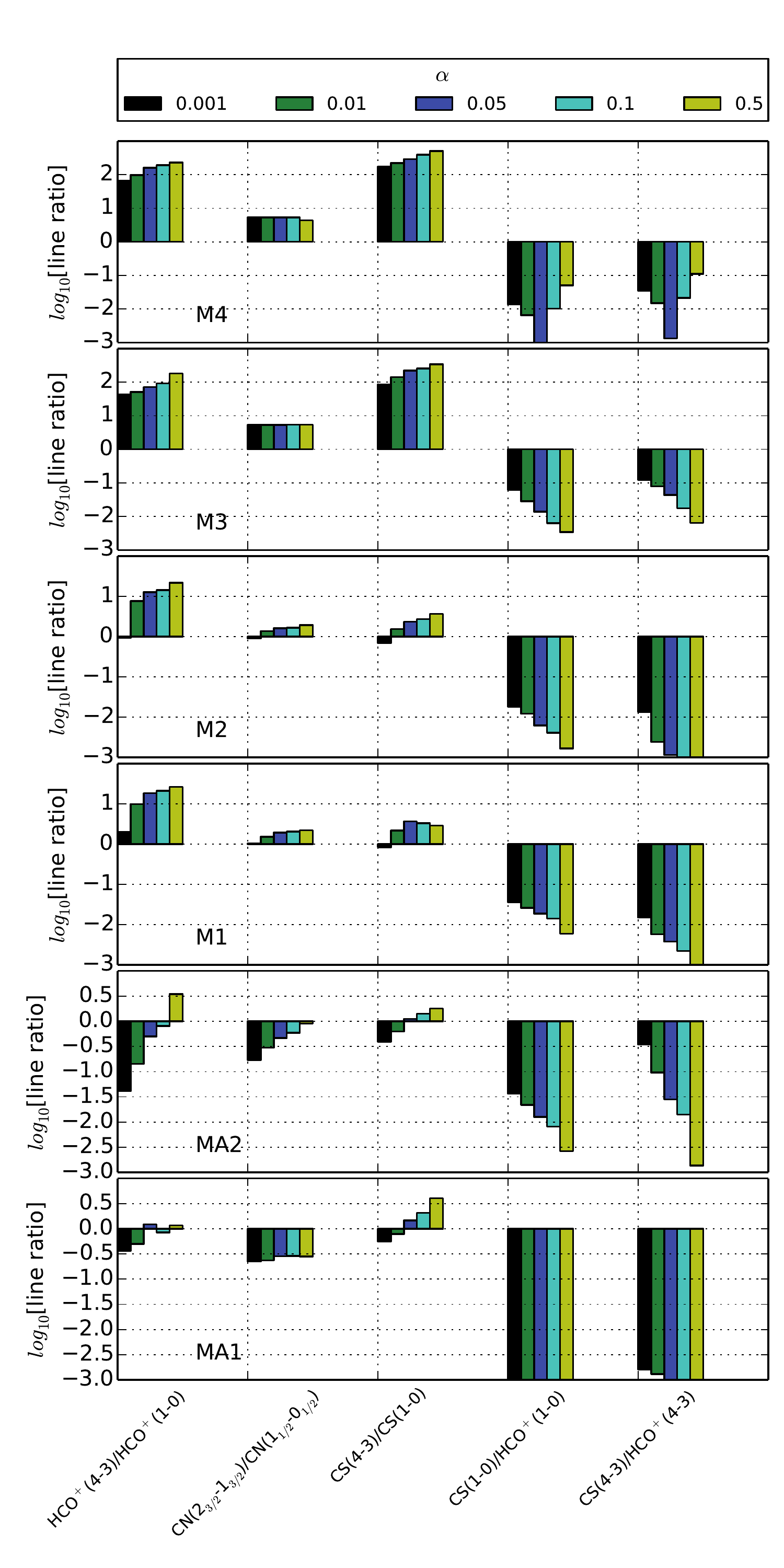}
  \includegraphics[scale=0.39]{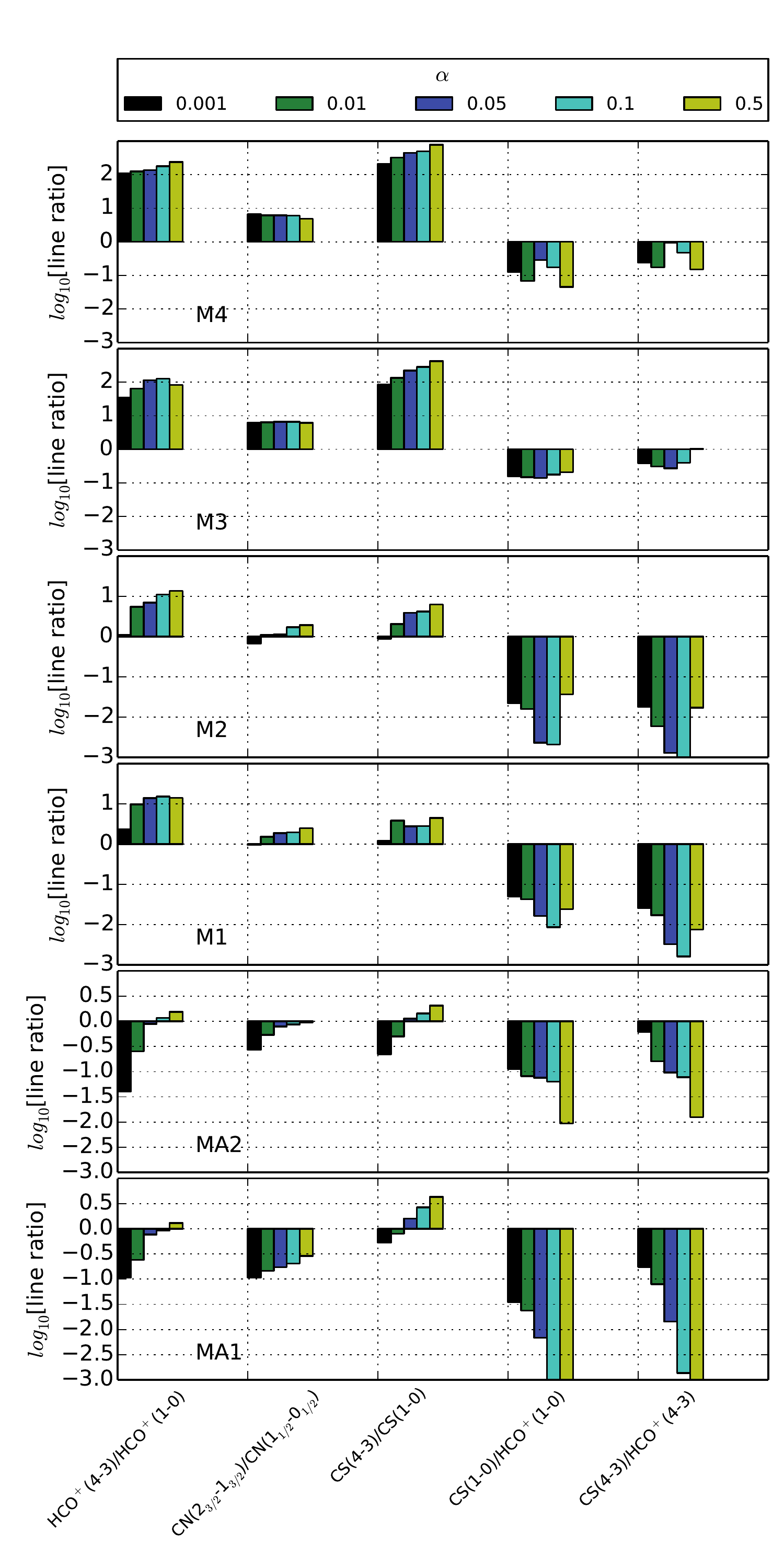}
  \end{minipage}
  \caption{Various line ratios of HCO$^+$, CS and CN for the reference models (see Table-\ref{tbl:refModels}) 
    as a function of \gm~for $Z=$~0.1~\zsun~(left), 
    0.5~\zsun~(middle) and 2.0~\zsun~(right).
    \label{fig:bar-plot-lineRatios-HCO+-CN-CS-z-other}}
\end{figure*}

\begin{figure*}[!tbh]
  \begin{minipage}[b]{1.5\linewidth} 
  \includegraphics[scale=0.39]{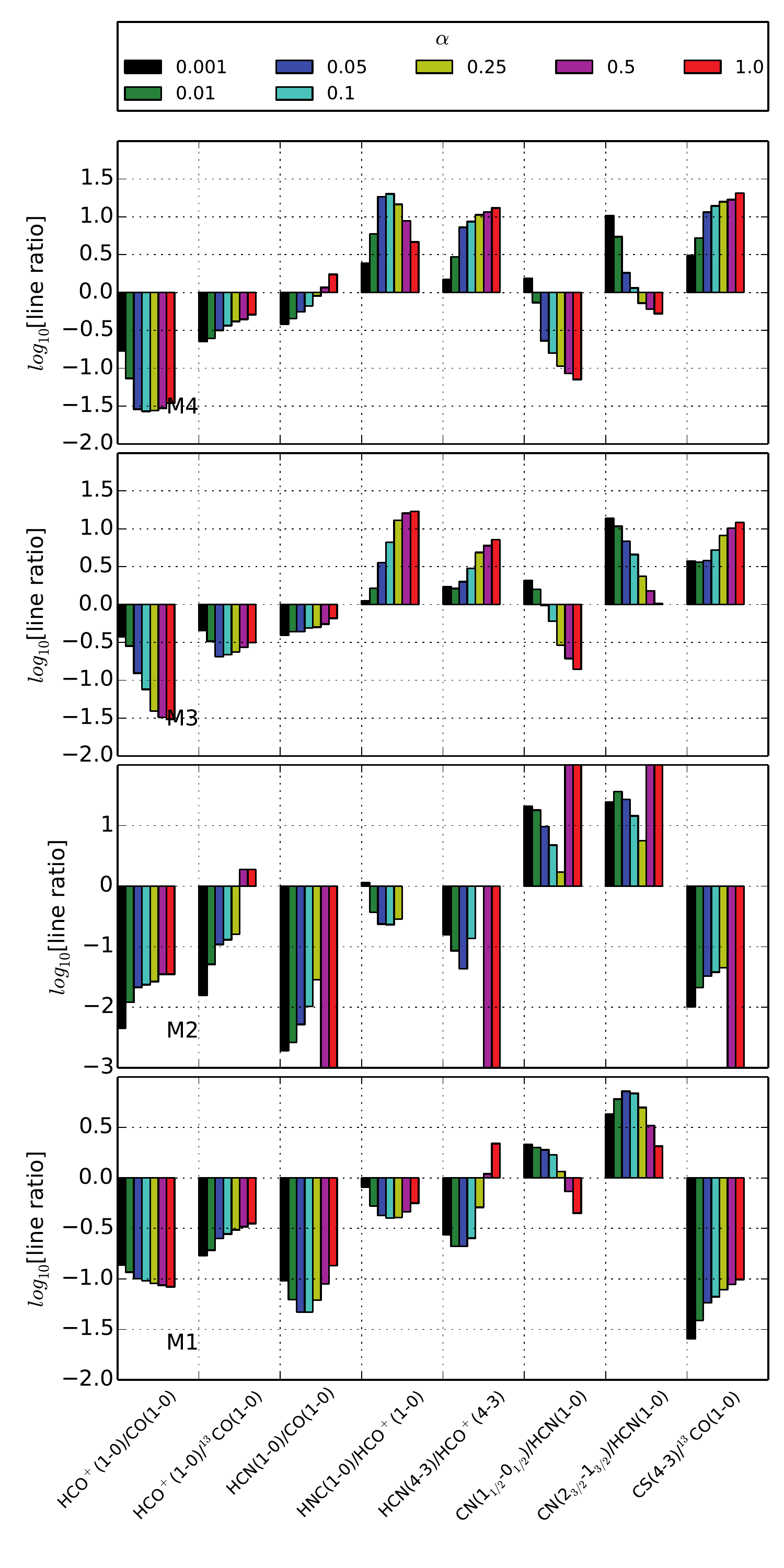}
  \includegraphics[scale=0.39]{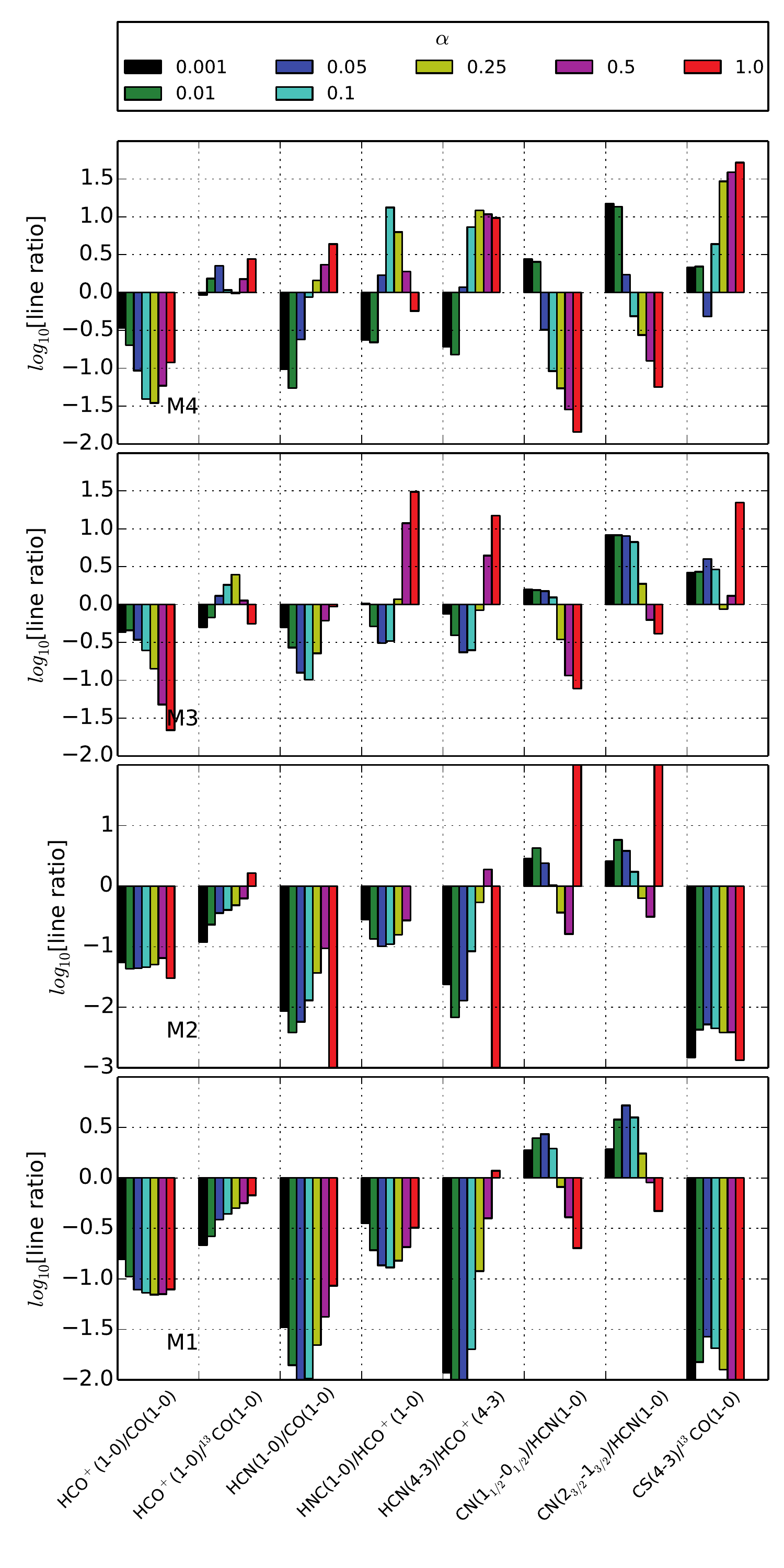}
  \includegraphics[scale=0.39]{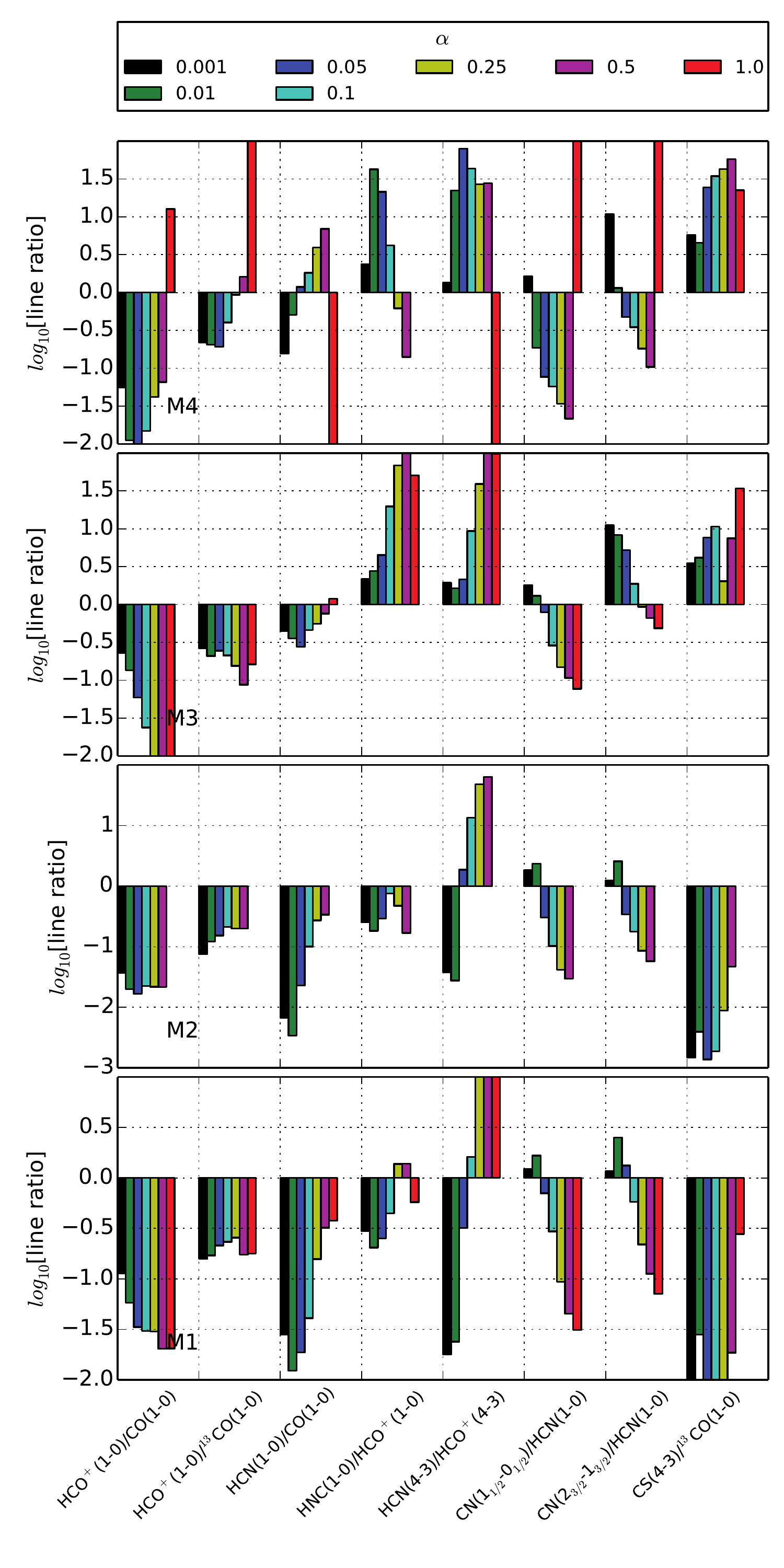}
  \end{minipage}
  \caption{Various line ratios (at $A_V = 10$~ mag) of HCN, HNC,
    HCO$^+$, CN, and CS which show a strong dependence on \gm~ for the
    reference models (see Table-\ref{tbl:refModels}) for different
    metallicities, $Z=$~0.1~\zsun~(left), 0.5~\zsun~(middle) and
    2.0~\zsun~(right).
\label{fig:bar-plot-lineRatios-misc1}}
\end{figure*}

\end{appendix}

\clearpage

\begin{appendix}{Appendix C: Several model grids}
\setcounter{figure}{0}
\makeatletter
\renewcommand{\thefigure}{C\@arabic\c@figure} 

\begin{figure*}[!tbh]
  \begin{minipage}[b]{1.0\linewidth} 
  \centering
  \includegraphics[scale=1.0]{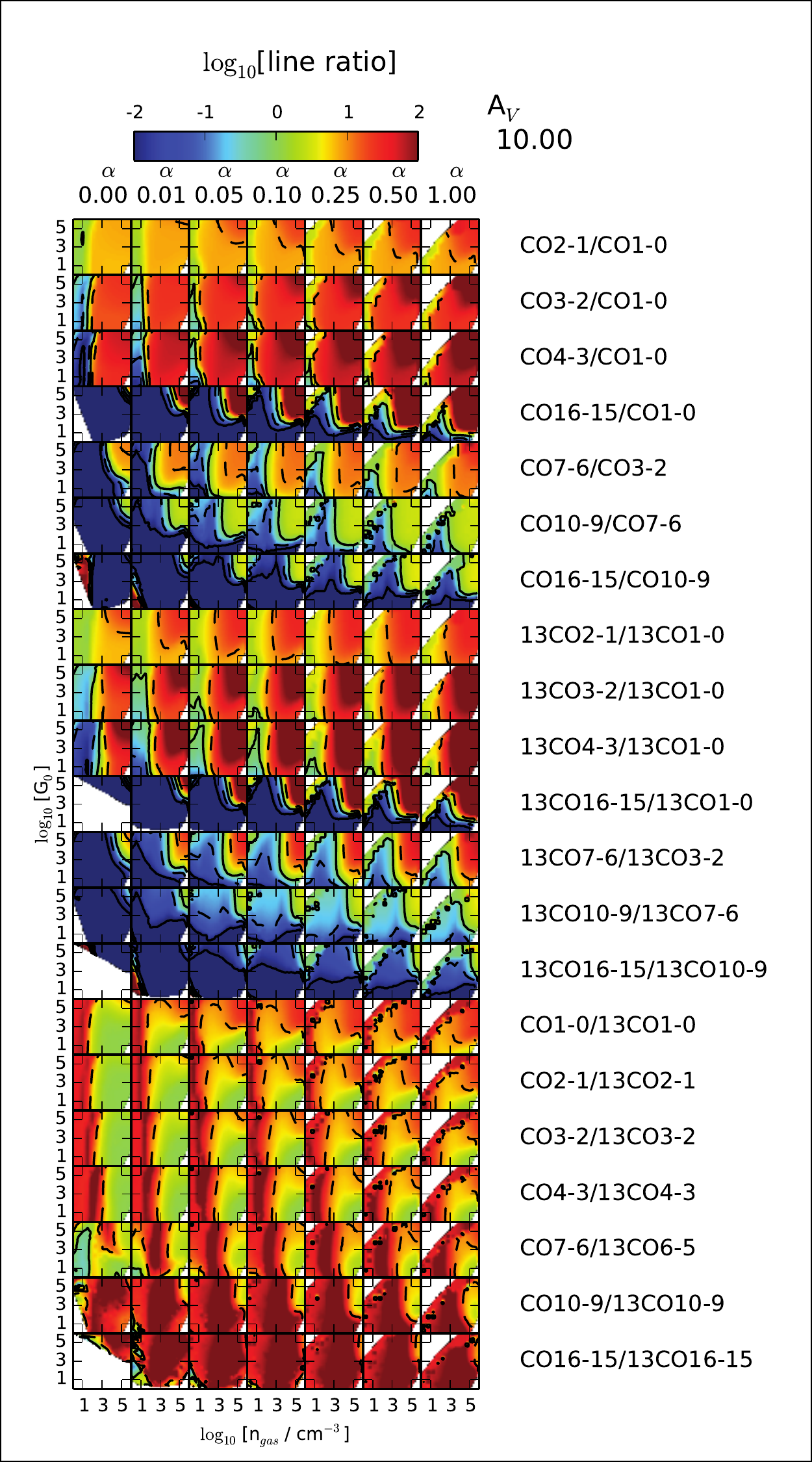}
  \end{minipage}
  \caption{Grids of CO and $^{13}$CO line ratios for different values
    of $\alpha$ for $A_V=10$~mag.\label{CO-13CO-grid-grids}}
\end{figure*}

\begin{figure*}[!tbh]
  \begin{minipage}[b]{1.0\linewidth} 
  \centering
  \includegraphics[scale=1.0]{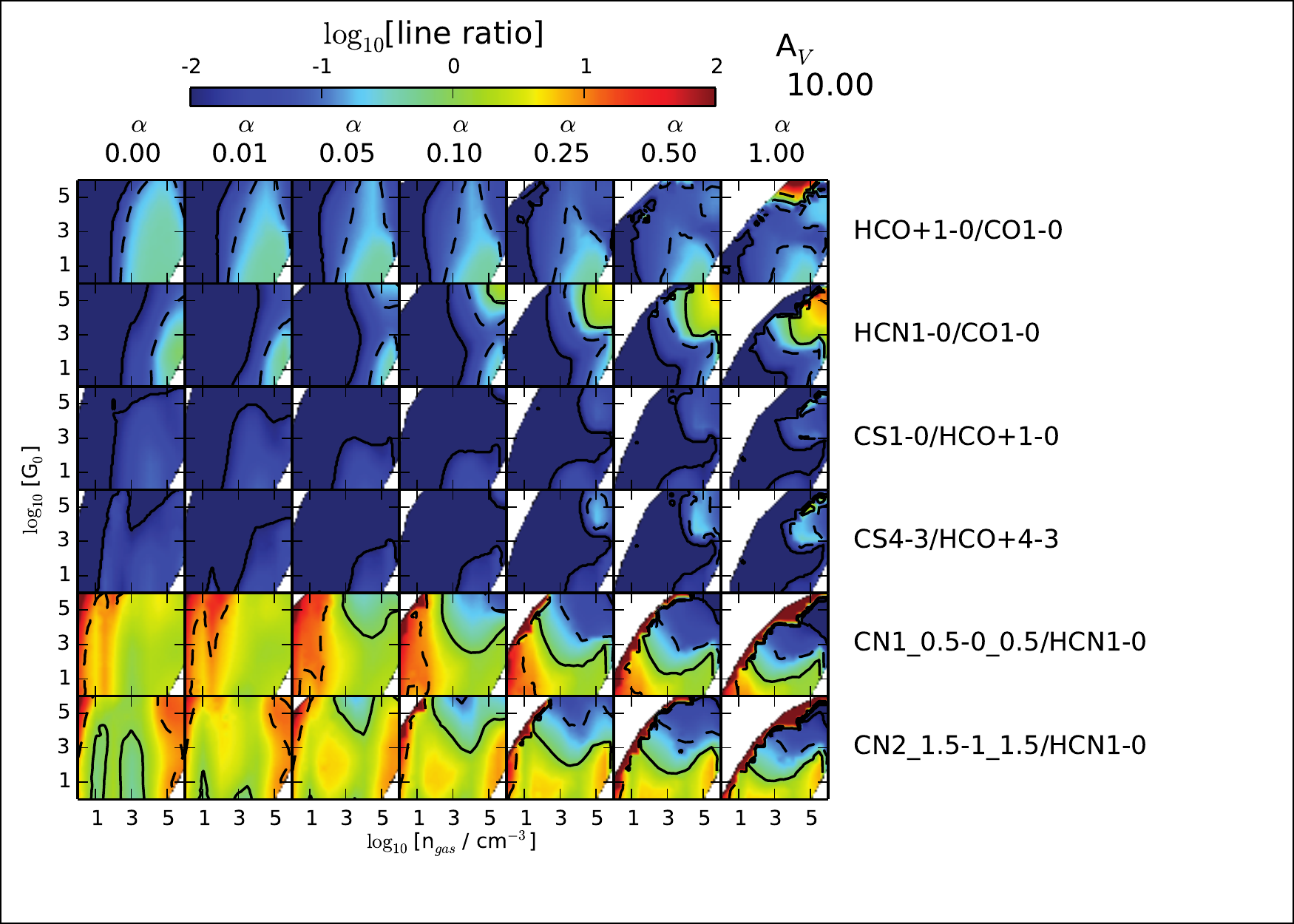}
  \end{minipage}
  \caption[angle=90]{Grids of various combination of molecular line
    ratios for different values of $\alpha$ and
    $A_V=10$~mag. \label{various-grid-grids}}
\end{figure*}

\begin{figure*}[!tbh]
  \begin{minipage}[b]{1.0\linewidth} 
    \centering
    \includegraphics[scale=0.75, angle=90]{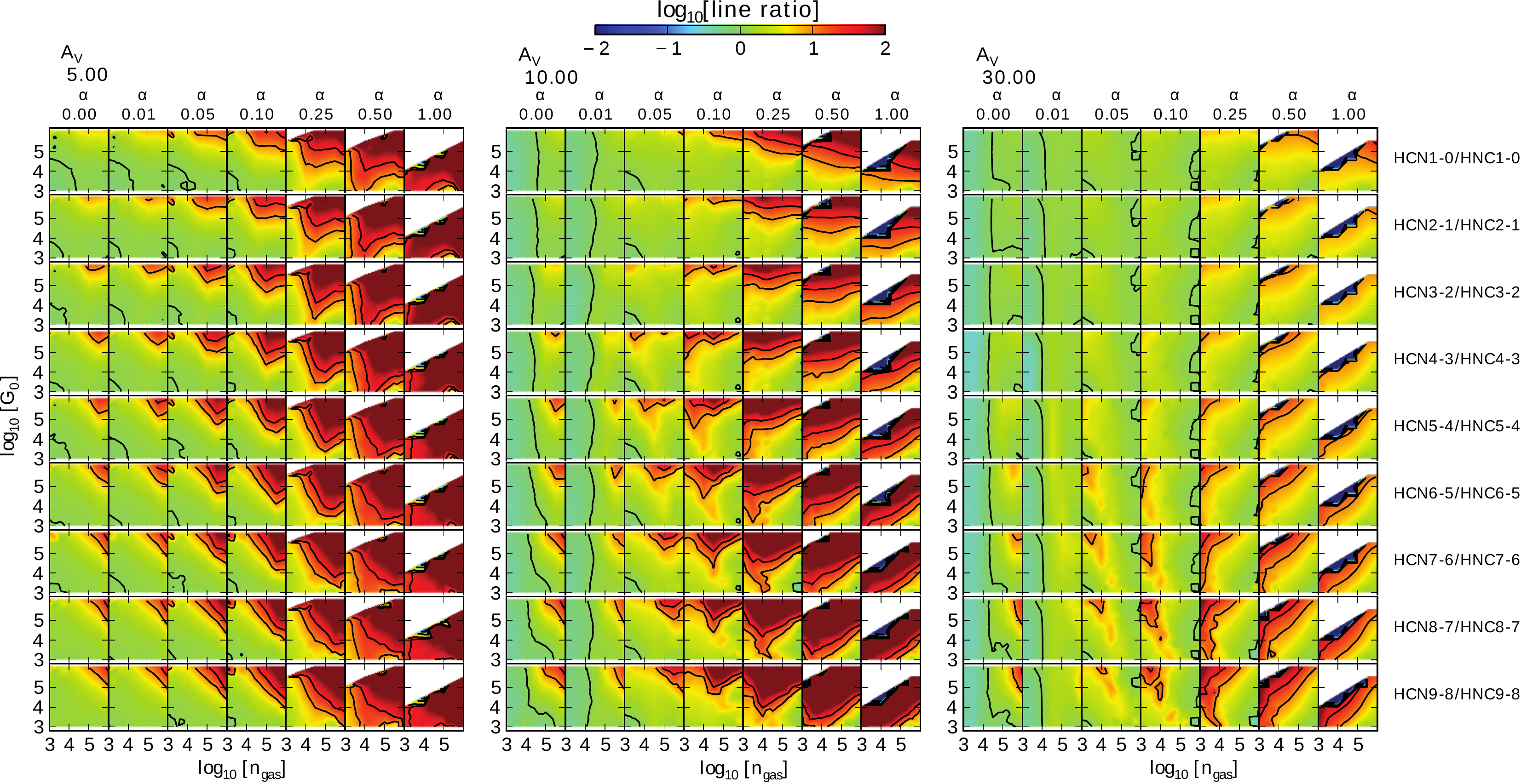}
  \end{minipage}
  \caption[angle=90]{Grids of HCN and HNC line ratios for different
    values of $\alpha$ and $A_V$. \label{HCN-HNC-grid-grids}}
\end{figure*}

\begin{figure*}[!tbh]
  \begin{minipage}[b]{1.0\linewidth} 
    \centering
    \includegraphics[scale=0.75, angle=90]{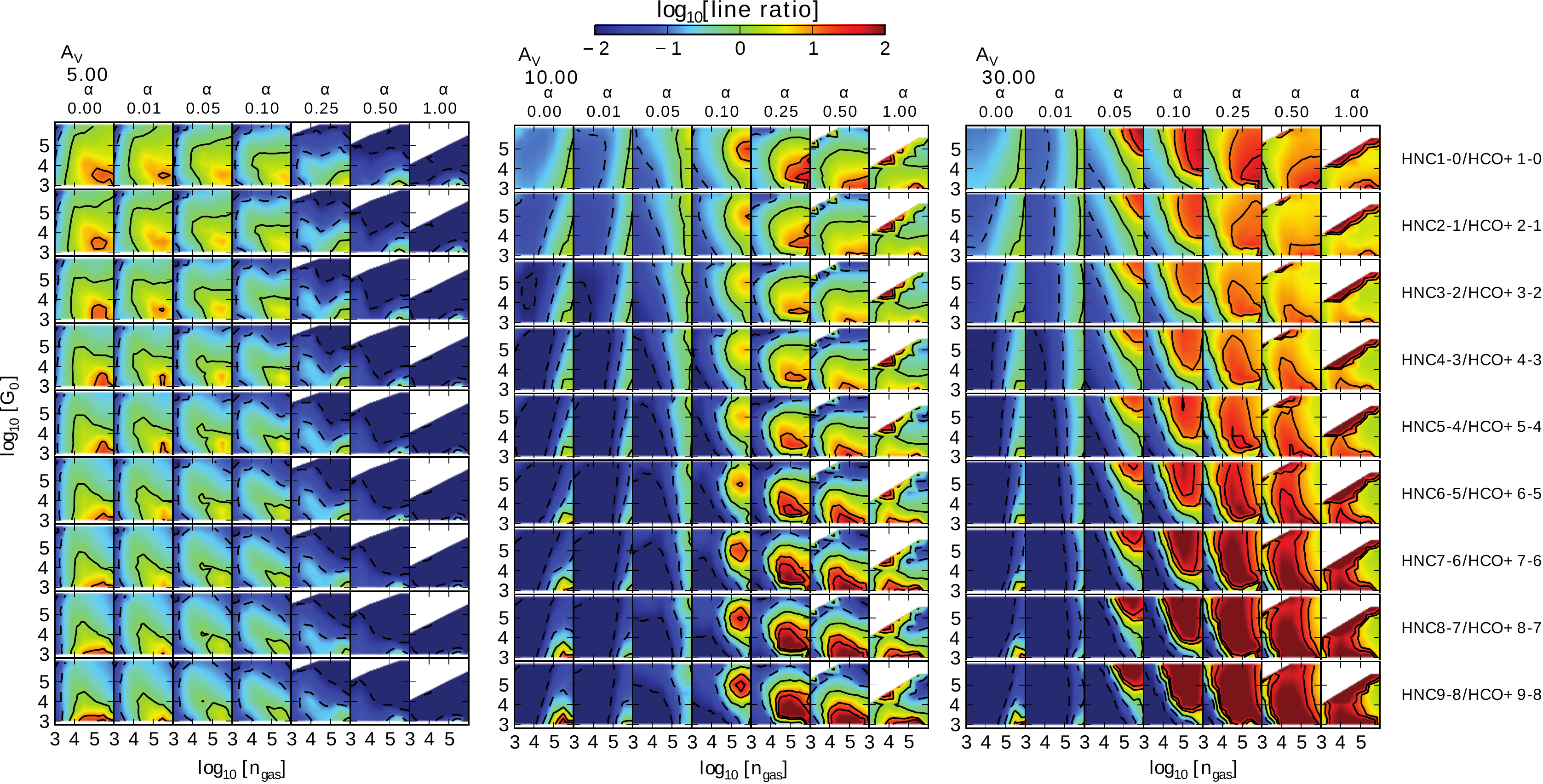}
  \end{minipage}
  \caption[angle=90]{Grids of HNC and HCO$^+$ line ratios for
    different values of $\alpha$ and
    $A_V$. \label{HNC-HCO+-grid-grids}}
\end{figure*}

\begin{figure*}[!tbh]
  \begin{minipage}[b]{1.0\linewidth} 
    \centering
    \includegraphics[scale=0.7, angle=90]{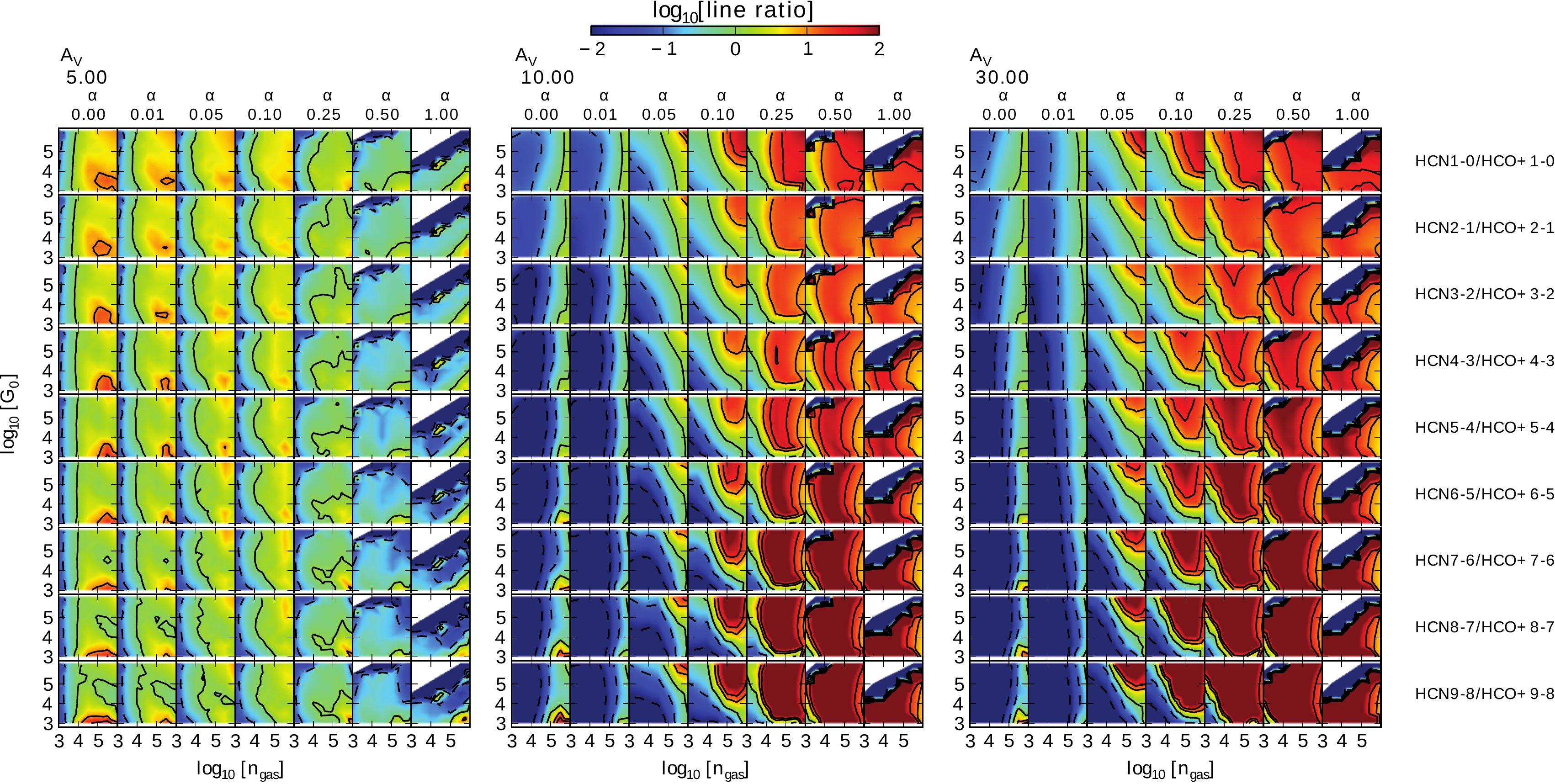}
  \end{minipage}
  \caption[angle=90]{Grids of HCN and HCO$^+$ line ratios for
    different values of $\alpha$ and
    $A_V$. \label{HCN-HCO+-grid-grids}}
\end{figure*}

\begin{figure*}[!tbh]
  \begin{minipage}[b]{1.0\linewidth} 
    \centering
    \includegraphics[scale=0.8, angle=90]{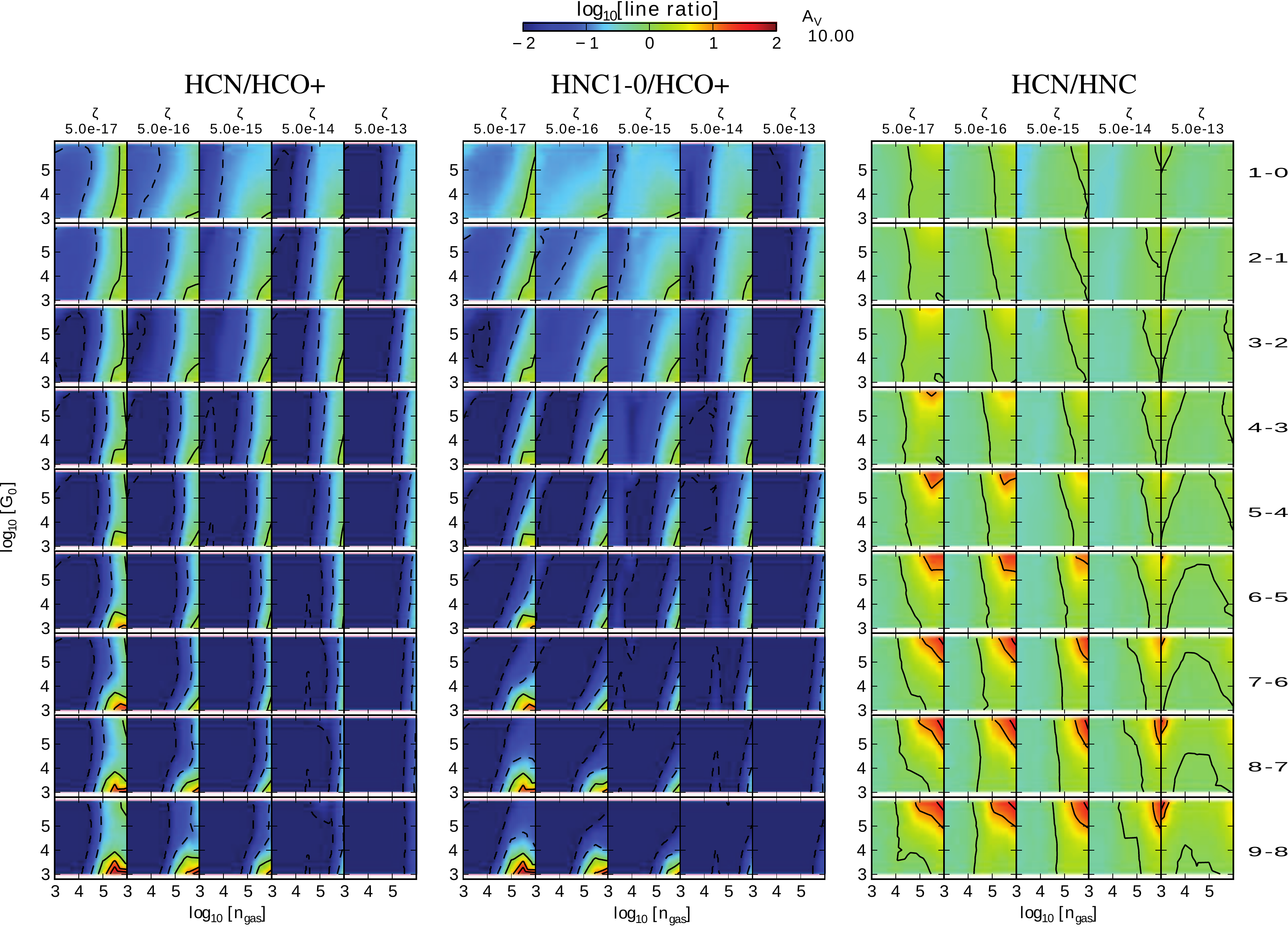}
  \end{minipage}
  \caption[angle=90]{Grids of HCN, HNC and HCO$^+$ line ratios
    ($J=1-0$ and $8-7$) for different cosmic ray ionization rates with
    $A_V=10$~mag. \label{CR-grid-grids}}
\end{figure*}

\begin{figure*}[!tbh]
  \begin{minipage}[b]{1.0\linewidth} 
    \centering
    \includegraphics[scale=0.9]{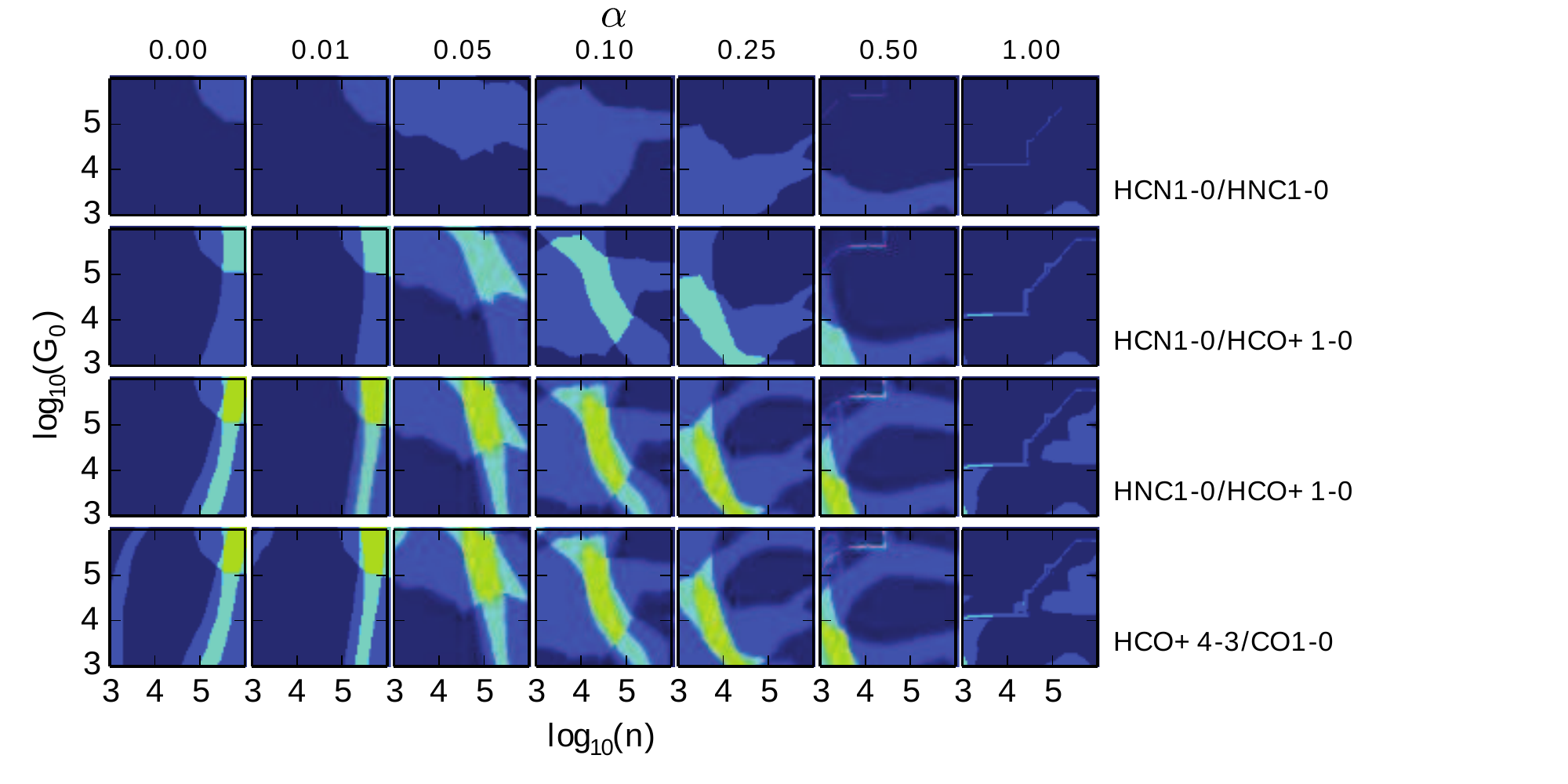} 
  \end{minipage}
  \caption{Constraining the \gm~, $A_V$, $n$ and \go~ for star-burst
    galaxies. Illustration for $A_v = 10$~mag. See caption of
    Figure-\ref{fig:constraining} for
    details\label{fig:constraining10}.}
\end{figure*}

\begin{figure*}[!tbh]
  \begin{minipage}[b]{1.0\linewidth} 
    \centering
    \includegraphics[scale=0.9]{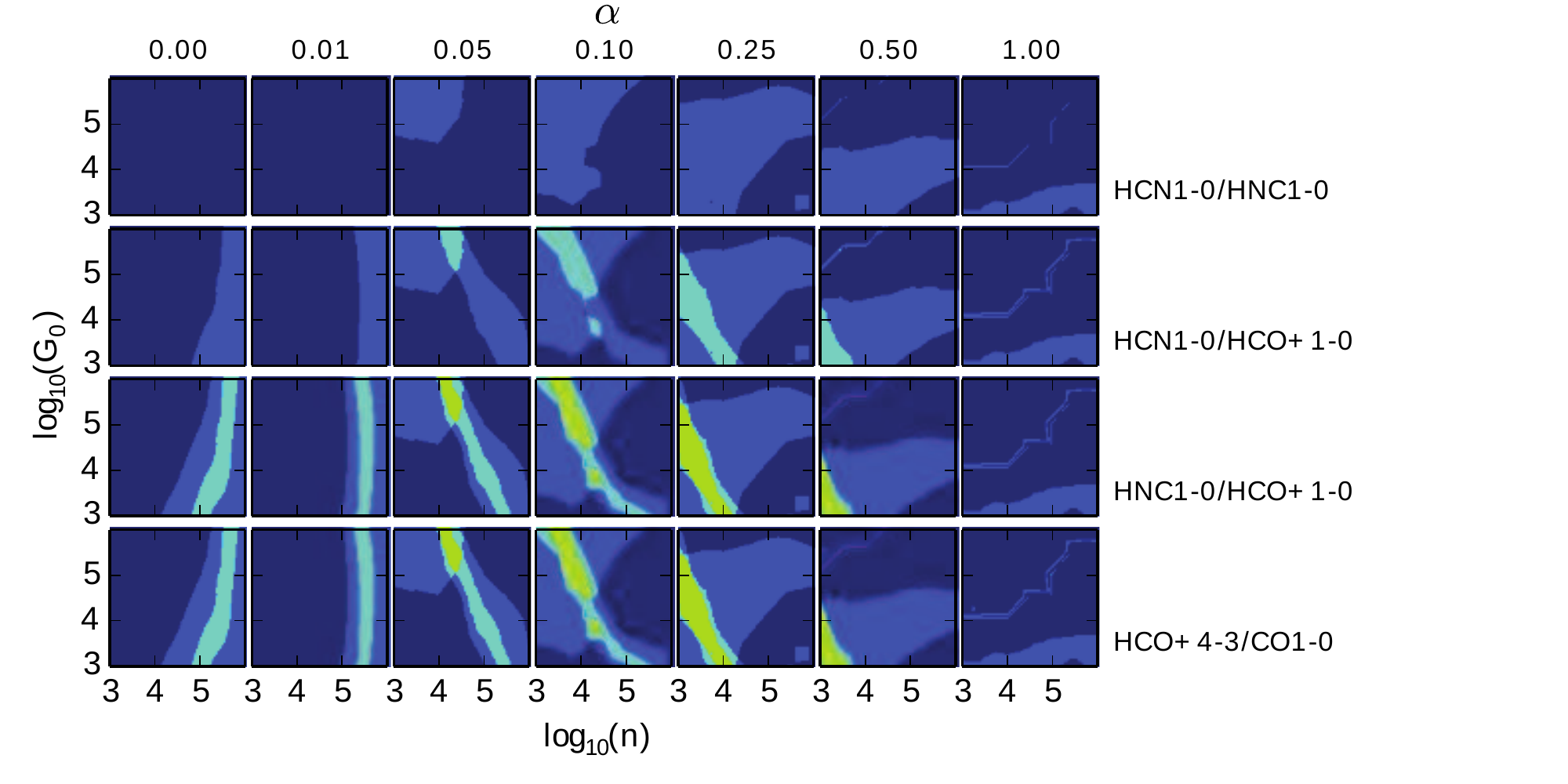} 
  \end{minipage}
  \caption{Constraining the \gm~, $A_V$, $n$ and \go~ for star-burst
    galaxies. Illustration for $A_v = 30$~mag. See caption of
    Figure-\ref{fig:constraining} for
    details\label{fig:constraining30}.}
\end{figure*}

\end{appendix}

\end{document}